\documentclass[3p]{elsarticle}

\usepackage{lineno}
\modulolinenumbers[5]

\journal{Journal}
\usepackage[fleqn]{amsmath}
\usepackage{mathtools}
\usepackage{stmaryrd}
\usepackage{amsthm}
\usepackage{amsfonts}
\usepackage{amstext}
\usepackage{amssymb} 
\usepackage{epstopdf}

\usepackage{subcaption}
\usepackage{float}
\usepackage{multicol}
\usepackage{multirow}
\usepackage[titletoc,toc,title]{appendix}
\usepackage[colorlinks=true]{hyperref} 
\hypersetup{bookmarksnumbered,linkcolor = blue}  
\usepackage{threeparttable}

\newcommand{\heqref}[1]{Eq.~\hyperref[#1]{\eqref{#1}}} 
\newcommand{\heqsref}[2]{Eqs.~\hyperref[#1]{\eqref{#1}} and~\hyperref[#2]{\eqref{#2}}} 
\newcommand{\hfigref}[1]{\hyperref[#1]{Fig.~\ref{#1}}}
\newcommand{\hfigsref}[2]{\hyperref[#1]{Figs.~\ref{#1}} and~\hyperref[#2]{\ref{#2}}}
\newcommand{\happref}[2]{\hyperref[#1]{Appendix~#2}}
\newcommand{\htablref}[1]{\hyperref[#1]{Table~\ref{#1}}}
\newcommand{\htablsref}[2]{\hyperref[#1]{Tables~\ref{#1}} and~\hyperref[#2]{\ref{#2}}}
\newcommand{\hccite}[1]{\hyperref[#1]{\cite{#1}}}
\newcommand{\secref}[1]{Section~\ref{#1}}

\begin{document}

\begin{frontmatter}

\title{Fourier Analysis and Evaluation of DG, FD and Compact Difference Methods for Conservation Laws \addcontentsline{toc}{chapter}{Fourier Analysis and Evaluation of DG, FD and Compact Difference Methods for Conservation Laws}}

\author{Mohammad~Alhawwary\corref{cor1}} 
\ead{mhawwary@ku.edu}
\cortext[cor1]{Corresponding author}
\author{Z.J.~Wang\corref{cor2}}
\ead{zjw@ku.edu}
\address{Department of Aerospace Engineering, University of Kansas, Lawrence, KS~66045, USA}

\begin{abstract}
Large eddy simulation (LES) has been increasingly used to tackle vortex-dominated turbulent flows. In LES, the quality of the simulation results hinges upon the quality of the numerical discretizations in both space and time. It is in this context we perform a Fourier analysis of several popular methods in LES including the discontinuous Galerkin (DG), finite difference (FD), and compact difference (CD) methods. We begin by reviewing the semi-discrete versions of all methods under-consideration, followed by a fully-discrete analysis with explicit Runge-Kutta (RK) time integration schemes. In this regard, we are able to unravel the  true dispersion/dissipation behavior of DG and Runge-Kutta DG (RKDG) schemes for the entire wavenumber range. The physical-mode is verified to be a good approximation for the asymptotic behavior of these DG schemes in the low wavenumber range. After that, we proceed to compare the DG, FD, and CD methods in dispersion and dissipation properties. Numerical tests are conducted using the linear advection equation to verify the analysis. In comparing different methods, it is found that the overall numerical dissipation strongly depends on the time step. Compact difference (CD) and central finite difference (FD) schemes, in some particular settings, can have more numerical dissipation than the DG scheme with an upwind flux. This claim is then verified through a numerical test using the Burgers' equation. %
\end{abstract}

\end{frontmatter}

\linenumbers

\section{Introduction}\label{sec:Intro1}%
According to NASA's 2030 Vision on CFD~\cite{SlotnickCFDVision20302014}, scale-resolving simulations such as large eddy simulation (LES), will be increasingly used to compute challenging vortex-dominated turbulent flow problems. Multiple international workshops on high-order CFD methods~\cite{WangHighorderCFDmethods2013} have conclusively demonstrated the advantage of high-order methods over 1st and 2nd order ones in accuracy/efficiency for such scale-resolving simulations due to their lower dispersion and dissipation errors. Interested readers can refer to several review articles on high-order methods~\cite{WangHighordermethodsEuler2007,HuynhHighordermethodscomputational2014,Wangperspectivehighordermethods2016,Wangreviewfluxreconstruction2016}.

 It was previously shown that some upwind-biased FD schemes are too dissipative~\cite{MittalSuitabilityUpwindBiasedFinite1997,LarssonEffectnumericaldissipation2007,JohnsenAssessmenthighresolutionmethods2010,KawaiAssessmentlocalizedartificial2010} to be a viable numerical approach for LES. Unfortunately, some researchers extrapolated from this to dismiss any kind of "upwinding" in numerical methods including DG-type methods, and argue for non-dissipative methods such as central FD schemes for LES. However, time integration schemes such as the Rung-Kutta (RK) scheme do introduce numerical dissipation. It is therefore very important to analyze  the fully-discretized versions to obtain  an accurate description on the amount of dispersion and dissipation errors. The main objective of the present study is to compare the fully-discretized DG, FD and CD methods.   
 
 In the context of LES, there is physical dissipation associated with the molecular viscosity. In addition, there is dissipation associated with the sub-grid-scale (SGS) stress, and finally there is numerical dissipation associated with the selected numerical method. The actual SGS stress obviously depends on the quality of the SGS model. Sometimes, the SGS stress provided by a model does not really correlate well with the physical SGS stress~\cite{Liprioriposteriorievaluations2016}. In this case, the role of the SGS model is to stabilize the simulation. For a central FD scheme, a dissipative SGS model is essential in achieving a successful simulation. While for dissipative methods such as the DG, or compact difference (CD) scheme with a spatial filtering, it is often not necessary to include a SGS model since the numerical dissipation is sufficient to stabilize the simulation. There have been overwhelming evidence which shows that  adding a SGS model can be detrimental to the solution quality~\cite{VisbalLargeEddySimulationCurvilinear2002,KawaiAssessmentlocalizedartificial2010,BogeyLargeeddysimulations2006,GarmannComparativestudyimplicit2013} for dissipative methods. In practice, implicit LES (ILES) has been shown to perform very well for a variety of flow problems~\cite{BogeyLargeeddysimulations2006,Rizzettahighordercompactfinitedifference2008,UrangaImplicitLargeEddy2011,VermeireImplicitlargeeddy2016,ZhuImplicitLargeEddySimulation2016,Wangindustriallargeeddy2017}. 

In order to assess the dispersion/dissipation characteristics and resolution of a  numerical scheme, Fourier analysis~\cite{VichnevetskyFourierAnalysisNumerical1982} is often utilized either in a semi-discrete~\cite{HuAnalysisDiscontinuousGalerkin1999,VandenAbeeleDispersiondissipationproperties2007,VandenAbeeleStabilityAccuracySpectral2008,VincentInsightsNeumannanalysis2011,GassnerComparisonDispersionDissipation2011} or fully discrete setting~\cite{YangDispersionDissipationErrors2013,MouraLineardispersiondiffusion2015,Vermeirebehaviourfullydiscreteflux2017,VanharenRevisitingspectralanalysis2017}. In our present work, we start with a review of  semi-discrete schemes, and then proceed to analyze the fully discrete schemes assuming a periodic boundary condition. For high-order DG-type methods, most of the previous work studied the behavior of high-order schemes based on what is called the physical-mode~\cite{HuAnalysisDiscontinuousGalerkin1999,VandenAbeeleDispersiondissipationproperties2007,VincentInsightsNeumannanalysis2011} defined as the one that approximates the exact dispersion relation for a range of wavenumbers while regarding other modes as spurious. Recently, Moura et al.~\cite{MouraLineardispersiondiffusion2015} provided new interpretations on the role of spurious or secondary modes. In their work, these modes are replicates of the physical-mode along the wavenumber axis and they improve the accuracy of the scheme. Vanharen et al.~\cite{VanharenRevisitingspectralanalysis2017} concluded that after a large number of iterations, high-order schemes behave in dispersion and dissipation according to the physical-mode asymptotically, for wavenumbers less than $\pi$. Nevertheless, the complete behavior of DG-type high-order schemes in dispersion and dissipation based on all eigenmodes has not been studied before. In this paper, we provide a first attempt to achieve this goal. 

Whilst there exists abundant work on the analysis of both high-order and low-order schemes or classical finite difference/finite-volume schemes, little attention was given to comparing the DG, FD, and CD schemes of the same order of accuracy. The DG method, originally introduced by Reed and Hill~\cite{ReedTriangularMeshMethods1973} to solve the neutron transport equation, is chosen in this study as a representative of the high-order polynomial-based methods capable of handling unstructured grids including the spectral difference (SD)~\cite{LiuSpectraldifferencemethod2006}, and the flux reconstruction (FR) or correction procedure via reconstruction (CPR) methods~\cite{HuynhFluxReconstructionApproach2007}. LaSaint and Raviart~\cite{LasaintFiniteElementMethod1974} performed an error analysis for the DG method. It was then further developed for convection-dominated problems and fluid dynamics by many researches, see for example(\cite{CockburnDiscontinuousGalerkinMethods2000,CockburnRungeKuttaDiscontinuous2001,BassiHighOrderAccurateDiscontinuous1997a,BassiHighOrderAccurateDiscontinuous1997,BassiHigherorderaccuratediscontinuous1997,HesthavenNodalDiscontinuousGalerkin2010,ShuHighorderWENO2016}) and the references therein. In addition, the compact difference (CD) method of Lele~\cite{LeleCompactfinitedifference1992} is also analyzed for comparison purposes. This method was further developed by Gaitonde et al.~\cite{GaitondeFurtherdevelopmentNavierStokes1999} and Visbal et al.~\cite{VisbalHighOrderAccurateMethodsComplex1999}, and applied successfully to perform ILES by Visbal et al.~\cite{VisbalUseHigherOrderFiniteDifference2002,VisbalLargeEddySimulationCurvilinear2002} and Rizzetta et al.~\cite{Rizzettahighordercompactfinitedifference2008}. Recently, a comparative study of the suitability of the method for ILES versus SGS was conducted by Garmann et al.~\cite{GarmannComparativestudyimplicit2013} and San has utilized the method for an analysis of low-pass filters for the approximate deconvolution closure~\cite{SanAnalysislowpassfilters2016}. Finally, we have also included central and upwind-biased FD schemes in the present comparison to illustrate the performance of a fundamental classical method.  

For time integration, we focus on the explicit Runge-Kutta (RK)~\cite{ButcherNumericalAnalysisOrdinary1987,GottliebStrongStabilityPreservingHighOrder2001,Spiterinewclassoptimal2002} method to demonstrate the importance of analyzing the fully discrete version.  Runge-Kutta schemes are easily incorporated with high-order methods such as the RKDG~\cite{CockburnRungeKuttaDiscontinuous2001} due to their ease of implementation and parallelization. 

In the present study, we first review the analysis of semi-discrete schemes, followed by a fully-discrete analysis for three classes of methods, namely, the DG, FD and CD methods. We clarify the relative efficiency and robustness of each method in terms of wave propagation properties. In addition, using a more detailed approach, we are able to compute the true dispersion/dissipation properties of DG schemes. It is verified that, at least in the low wavenumber range, the physical-mode (defined by Hu et al.~\cite{HuAnalysisDiscontinuousGalerkin1999}) can serve as a good approximation for the complete behavior of a high-order scheme.  

This paper is organized as follows.~\secref{sec:num_methods_2} introduces the basic formulations of all numerical methods under-consideration in the present study. After that, we present a semi-discrete analysis followed by a fully discrete one in~\secref{sec:Fourier_3} .~\secref{sec:DG_FD_CD_comp_4} presents the comparison of dispersion/dissipation  behavior of the DG, FD and CD schemes coupled with RK schemes. Numerical  verifications and test cases are presented in Section~\ref{sec:num_results_5}. Finally, conclusions are summarized in~\secref{sec:conclusions}.%
%
\section{Numerical Methods} \label{sec:num_methods_2}%
In this section we present the basic formulation of all the methods considered in the present study for a one-dimensional conservation law of the following form%
\begin{align}
	\frac{\partial u}{\partial t}+\frac{\partial f(u)}{\partial x} = 0,  \label{eqn:cons_law_1d} \\
    u(x,0) = u_{o}\left(x\right),  & \quad \text{I.C.} 
\end{align}%
with a periodic boundary condition. %
%
\subsection{Discontinuous Galerkin Method (DG)} \label{subsec:DG_2.1} %
%
In the DG framework, the domain $\mathcal{D}$ in one-dimension is discretized into $N_{e}$ number of non-overlapping elements $\Omega_{e} = \left[x_{e-1/2}, x_{e+1/2}\right]$, such that $\mathcal{D}\:=\: \cup_{e=1}^{N_{e}} \Omega_{e}$, and each element has a variable width of $h_{e}$ and a center point $x_{e}$. In addition, DG assumes a reference element with local coordinate $\xi \in [-1,1]$, and defines a linear mapping between the physical and reference element as follows%
\begin{equation}
  \xi =2(x-x_{e})/h_{e}.
  \label{eqn:xi_mapping}
\end{equation}

On element $\Omega_{e}$, the solution is approximated by a polynomial $u^{e}(x,t)$ of degree $p$ in space, i.e., $u^{e} \in \mathcal{P}^{p}$ which is a finite dimensional space of polynomials of degree $p$ or less. For a reference element $\Omega_{r} = [-1,1]$, the solution polynomial $u^{e}$ can be constructed as a weighted sum of some specially chosen local basis functions $\phi \in \mathcal{P}^{p}$ defined on the interval $[-1,1]$%
\begin{equation}
u^{e}(\xi,t) = \sum_{j=0}^{p} U^{e}_{j}(t) \phi_{j}(\xi)
\label{eqn:u_h_xi_phi},
\end{equation}%
where the coefficients $U_{j}$ are the unknown degrees of freedom (DOFs). In a DG formulation, the integration of the conservation law~\heqref{eqn:cons_law_1d} against a test function (that is the same as the solution basis) $\phi \in \mathcal{P}^{p}$ is required to vanish locally over any element $\Omega_{e} \in \mathcal{D}$%
\begin{equation}
	\int_{\Omega_{e}} \left(\frac{\partial u^{e}}{\partial t}+\frac{\partial f(u^{e})}{\partial x} \right) \: \phi_{l} \: dx = 0, \quad l=0, ..., p.
    \label{eqn:weak_form}
\end{equation}%
Applying integration by parts to the second term in~\heqref{eqn:weak_form}, we obtain%
\begin{equation}
	 \frac{\partial }{\partial t} \int_{x_{e-1/2}}^{x_{e+1/2}} u^{e} \: \phi_{l} \: dx \: + \: \left[f(u^{e}) \: \phi_{l} \right]^{x_{e+1/2}}_{x_{e-1/2}} - \int_{x_{e-1/2}}^{x_{e+1/2}} f(u^{e}) \: \frac{d \phi_{l}}{d x} \: dx  = 0.
    \label{eqn:DG_form_physical}
\end{equation}

In the present work, the orthogonal Legendre polynomials are employed as the basis functions, and $\phi_{l}$ is a Legendre polynomial of degree $l$. The flux $f(u^{e})|_{\partial \Omega}$ at the element boundary $\partial \Omega_{e}$ is approximated by a numerical flux function $\hat{f}\left(\hat{u}_{-},\hat{u}_{+} \right)$ to be defined later. Writing~\heqref{eqn:DG_form_physical} for a reference element results in%
\begin{equation}
	\frac{h_{e}}{2} \frac{\partial }{\partial t} \int_{-1}^{1} u^{e} \: \phi_{l} \: d\xi \: + \: \left( \hat{f} \phi_{l} \right) |_{1} -   \left( \hat{f} \phi_{l}\right) |_{-1} - \: \int_{-1}^{1} f\left(u^{e}\right)  \: \frac{d \phi_{l}}{d \xi} \: d\xi = 0,
    \label{eqn:DG_form_xi_Riemann}
\end{equation}%
which by using the definition in~\heqref{eqn:u_h_xi_phi} is essentially a system of $p+1$ equations to be solved for the unknown DOFs $U^e_{j}, \:\: j=0,...,p$. For each unknown $U^{e}_{j}$,~\heqref{eqn:DG_form_xi_Riemann} reads due to the orthogonality of the basis%
\begin{equation}
\frac{h_{e}}{2} \frac{\partial U^{e}_{l}}{\partial t} \int_{-1}^{1} \phi_{l} \: \phi_{l} \: d\xi \: + \: \left[ \hat{f} \phi_{l} \right]_{-1}^{1} - \: \int_{-1}^{1} f\left(u^{e}\right)  \: \frac{d \phi_{l}}{d \xi} \: d\xi = 0.
    \label{eqn:DG_form_xi_Riemann_oneDOF}
\end{equation}%
Since we focus on the Fourier analysis of the linear-advection equation, in which $f=au$,~\heqref{eqn:DG_form_xi_Riemann_oneDOF} can be written as%
\begin{equation}
\frac{h_{e}}{2} \frac{\partial U^e_{l}}{\partial t} L_{ll} \: + \: \left[ \hat{f} \phi_{l} \right]_{-1}^{1} - \: a \sum_{j=0}^{p} S_{lj} U^{e}_{j} = 0,
    \label{eqn:DG_form_xi_Riemann_oneDOF_au}
\end{equation}%
where $S_{lj}, L_{ll}$ are given by%
\begin{equation}
S_{lj} \:= \: \int_{-1}^{1} \: \phi_{j} \frac{d \phi_{l}}{d \xi} \: d\xi, \quad l,j=0,...,p \:,\quad L_{ll} = \int_{-1}^{1} \phi_{l} \: \phi_{l} \: d\xi, \quad l=0,...,p.
\label{eqn:Sj}
\end{equation}

It remains to define the numerical flux function $\hat{f}\left(\hat{u}_{-},\hat{u}_{+} \right)$ at a certain interface for the DG formulation to be complete. We define a general numerical flux function that encompasses both upwind (traditionally used with DG) and central numerical fluxes. The numerical flux function takes the following form%
\begin{equation}
\hat{f} = \beta f^{upwind} + (1-\beta) f^{central},
\label{eqn:num_flux1}
\end{equation}%
where $\beta$ is the upwind parameter. For the case of $\beta=0$, the numerical flux is of central type (simply the average of the left and right fluxes) while for $\beta=1$ we recover the fully upwind flux. The upwind flux could be of any type either exact upwinding for the $1$D linear case or any other approximate Reimann solver such as Roe~\cite{RoeApproximateRiemannsolvers1981} for a more general/nonlinear problem. Considering the linear case, the above numerical flux function for a given interface, can be written in a more compact form as%
\begin{equation}
\hat{f}\left(\hat{u}_{-},\hat{u}_{+} \right) = a \hat{u} = a \left( \mathcal{B}^{+} \hat{u}_{-} + \mathcal{B}^{-} \hat{u}_{+} \right),
\label{eqn:num_flux2}
\end{equation}%
and %
\begin{equation}
\mathcal{B}^{+} = \frac{1 + \beta \tilde{a} }{2}, \quad \mathcal{B}^{-} = \frac{1 - \beta \tilde{a} }{2} ,
\end{equation}%
where $\tilde{a} = \frac{|a|}{a}$, and $\hat{u}_{-},\hat{u}_{+}$ are the interface numerical solution values from the left and right neighboring elements, respectively. As a result,~\heqref{eqn:DG_form_xi_Riemann_oneDOF_au} can now be written in a vector form for each element $e$ as%
\begin{equation}
\frac{\partial \vec{U}^{e}}{\partial t} = \frac{2a}{h_{e}} \left( \mathcal{K}^{-} \vec{U}^{e-1} + \mathcal{K} \vec{U}^{e} + \mathcal{K}^{+} \vec{U}^{e+1} \right) ,
\label{eqn:DG_lin_advec_vectorform}
\end{equation}%
where $\vec{U}^{e} = \left[ U^{e} _{0}, ..., U^{e} _{p} \right]$ is the vector of DOFs of element $e$. The matrices $\mathcal{K}^{-}, \mathcal{K}^{+}$ in~\heqref{eqn:DG_lin_advec_vectorform} are given by%
\begin{alignat}{2}
\mathcal{K}_{l,m} &=  \left( S_{lm} + \mathcal{B}^{-} (-1)^{l+m}  - \mathcal{B}^{+} \right) / L_{ll},   \\
\mathcal{K}_{l,m}^{-} &=  \left( \mathcal{B}^{+}  (-1)^{l} \right) / L_{ll},  \quad \mathcal{K}_{l,m}^{+} =  - \left( \mathcal{B}^{-} (-1)^{l}  \right) / L_{ll} ,
\end{alignat}%
where $\phi_{j} (-1) = (-1)^{j}, \: \phi_{j} (+1) = (+1)^{j}$. %
%
\subsection{Finite Difference Method (FD) } \label{subsec:FD_2.2}%
%
Finite difference (FD) formulas for a certain derivative $u^{\prime}$ can be derived from a Taylor series expansion of a function $u(x)$ around a point $x_{j}$. By virtue of this Taylor expansion, different schemes can be obtained depending on where the series is truncated and the points that are involved in the stencil. In this type of nodal schemes, the solution domain $\mathcal{D}$ is decomposed into an $N_{n}$ number of grid points $x_{j} \in \mathcal{D}$ with a uniform grid spacing $h = x_{j+1}-x_{j}, j=0,...,N_{n}-1$. The solutions at the grid points are the DOFs. A number of central FD schemes are analyzed in the present study such as the 2$^{nd}$ order $($FD$2)$, 4$^{th}$ order $($FD$4)$, and  the 6$^{th}$ order $($FD$6)$ central schemes. 
The 4$^{th}$ order central scheme $($FD$4)$  can be expressed as%
\begin{equation}
u_{j}^{\prime}  = \frac{-u_{j+2} + 8 u_{j+1} - 8 u_{j-1} + u_{j-2} }{ 12 h} + O(h)^{4},
\label{eqn:FD_cent_4th}
\end{equation}%
and the 6$^{th}$ order $($FD$6)$ central scheme is given by%
\begin{equation}
u_{j}^{\prime}  = \frac{u_{j+3} -9 u_{j+2} + 45 u_{j+1} - 45 u_{j-1} + 9 u_{j-2} - u_{j-3}  }{ 60 h} + O(h)^{6}.
\label{eqn:FD_cent_6th}
\end{equation}

In addition, we analyze several fully-upwind and upwind-biased FD schemes up to 7$^{th}$ order of accuracy (OA=$7$). More FD formulas can be found in~\hyperref[appx:A]{Appendix~A}. The number of biased points is defined as the difference between the number of upstream and downstream points of the current point $j$. For instance, the fully-upwind  3$^{rd}$ order scheme $($FD$3)$, assuming right running waves ($a>0$), has an FD stencil of points $[ j,j-1,j-2,j-3 ]$,%
\begin{equation}
u_{j}^{\prime} = \frac{11 u_{j} -18 u_{j-1} + 9 u_{j-2} -2 u_{j-3} }{ 6 h} + O(h)^{3},
\label{eqn:FD_upwind_3rd}
\end{equation}%
while for the $1$-point upwind-biased FD$3$, the FD stencil utilizes one point upstream of $j$, namely $j+1$ such that %
\begin{equation}
u_{j}^{\prime} = \frac{2 u_{j+1} +3 u_{j} -6 u_{j-1} + u_{j-2} }{ 6 h} + O(h)^{3}.
\label{eqn:FD_upwind-biased_3rd}
\end{equation}

The derivative of the flux $f(u)$ in~\heqref{eqn:cons_law_1d} is discretized with one of the above formulas. Afterwards, the time derivative can be obtained directly at each node according to the following semi-discrete equation%
\begin{equation}
\frac{\partial u_{j}}{\partial t} = \mathbb{D} \left( f_{j}(u) \right), \quad j=0,...,N_{n}-1,
\label{eqn:FD_spatial_disc_consv_law}
\end{equation}%
where $\mathbb{D}$ is the FD spatial discretization operator. %
%
\subsection{Compact Difference Method (CD)  } \label{subsec:CD_2.3}%
%
Compact-difference (CD) utilizes a nodal stencil to approximate the derivative $u^{\prime}$, similar to FD schemes. Nevertheless, unlike classical FD schemes, CD schemes define a compact-central-stencil where all the derivatives in the mesh are tied together in a tridiagonal system as follows%
\begin{equation}
\alpha u^{\prime}_{j-1} + u_{j}^{\prime} + \alpha u_{j+1}^{\prime} = c \frac{u_{j+2}-u_{j-2}}{4 h} + d \frac{u_{j+1}-u_{j-1}}{2 h},
\label{eqn:CD_formula}
\end{equation}%
where $\alpha, c,$ and $ d $ are constants whereby different central schemes with different orders, can be obtained. In this study, we focus on the $6^{th}$ order CD$6$ scheme where $\alpha=1/3$, $c=1/9$, and $d=14/9$. The coefficients of other schemes including the ones used in this study can be found in~\cite{LeleCompactfinitedifference1992,VisbalUseHigherOrderFiniteDifference2002}. In order to evaluate the derivative of the flux $f(u)$ in~\heqref{eqn:cons_law_1d}, the CD formula~\heqref{eqn:CD_formula} requires the solution of a tridiagonal system~\cite{LeleCompactfinitedifference1992} for the derivatives at all the nodes $x_{j}, j=0,...,N_{n}-1$. For a general BC case, near boundary formulas are needed, whereas, for a periodic BC, a slightly modified tridiagonal system is solved. Afterwards, the time derivative can be obtained directly at each node according to the same semi-discrete~\heqref{eqn:FD_spatial_disc_consv_law} as in FD, with $\mathbb{D}$ in this case as the CD spatial discretization operator. %
%
\subsubsection{Compact filters} \label{subsubsec:CF_2.3.1}%
%
Owing to the natural non-dissipative property of central stencils, a spatial filter is usually needed in order to add some stabilization to the scheme, especially for non-uniform grids. In the present study, we analyze the classical compact Pad\'{e} filter used with CD schemes by Lele~\cite{LeleCompactfinitedifference1992} and further developed and analyzed by Gaitonde and Visbal and their co-authors~\cite{GaitondeFurtherdevelopmentNavierStokes1999,VisbalUseHigherOrderFiniteDifference2002}. After each complete time step (RK final stage), the filter is applied to the solution $u$ in order to obtain a filtered solution $\tilde{u}$. This is accomplished by solving the following tridiagonal system%
\begin{equation}
\alpha_{f} \tilde{u}_{j-1} + \tilde{u}_{j} + \alpha_{f} \tilde{u}_{j+1} = \sum_{l=0}^{N} \frac{d_{l}}{2} \left( u_{j+l} + u_{j-l} \right),
\label{eqn:CD_filter_eqn}
\end{equation}%
for the filtered solution $\tilde{u}$, where $\alpha_{f}, d_{l}(\alpha_{f})$ are constant coefficients that determine the order of the filter, and $N$ is the number of nodes where the solution is required to be filtered. The amount of added filtering/dissipation is controlled by adjusting the filter parameter $\alpha_{f}$ which satisfies $0<|\alpha_{f}|\leq 0.5$. While $\alpha_{f}=0.5$ corresponds to the no-dissipation/filtering case, more dissipation is achieved as $\alpha_{f}$ goes towards its negative limit$-0.5$ and for $\alpha_{f}=0.0$ an explicit filter is obtained. The coefficients of different filters can be found in~\cite{GaitondeFurtherdevelopmentNavierStokes1999,VisbalUseHigherOrderFiniteDifference2002} and in~\hyperref[appx:B]{Appendix~B} we provide them for the $8^{th}$ order filter. %
%
\subsection{Runge-Kutta Time Integration Schemes (RK)}\label{subsec:RK_2.4}%
%
Applying one of the spatial methods discussed in the previous sections to discretize~\heqref{eqn:cons_law_1d} results in an ordinary differential equation (ODE), which for the particular case of linear-advection can be written as%
\begin{equation}
\frac{d \vec{u}}{d t}= \mathcal{A} \vec{u},
\label{eqn:time_ODE_linadvec}
\end{equation}%
where $\vec{u}$ is the vector of all unknown global DOFs (either nodal solution values for FD and CD schemes, or element-wise DOFs for DG-type high-order methods), and $\mathcal{A}$ is the space discretization operator. This ODE can be solved using any time marching scheme. In the present work, we employ two strong-stability-preserving Runge-Kutta (SSPRK) schemes~\cite{GottliebStrongStabilityPreservingHighOrder2001}, namely, the second-order (RK2), and the third-order (RK3) schemes, in addition to the classical fourth-order (RK4)~\cite{ButcherNumericalAnalysisOrdinary1987} scheme. Applying a RK scheme to the ODE~\heqref{eqn:time_ODE_linadvec} results in an update formula for the solution  $\vec{u}$ at $t=t+\Delta t$ of the following form%
\begin{equation}
\vec{u}(t+\Delta t) = \left[ I + \sum_{m=1}^{s} \frac{\left(\Delta t \: \mathcal{A} \right)^{m}}{m!} \right] \vec{u}(t)  = \mathcal{P}(\Delta t \mathcal{A}) \vec{u}(t) ,
\label{eqn:RK_update_form_DG}
\end{equation}%
where $s=2$ for RK2, $s=3$ for RK3, and $s=4$ for RK4, and $\mathcal{P}$ is a polynomial of degree $s$. %
%
\section{Fourier Dispersion/Dissipation Analysis}\label{sec:Fourier_3}%
%
Consider the linear-advection problem defined in an unbounded domain $-\infty < x < \infty$, that takes the form%
\begin{equation}
\frac{\partial u}{\partial t} + a \frac{\partial u}{\partial x} = 0, \quad \text{with periodic B.C.} ,
\label{eqn:lin_advec} 
\end{equation}%
where $a$ is a positive constant wave speed. For an initial wave solution%
\begin{equation}
 u(x,0) = u_{o}(x) = e^{ikx}, 
 \label{eqn:linadvec_IC}
\end{equation}%
\heqref{eqn:lin_advec} admits a wave solution of the form %
\begin{equation}
u(x,t) = e^{i\left(kx-\omega t\right)},
\label{eqn:lin_advec_exact_sol}
\end{equation}%
where $k$ is the spatial wavenumber, and $\omega$ denotes the frequency that admits the exact dispersion relation $\omega=ka$. In a temporal Fourier-analysis~\cite{VichnevetskyFourierAnalysisNumerical1982}, a prescribed wavenumber $k$ is assumed for the IC~\heqref{eqn:linadvec_IC} and different spatial and temporal schemes are applied to~\heqref{eqn:lin_advec} in order to study their dispersion/dissipation properties based on the numerical frequency $\tilde{\omega}$. All the analysis performed in this section were implemented in a MATLAB set of functions/scripts as a toolbox for Fourier analysis of the considered methods.%
%
\subsection{Semi-discrete analysis of DG schemes} \label{subsec:DGsd_3.1}%
%
In the semi-discrete type of analysis, only spatial discretization is applied to~\heqref{eqn:lin_advec} so that dispersion and dissipation properties can be studied, being solely dependent on the characteristics of the spatial scheme. This case can also be interpreted as the limiting case of a fully-discrete scheme (in space and time) when the time-step $\Delta t \rightarrow 0$. 

The method used in this section for DG is similar to the one previously presented by Hu et al.~\cite{HuAnalysisDiscontinuousGalerkin1999} and Moura et al.~\cite{MouraLineardispersiondiffusion2015} among others. For DG schemes, applying the spatial discretization to~\heqref{eqn:lin_advec} results in a system of semi-discrete equations of the form~\heqref{eqn:DG_lin_advec_vectorform}. We assume a uniform mesh in the following analysis. The initial DG solution is the projection of the initial condition to the DG solution space, and for element $\Omega_{e}$, the element-wise DOFs $U^{e}_{l}$ are computed as%
\begin{equation}
U_{l}^{e}(0) = \frac{\int_{\Omega_{e}} u^{e}(x,0) \phi(x) dx }{ \int_{\Omega_{e}} \phi(x) \phi(x) dx }= \frac{\int_{\Omega_{r}} u(x_{e}+\xi h/2,0) \phi_{l}(\xi) d\xi }{L_{ll}}. 
\end{equation}%
For the initial wave form ~\heqref{eqn:linadvec_IC}, these DOFs can be written as%
\begin{equation}
U_{l}^{e}(0) = {\mu}_{e,l} \: e^{i k x_{e} },
\label{eqn:lin_advec_sol_coef_blochwave}
\end{equation}%
where ${\mu}_{e,l}$ is defined as%
\begin{equation}
{\mu}_{e,l} = \frac{\int_{-1}^{1}  e^{ik \left(\xi h/2\right) } \phi_{l}(\xi) d\xi }{L_{ll}}.
\label{eqn:mu_projection}
\end{equation}%
It is easy to see that the exact DG solution can be expressed as %
\begin{equation}
U_{l}^{e}(t) =  \mu_{l} e^{i(k x_{e} - \omega t)}, 
\label{eqn:DG_sdisc_waveform}
\end{equation}%
where $\mu_{l} = {\mu}_{e,l}$. We note that different high-order methods result in different projection coefficients $\mu_{l}$ according to the choice of the basis functions $\phi$ and the expansion form of the solution~\heqref{eqn:u_h_xi_phi}, i.e., whether admitting a nodal or a modal form. However, the rest of the analysis
steps can be applied in a similar manner to any high-order method. 

By seeking a solution in the form of~\heqref{eqn:DG_sdisc_waveform} and substituting it into~\heqref{eqn:DG_lin_advec_vectorform}, we get%
\begin{equation}
\left( \frac{h}{a} \right) \frac{\partial \vec{U}^{e}}{\partial t} = 2 \left( \mathcal{K}^{-} e^{-ikh} + \mathcal{K} + \mathcal{K}^{+} e^{ikh}  \right) \vec{U}^{e} = \mathcal{A} \vec{U}^{e},
\label{eqn:DG_semi_disc}
\end{equation}%
and by differentiating this equation, we get the semi-discrete relation%
\begin{equation}
\left(-i \tilde{\omega} \frac{h}{a} \right) \vec{\mu} = \mathcal{A} \vec{\mu},
\label{eqn:DG_semi_disc1}
\end{equation}%
where $\tilde{\omega}$ is the numerical frequency, and  $\vec{\mu}=\left[\mu_0,...,\mu_{p}\right]^{T}$. The semi-discrete system~\heqref{eqn:DG_semi_disc1} constitutes an eigenvalue problem. The matrix $\mathcal{A}$ has $p+1$ eigenvalues $\lambda_{j}$ and $p+1$ eigenvectors $\vec{\mu}_{j}$ for a given value of $k$. As a result, the general solution~\heqref{eqn:DG_sdisc_waveform} can be written as a linear expansion in the eigenvector space%
\begin{equation}
\vec{U}^{e}(t) = \sum_{j=0}^{p} \vartheta_{j} \vec{\mu}_{j} \: e^{i\left( k x_{e} - \tilde{\omega}_{j} t \right)} .
\label{eqn:semi-disc_gen_sol_eigenvector_space}
\end{equation}%
The expansion coefficients $\vartheta_{j}$ are obtained from the initial condition, i.e.,%
\begin{equation}
\vec{{\mu}}_{e} = \sum_{j=0}^{p} \vartheta_{j} \vec{\mu}_{j}, \quad \text{or} \quad \vec{\vartheta} = \mathcal{M}^{-1}  \vec{{\mu}}_{e},
\label{eqn:eta_relation1}
\end{equation}%
where $\mathcal{M}=\left[\vec{\mu}_{0},...,\vec{\mu}_{p}\right]$ is the matrix of eigenvectors, and $\vec{\vartheta}=\left[\vartheta_{0},...,\vartheta_{p}\right]^{T}$. In addition, the solution coefficients can be written in a more compact form as%
\begin{equation}
\vec{U}^{e}(t) = \mathcal{M} \vec{\Theta} \: e^{ik x_{e}}, \quad \vec{\Theta}=[\Theta_{0},...,\Theta_{p}]^{T} \: \text{with} \: \Theta_{j} = \vartheta_{j} e^{- i \tilde{\omega}_{j} t} ,
\label{eqn:elem_coeff_eigen_vectorform}
\end{equation}%
and finally the element-wise polynomial solution is expressed as a dot product of the form%
\begin{equation}
u^{e}(\xi,t) = \langle \vec{\phi}, \vec{U^{e}} \rangle = \langle \vec{\phi} , \mathcal{M} \vec{\Theta} \rangle e^{i k x_{e}},
\label{eqn:DG_sdisc_numsol_compact_form}
\end{equation}%
in which $\vec{\phi}=[\phi_{0},...,\phi_{p}]^{T}$. Similarly, the exact solution can be written as%
\begin{equation}
u^{e}_{ex}(\xi,t) = \langle \vec{\phi}, \vec{U^{e}_{ex}} \rangle = \langle \vec{\phi} , \vec{\mu}_{e} \rangle e^{i \left(k x_{e} - \omega t \right)} , \quad \omega = ka .
\label{eqn:DG_sdisc_exactsol_compact_form}
\end{equation}

The numerical solution~\heqref{eqn:DG_sdisc_numsol_compact_form} is essentially a linear combination of $p+1$ waves, each having its own dispersion and dissipation behavior. The eigenvalues, $\lambda_{j}, \; j=0,...,p$, are generally complex and hence numerical dispersion and dissipation are induced by each eigenvalue through the numerical frequency $\frac{\tilde{\omega}_{j}}{a} h = k_{m,j} h= i \lambda_{j}$, where $k_{m}$ is defined as the modified wavenumber.

In a DG-type method $h/(p+1)$ can be considered as the smallest length-scale that can be captured by the scheme~\cite{MouraLineardispersiondiffusion2015}. In order to have a fair way to compare multi-degree of freedom methods (such as DG) with single degree of freedom methods (such as FD and CD), we define the non-dimensional wavenumber to be $K= kh/(p+1)$. Consequently, the modified non-dimensional wavenumber is $K_{m}=k_{m} h /(p+1)=  \frac{\tilde{\omega} }{a} h /(p+1)$ and the numerical dispersion relation requires that%
\begin{equation}
\operatorname{\mathcal{R}e}(K_{m}(K)) = - \operatorname{\mathcal{I}m}(\lambda)  \approx K ,
\label{eqn:DG_num_sdisc_disper_rel}
\end{equation}%
while for stability, the numerical dissipation behavior should satisfy%
\begin{equation}
\operatorname{\mathcal{I}m} (K_{m}(K) ) =  \operatorname{\mathcal{R}e}(\lambda) \leq 0 .
\label{eqn:DG_num_sdisc_dissip_rel}
\end{equation}%
This kind of non-dimensionalization helps in quantifying the accuracy of DG schemes per DOF since for the same nDOF, FD and CD have a length-scale of $h = h_{DG}/(p+1)$. %
\begin{figure}[H]
    \centering
    \begin{subfigure}[h]{0.48\textwidth}
    \includegraphics[width=\textwidth]{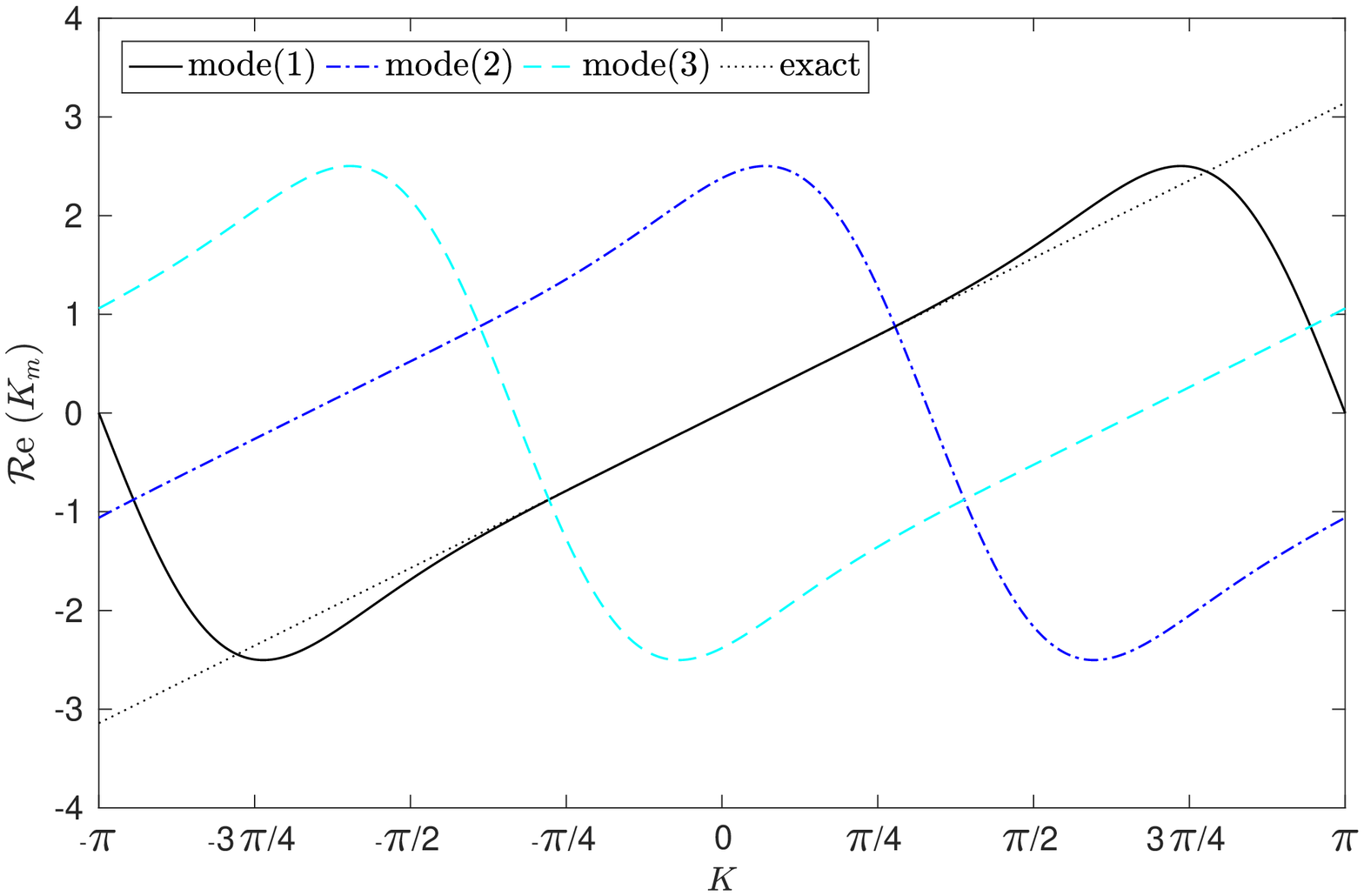} 
    \caption{Dispersion}
    \end{subfigure}
    \hspace{0.015\textwidth}
    \begin{subfigure}[h]{0.48\textwidth}
    \includegraphics[width=\textwidth]{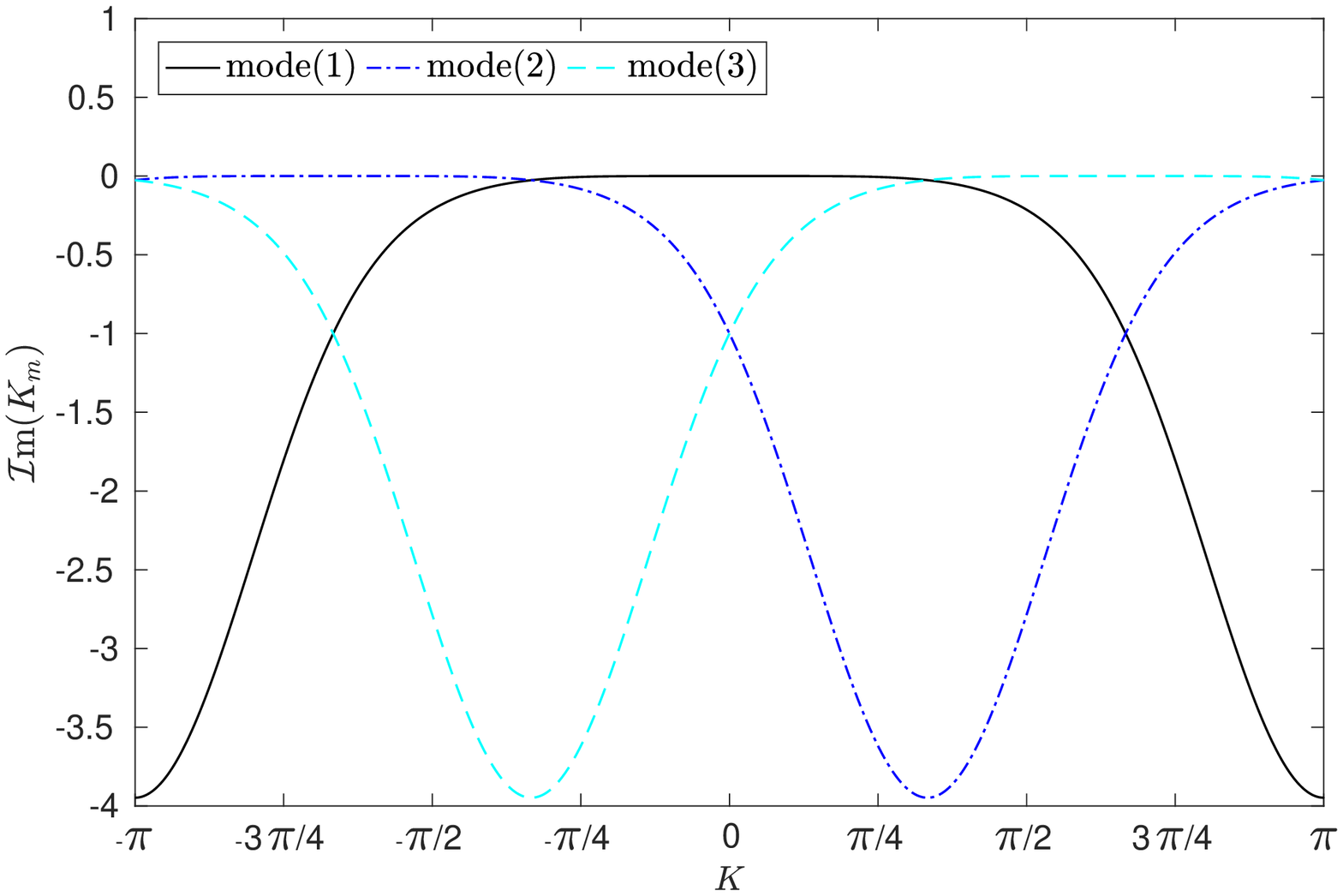}  
    \caption{Dissipation}
    \end{subfigure}
	\caption{Semi-discrete dispersion/dissipation of the DGp$2$ scheme with $\beta=1$(upwind). This plot contains all the three modes of the scheme with mode$(1)$ as the physical-mode.}
    \label{fig:p2_beta1_sdisc}
\end{figure}%
The dispersion and dissipation parts of $K_{m}$ for the DGp$2$ scheme with an upwind flux ($\beta=1.0$) are presented in~\hfigref{fig:p2_beta1_sdisc}, where we can see the three different eigenmodes associated with each base wavenumber $K$. In studying these curves, many authors interpreted them in different ways. One popular idea is to consider only one of them as the physical-mode~\cite{HuAnalysisDiscontinuousGalerkin1999,VandenAbeeleDispersiondissipationproperties2007,VincentInsightsNeumannanalysis2011,GassnerComparisonDispersionDissipation2011,YangDispersionDissipationErrors2013} while regarding the others as parasites. Van Den-Abeele et al.~\cite{VandenAbeeleDispersiondissipationproperties2007} considered each one of the $(p+1)$ solutions as corresponding to another wavenumber and reassigned them graphically~\cite{VandenAbeeleDevelopmenthighorderaccurate2009}, whereas Vincent et al.~\cite{VincentInsightsNeumannanalysis2011}  adopted the same idea and proposed an automated way to identify the corresponding wavenumber. The physical-mode is defined to be the one that approximates the exact dispersion relation for a range of wavenumbers~\cite{HuAnalysisDiscontinuousGalerkin1999}, mode$(1)$ in~\hfigref{fig:p2_beta1_sdisc}. On the other hand, Vanharen et al.~\cite{VanharenRevisitingspectralanalysis2017} utilized the matrix power method to identify the asymptotic behavior of fully-discrete SD schemes coupled with RK time integration schemes for $kh \leq \pi$, and proposed new definitions of dispersion/dissipation for wavenumbers $kh>\pi$.

\begin{figure}[H]
 \begin{subfigure}[h]{0.475\textwidth}
    \centering
    \includegraphics[width=0.95\textwidth]{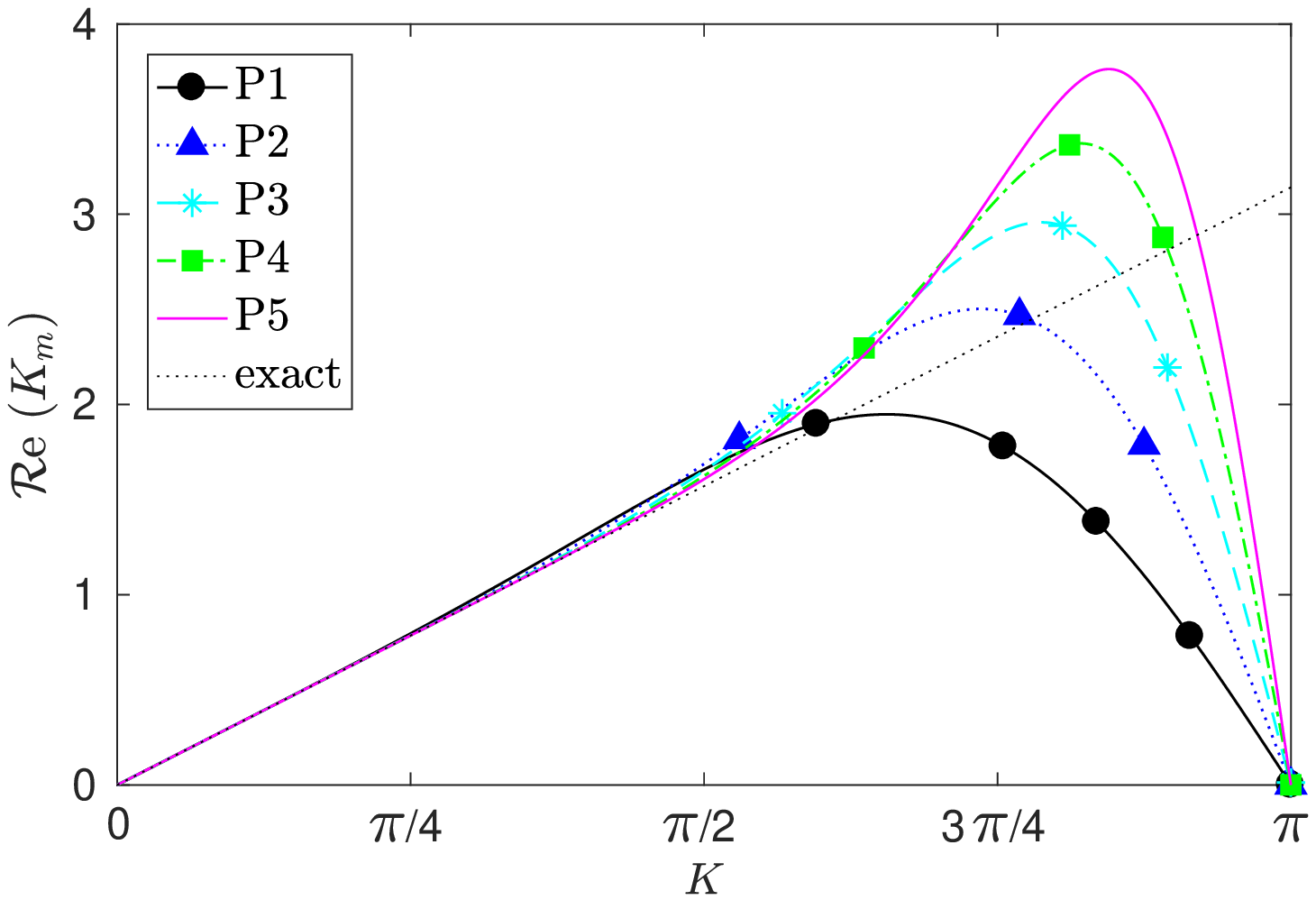} 
    \caption{Dispersion}
    \label{fig:DGp2_sdisc_compare_order_wp}
    \end{subfigure}
    \hspace{0.0125\textwidth}
    \begin{subfigure}[h]{0.475\textwidth}
    \centering
    \includegraphics[width=0.95\textwidth]{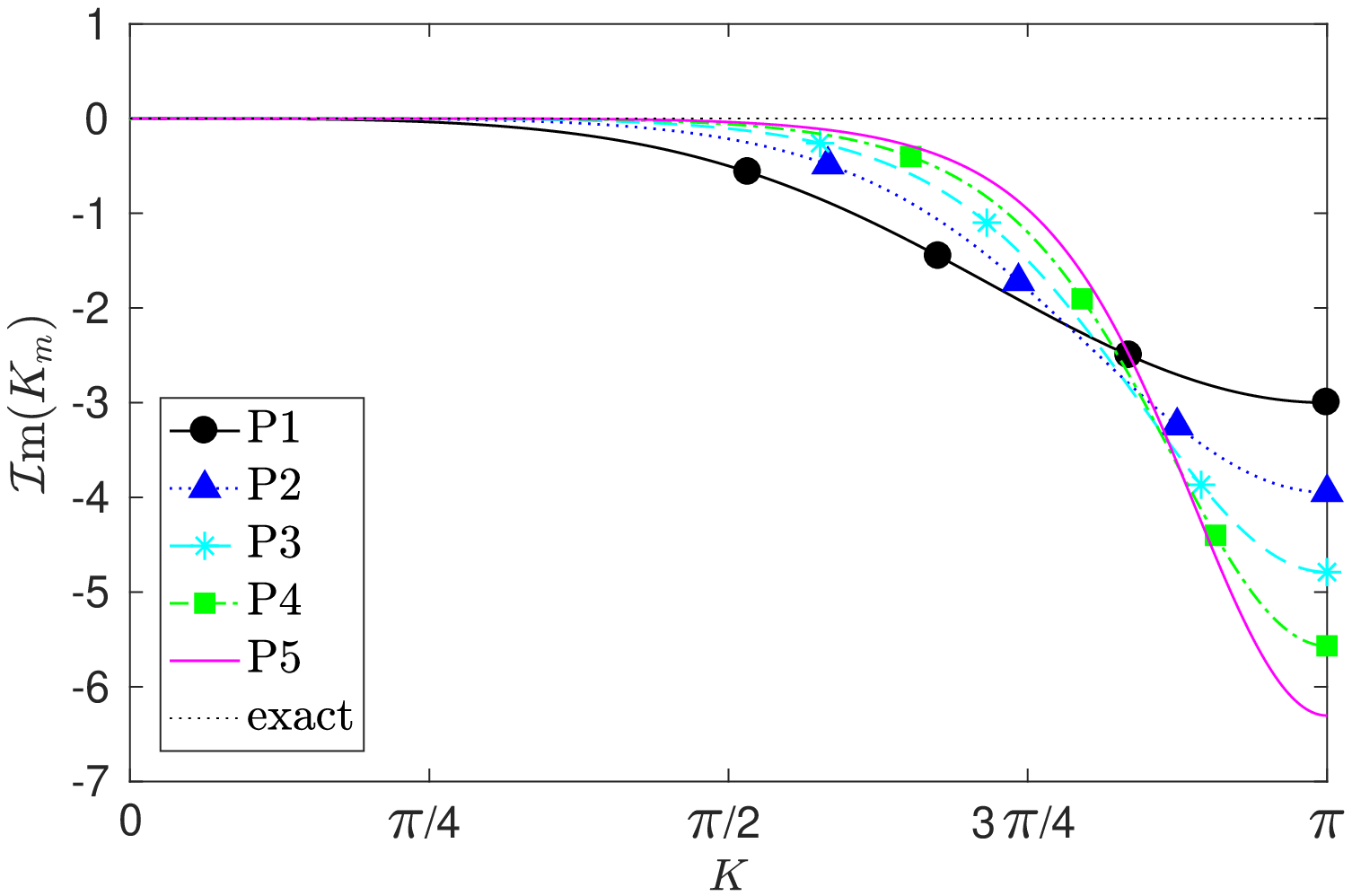}  
    \caption{Dissipation}
    \label{fig:DGp2_sdisc_compare_order_wd}
    \end{subfigure} 
	\caption{Comparison of the semi-discrete dispersion/dissipation of DG schemes with different orders, based on the physical-mode behavior. }
    \label{fig:DGp2_sdisc_compare_order}
\end{figure}

The changes of dispersion and dissipation curves with the order of the scheme for the physical-mode are presented in~\hfigref{fig:DGp2_sdisc_compare_order}. From this figure it is evident that increasing the order of the scheme continuously improves its dissipation in the low wavenumber range while adding more dissipation in the high wavenumber range. In addition, increasing the order of the scheme improves its dispersion for the same range where dissipation were reduced. These results agree well with the results previously presented in~\cite{HuAnalysisDiscontinuousGalerkin1999,VincentInsightsNeumannanalysis2011,GassnerComparisonDispersionDissipation2011,MouraLineardispersiondiffusion2015} and serve as a verification of our analysis. %
%
\subsubsection{Discussions on the behavior of secondary modes} \label{subsubsec:DGsd_secondmode_3.1.1}%
%
Recently, Moura et al.~\cite{MouraLineardispersiondiffusion2015} suggested a new interpretation for DG-type methods in admitting more than one eigensolution. They considered secondary modes to be replicates of the primary/physical one and that they behave like the primary mode but at a different wavenumber. This is verified by our analysis as well, see~\hfigref{fig:p2_beta1_sdisc}. However, when it comes to the case of $\beta=0$ using the central flux, it becomes complicated to identify the physical-mode for the entire range of wavenumbers and the replication property may be lost.

\begin{figure}[H]
\centering
    \begin{subfigure}[h]{0.48\textwidth}
    \includegraphics[width=\textwidth]{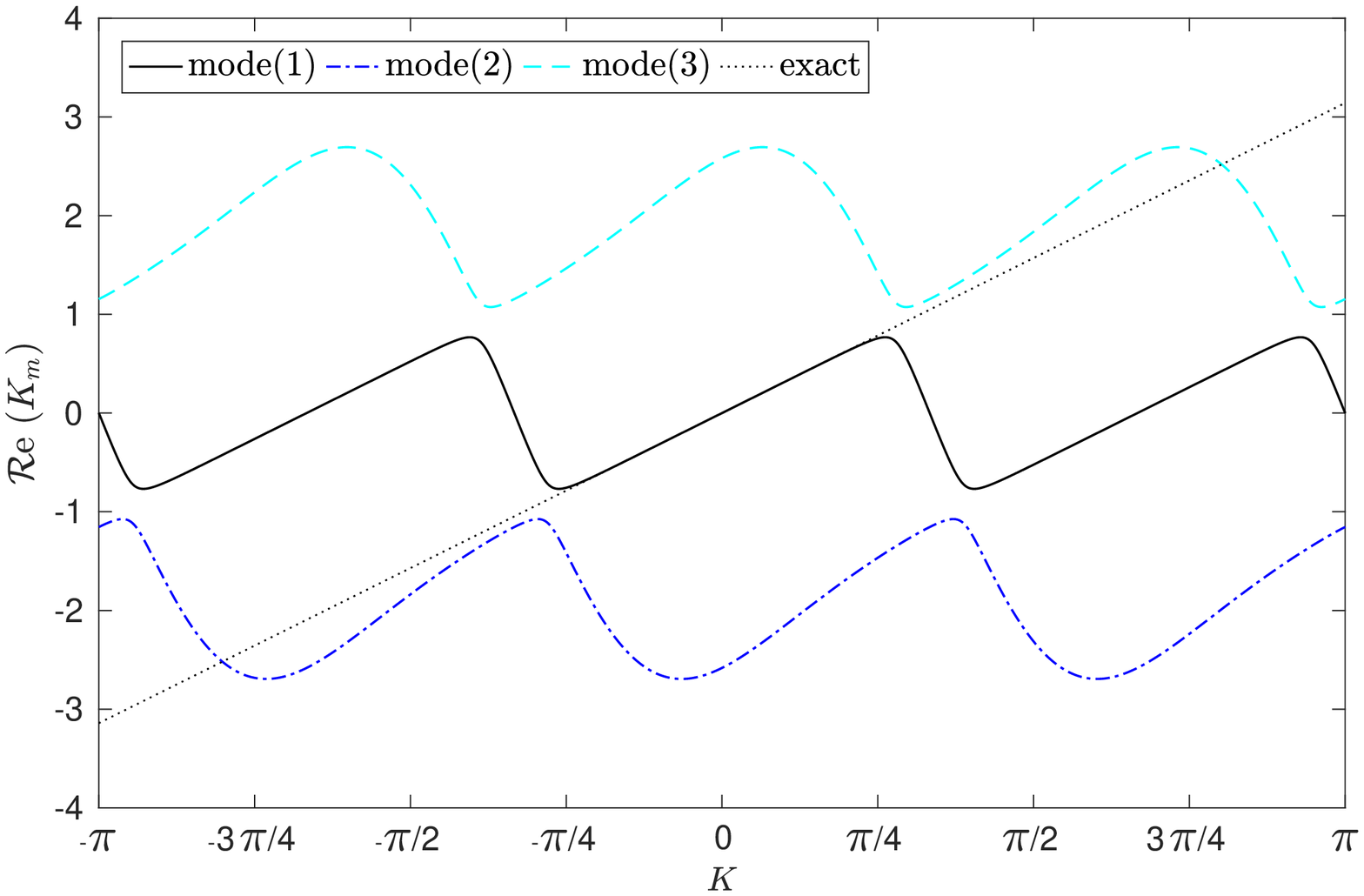} 
    \caption{option($1$)}
    \label{fig:p2_beta0_wp_compare_sdisc1}
    \end{subfigure}
    \hspace{0.015\textwidth}
    \begin{subfigure}[h]{0.48\textwidth}
    \includegraphics[width=\textwidth]{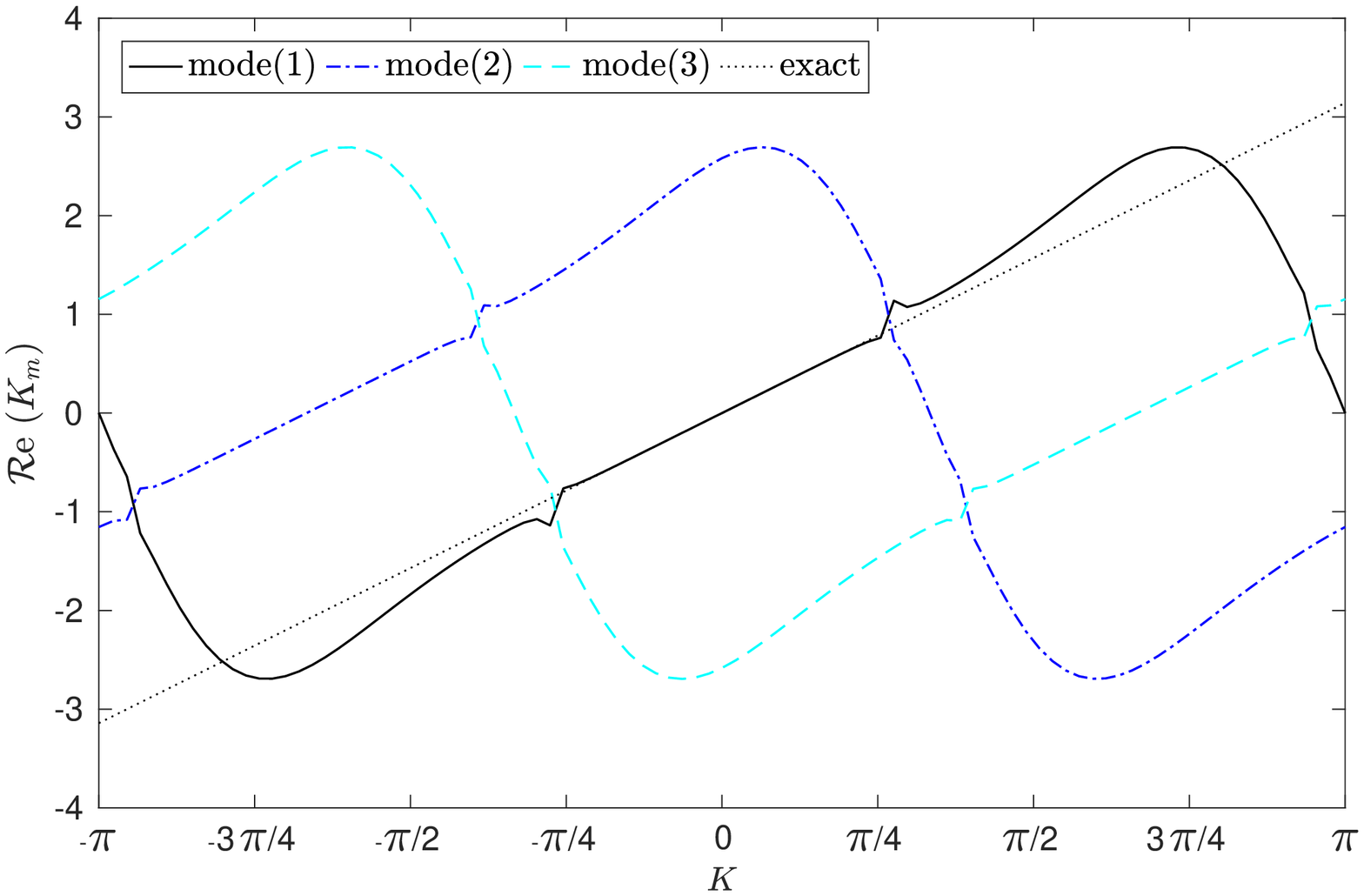} 
    \caption{option($2$)}
    \label{fig:p2_beta0_wp_compare_sdisc2}
    \end{subfigure}
	\caption{Semi-discrete dispersion behavior for the DGp$2$ scheme with central fluxes, $\beta=0.00$. This figure includes all the three modes of the scheme.}
    \label{fig:p2_beta0_wp_compare_sdisc}
\end{figure}

\hfigref{fig:p2_beta0_wp_compare_sdisc} displays the dispersion curves for DG with $\beta=0$ (central flux) using two options to identify the physical-mode. If option (1) is adopted as in~\cite{HuAnalysisDiscontinuousGalerkin1999,AsthanaHighOrderFluxReconstruction2015},  no replication can be seen as in~\hfigref{fig:p2_beta0_wp_compare_sdisc1} and there are some discontinuities between the curves that approximate the exact dispersion relation. However, if we connect the curves as in option (2) discussed in~\cite{MouraEigensolutionanalysisspectral2016}, the replication property is restored as shown in~\hfigref{fig:p2_beta0_wp_compare_sdisc2}. Moreover, Asthana et al.~\cite{AsthanaHighOrderFluxReconstruction2015} compared the energy distribution among different modes in the central flux case. They concluded that the physical-mode, mode$(1)$ in~\hfigref{fig:p2_beta0_wp_compare_sdisc1}, has the highest energy until it falls around $(K\approx \pi /4$) causing large dispersion errors. After that, the highest energy is contained by another mode (mode$(2)$ in~\hfigref{fig:p2_beta0_wp_compare_sdisc1}) until near ($K \approx 3.0$) and mode $(1)$ contains the highest energy again. The third mode has almost zero energy for the entire range of wavenumbers. It is worth noting that they only presented the results for the positive wavenumber range, and it is concluded that a similar energy distribution exists for the negative part but with mode $(3)$ being more energetic than mode $(2)$ in this case. This suggests that option (2) may be a more appropriate choice. In this case, the central flux behaves in dispersion similar to the upwind one and it is only the jump/gap areas between different modes where the physical-mode is not defined. From our analysis we note that the vertical distance between the modes becomes larger as the order of the scheme increases, making it more complicated to identify the physical-mode for this case, see also~\cite{MouraEigensolutionanalysisspectral2016}. 

We show next, using a combined-mode analysis approach, that the true behavior of DG schemes with a central flux follows the behavior of the physical-mode identified in option ($2$) except near the gap between the curves. In this approach, we follow the ideas proposed in previous studies~\cite{HuAnalysisDiscontinuousGalerkin1999,MouraLineardispersiondiffusion2015,VanharenRevisitingspectralanalysis2017}. In addition, we provide verifications that help to explain the behavior of DG schemes with the central flux ($\beta=0$) in particular, among other interesting results. We refer to this approach as the "combined-mode analysis", and to the behavior of DG schemes under this approach as the "true" behavior in contrast to the one that is based on the physical-mode which we refer to as "asymptotic/physical-mode" behavior. %
%
\subsubsection{True behavior of DG schemes through a combined-mode semi-discrete analysis}\label{subsubsec:DGsd_true_3.1.2}%
%
The dissipation of a certain scheme can be defined as the loss of energy of the initial wave, while the dispersion is the phase shift between the exact and numerical wave solutions~\cite{VanharenRevisitingspectralanalysis2017}. Considering the numerical and exact solution polynomials resulted from the semi-discrete Fourier analysis,~\heqsref{eqn:DG_sdisc_numsol_compact_form}{eqn:DG_sdisc_exactsol_compact_form}, the energy based on the $L_{2}$ norm of a complex function is given by%
\begin{equation}
E^{e} ( k,t ) = \sqrt { \frac{\int_{-1}^{1} |u^{e}(\xi,t)|^{2} \: d\xi }{\int_{-1}^{1}  d\xi} }.
\label{eqn:DG_numEnergy_rel}
\end{equation}%
A similar relation can be written for the exact solution energy $E^{e}_{ex}( k,t )$, i.e., the projected energy distribution for each prescribed wavenumber $k$. In order to quantify dissipation, the true amplification factor $G^{true}(k,t)$ is defined as the ratio between the numerical energy and the exact one%
\begin{equation}
G^{true}(k,t) = \frac{E^{e}(t)}{E^{e}_{ex}(t)},
\label{eqn:sdisc_true_G}
\end{equation}%
whereas $G$, based on the physical-mode solely, is given by%
\begin{equation}
G^{phys}(k,t) = e^{\mathcal{I}m(\tilde{\omega}) t} = e^{\mathcal{I}m(K_{m}) \frac{(p+1)}{(h/a)} t}.
\label{eqn:sdisc_phys_G}
\end{equation}%
In addition, the phase shift between two complex signals is%
\begin{equation}
\psi(k,t) = \text{angle} \left( \int_{-1}^{1} u^{e}(\xi)  \times  \left( u^{e}_{ex}(\xi)  \right)^{*} d\xi \right),
\label{eqn:sdisc_true_phase_shift}
\end{equation}%
where $\left( u^{e}_{ex} \right)^{*} $ is the complex conjugate of the exact solution. As a result, for dispersion quantification we define the phase error (non-dimensionalized with respect to the effective length-scale) between the two true signals as%
\begin{equation}
\Delta \psi^{true}(k,t) =  |\psi(k,t)| /(p+1),
\label{eqn:sdisc_true_S_err}
\end{equation}%
and the phase error based on the physical-mode solely, is defined as
\begin{equation}
\Delta \psi^{phys}(k,t) =  | ( \mathcal{R}e(K_{m}) - K ) | \left(\frac{a}{h}\right) t,
\label{eqn:sdisc_phys_S_err}
\end{equation}%
where the multiplication by $(a/h)$ and division by $(p+1)$ are results of non-dimensionalization of $K,K_{m}$. In this study we assume $(a/h)=1$ for simplicity. Using the above defined quantities we can compare the behavior of the DG scheme based on combining all the eigenmodes versus one single physical mode. %
\begin{figure}[H]
\begin{subfigure}[h]{0.475\textwidth}
\centering
 \includegraphics[width=\textwidth]{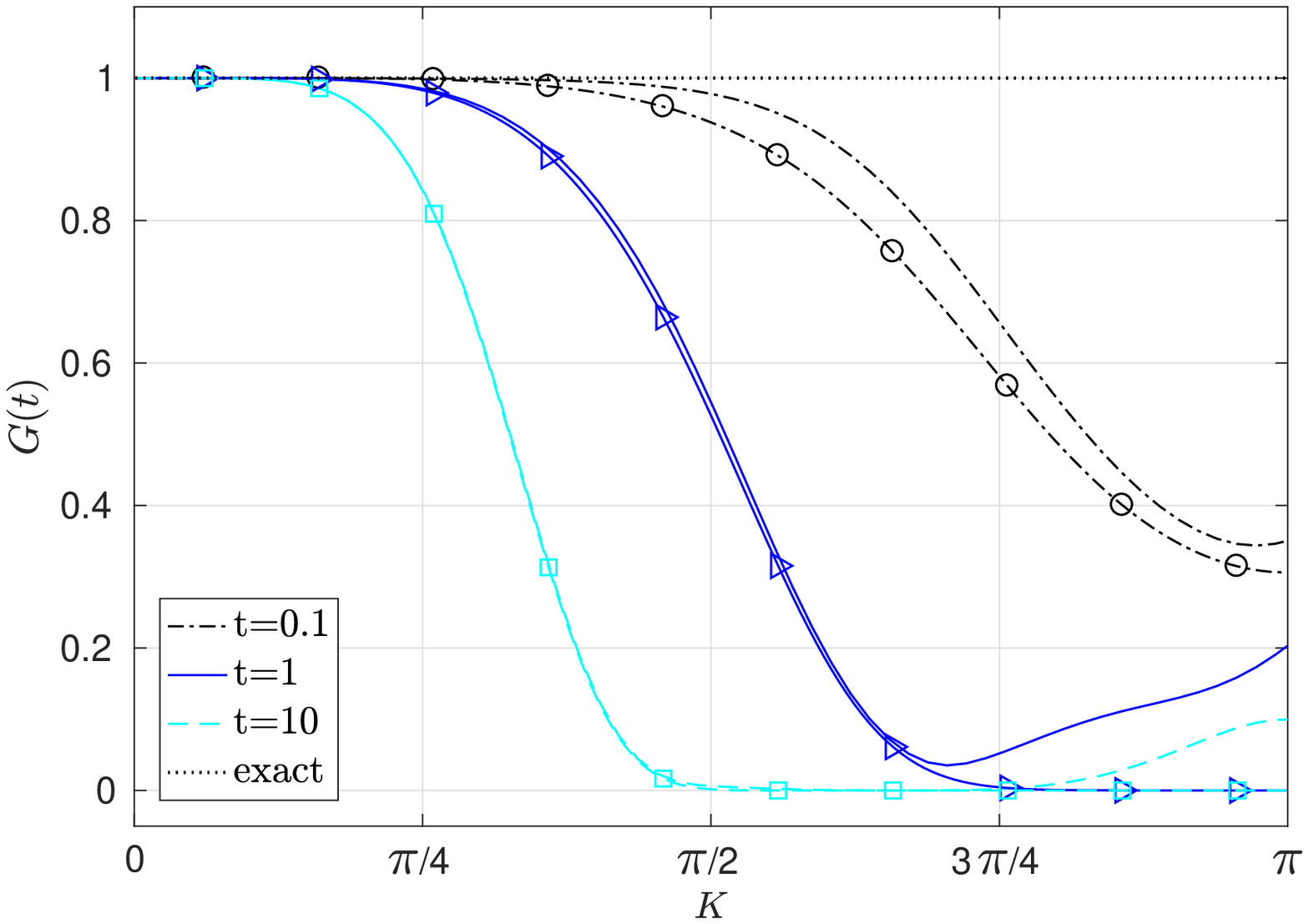}  
 \caption{Amplification factor}
 \label{p2_beta1_sdisc_G_truephys}
 \end{subfigure}
 \hspace{0.0125\textwidth}
 \begin{subfigure}[h]{0.475\textwidth}
 \centering
 \includegraphics[width=\textwidth]{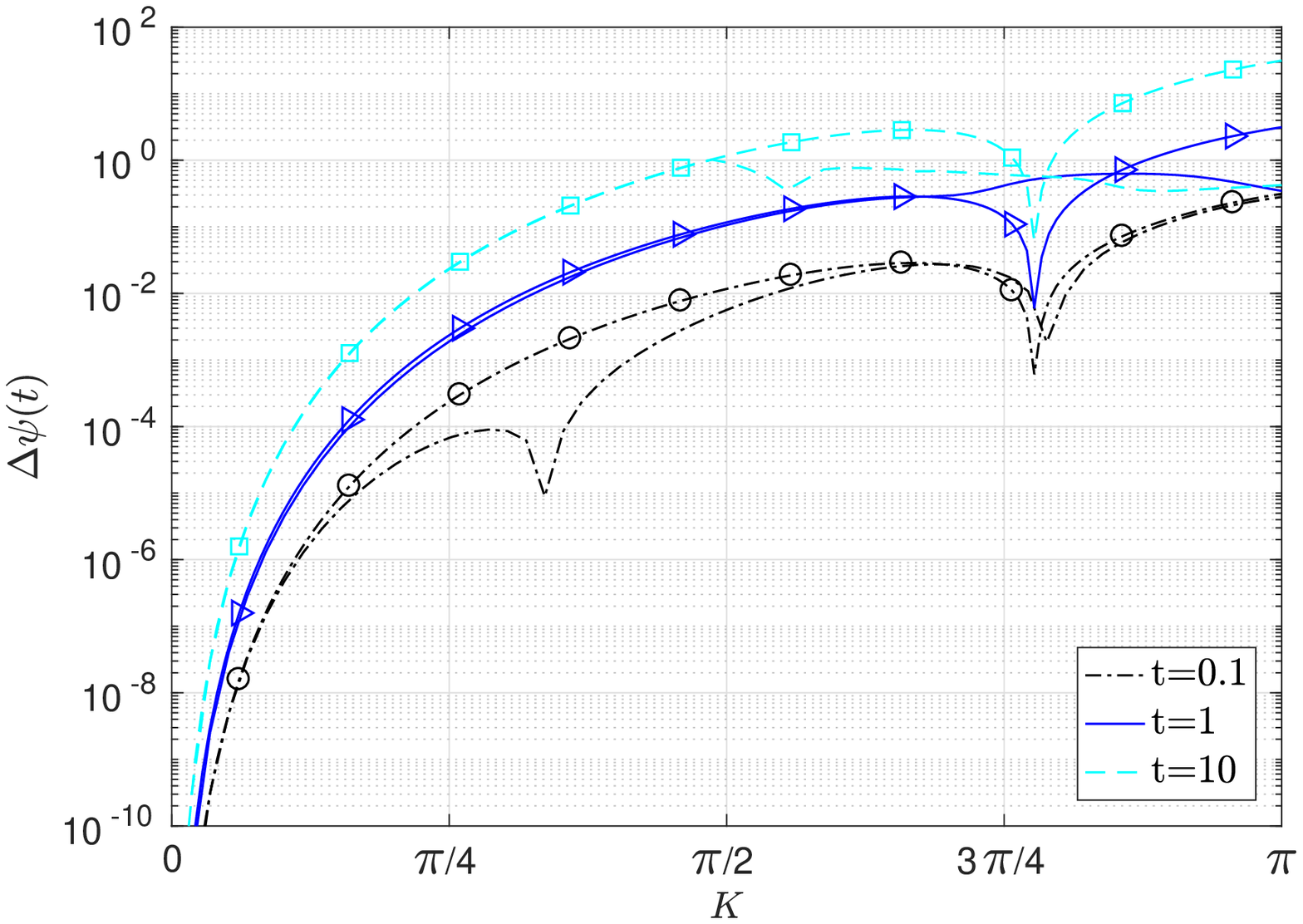}  
 \caption{Phase error}
 \label{p2_beta1_sdisc_S_err}
 \end{subfigure}%
 \caption{Comparison of the semi-discrete true behavior (indicated by "plain" curves) and physical-mode behavior (indicated by "lines-with-symbols") for the DGp$2$ -$\beta1.0$ scheme at different time points. }
    \label{fig:p2_beta1_sdisc_truephys}
\end{figure}%
\begin{figure}[H]
\begin{subfigure}[h]{0.475\textwidth}
\centering
 \includegraphics[width=\textwidth]{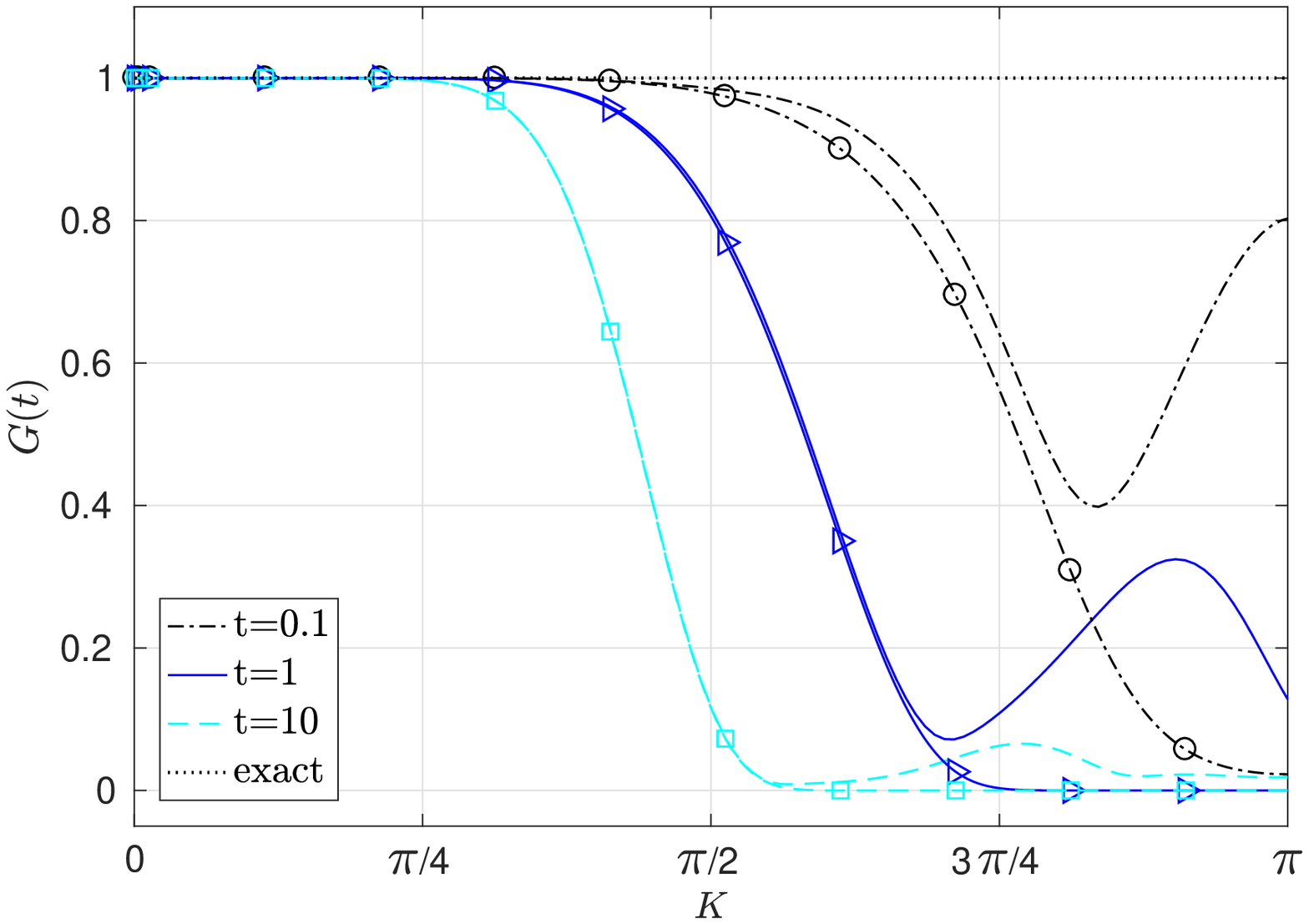}  
  \caption{Amplification factor}
 \label{p5_beta1_sdisc_G_truephys}
 \end{subfigure}
\hspace{0.0125\textwidth}
 \begin{subfigure}[h]{0.475\textwidth}
 \centering
 \includegraphics[width=\textwidth]{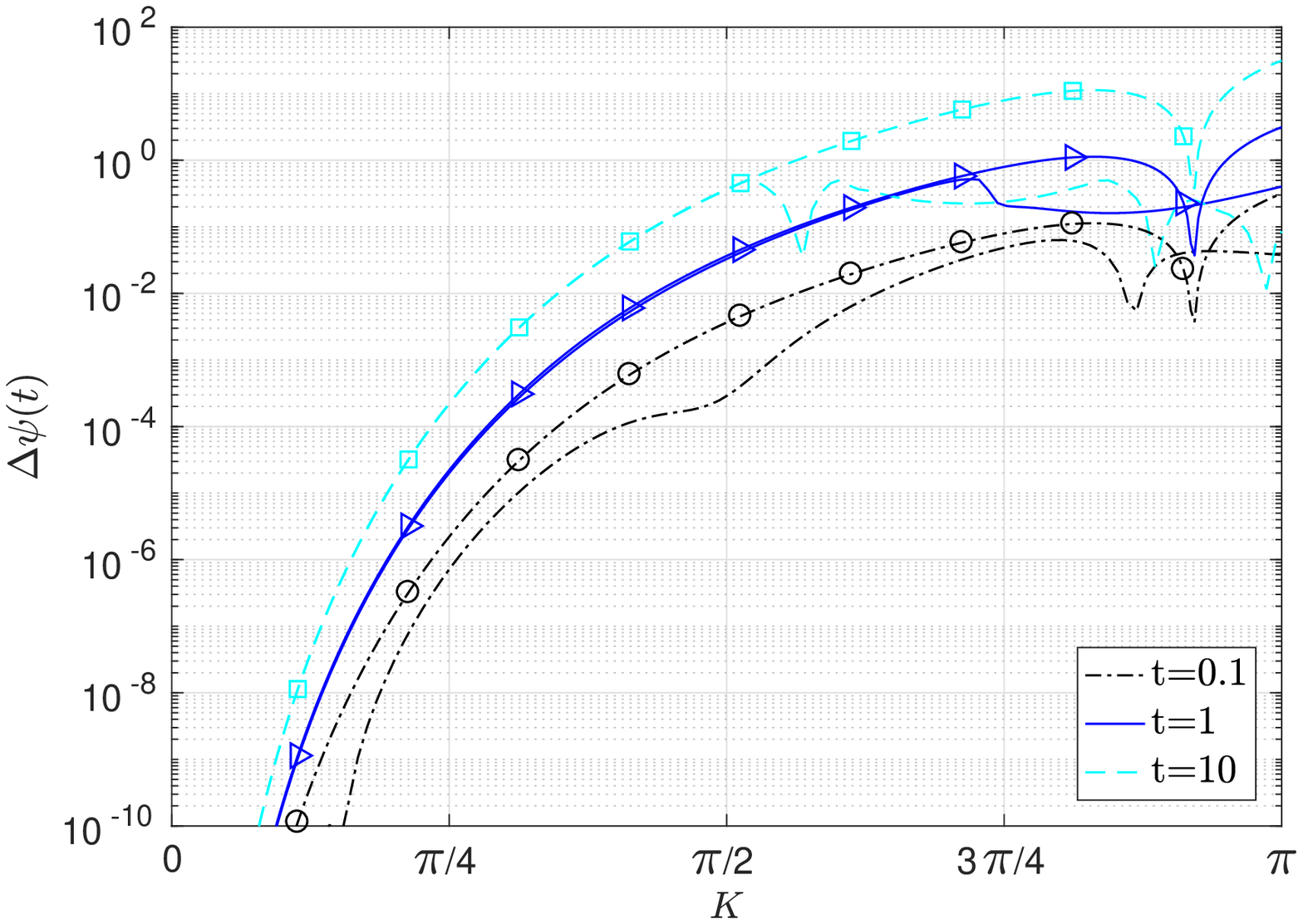}  
  \caption{Phase error}
 \label{p5_beta1_sdisc_S_err}
 \end{subfigure}
 \caption{Comparison of the semi-discrete true behavior (indicated by "plain" curves) and physical-mode behavior (indicated by "lines-with-symbols") for the DGp$5$ -$\beta1.0$ scheme at different time points. }
    \label{fig:p5_beta1_sdisc_truephys}
\end{figure}%
\hfigref{fig:p2_beta1_sdisc_truephys} shows the comparison of both $G$ and $\Delta \psi$ for DGp$2$-$\beta1.0$ scheme, i.e., DGp$2$ with upwind flux. In this figure the true behavior curves are the "plain" curves without any symbols, and the physical-mode curves are the ones with symbols. It is observed that for the low wavenumber range, the physical-mode  is a good approximation for the dispersion/dissipation behavior. In addition, the true behavior curve is usually more accurate, which indicates that secondary modes appear to reduce the dispersion/dissipation error.

In the high wavenumber range, the  dispersion errors are very high, and the behavior of the scheme does not simply follow the physical mode. Similar results are demonstrated by the DGp$5$-$\beta1.0$ scheme in~\hfigref{fig:p5_beta1_sdisc_truephys}, where the difference between the true behavior and physical-mode behavior becomes more remarkable in the high-wavenumber range. For nonlinear problems, energy piles up in the high wavenumber regime without sufficient numerical dissipation, which may cause the simulation to diverge. It is therefore desirable to have numerical dissipation in the high wavenumber regime. 

\begin{figure}[H]
\begin{subfigure}[h]{0.475\textwidth}
\centering
 \includegraphics[width=\textwidth]{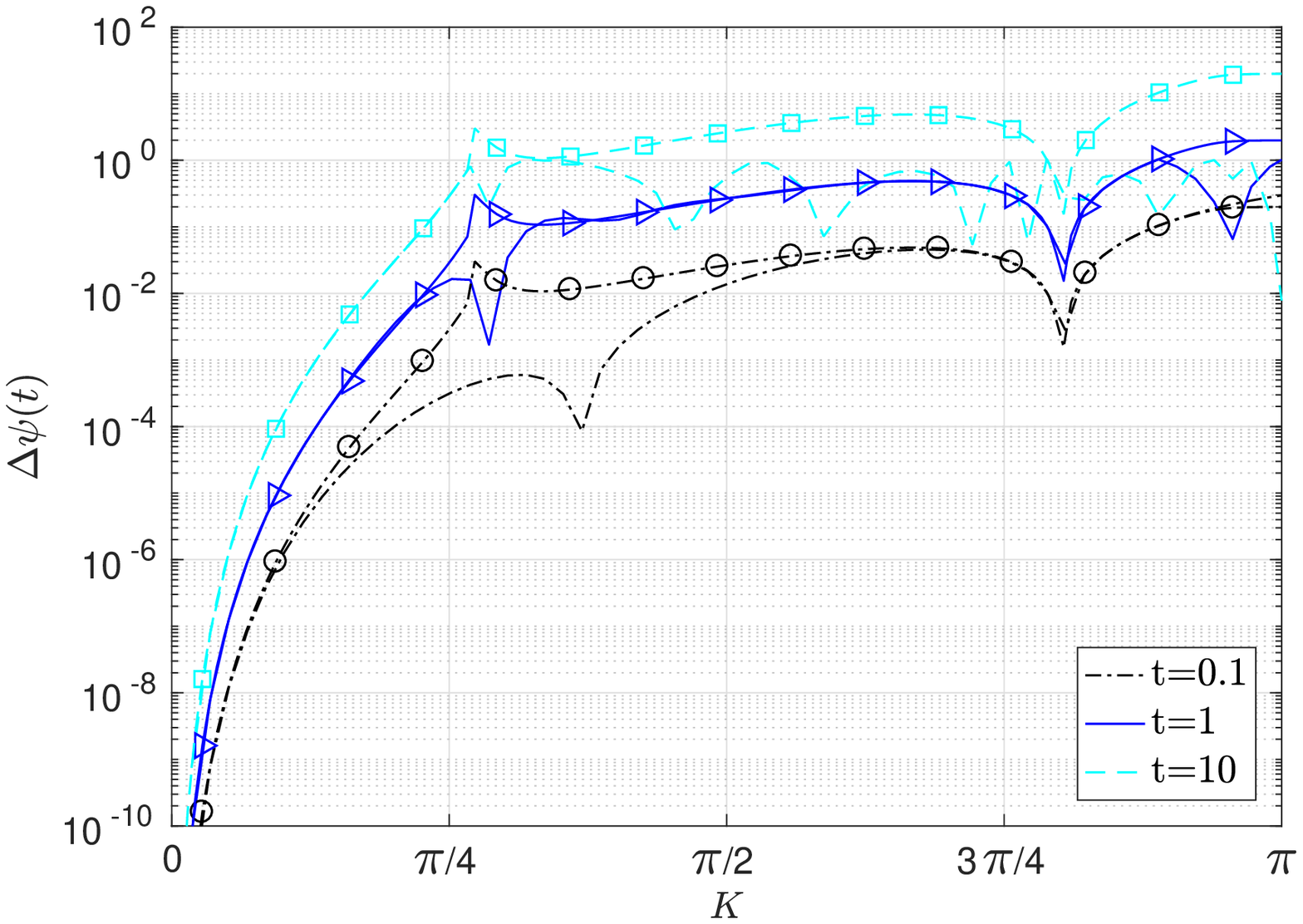} 
 \caption{DGp$2$-$\beta0.0$}
 \label{fig:p2_beta0_sdisc_S_err}
 \end{subfigure}
 \hspace{0.0125\textwidth}
 \begin{subfigure}[h]{0.475\textwidth}
\centering
 \includegraphics[width=\textwidth]{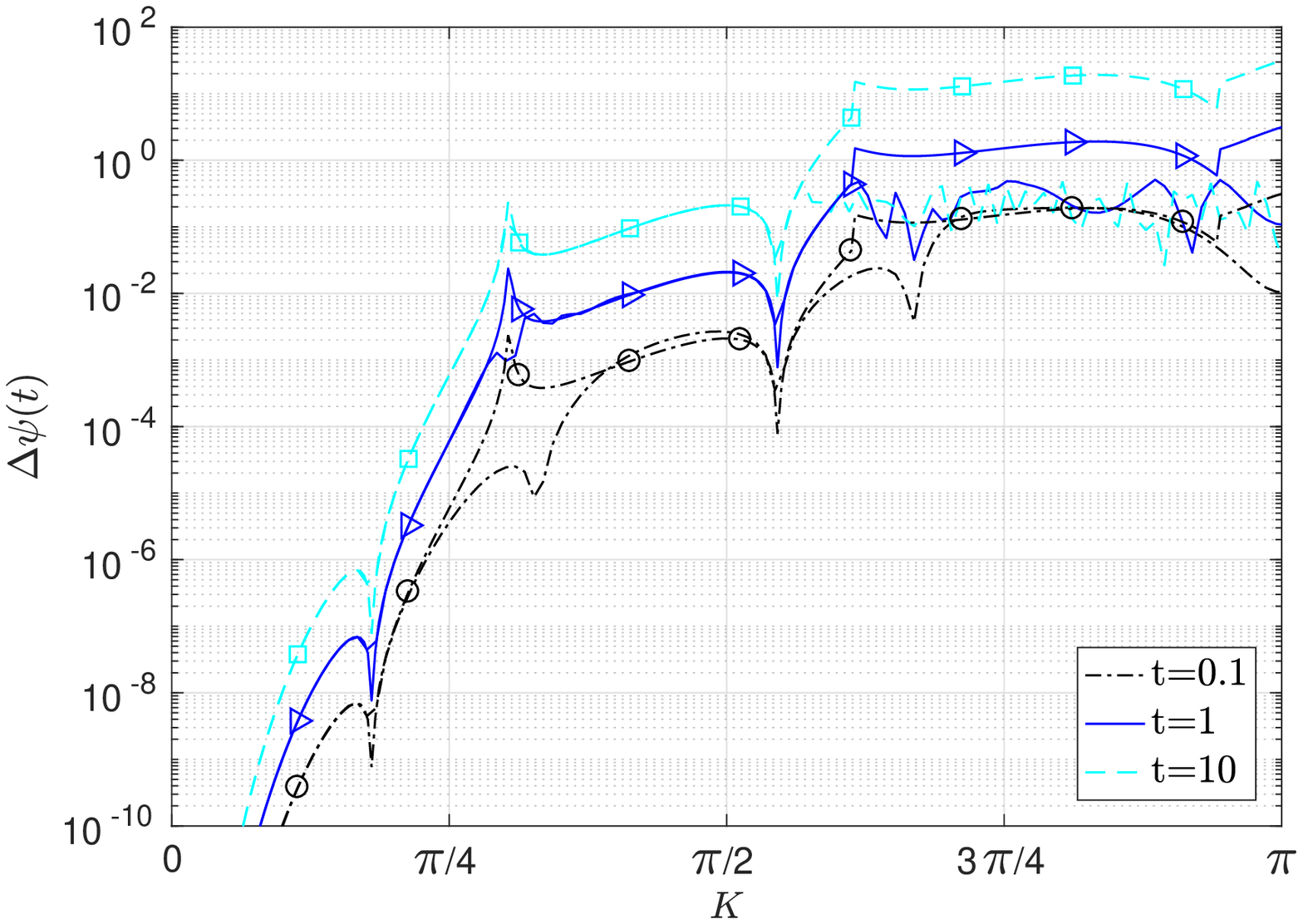}  
 \caption{DGp$5$-$\beta0.0$}
 \label{fig:p5_beta0_sdisc_S_err}
 \end{subfigure}
 \caption{Semi-discrete phase error for the DGp$2$-$\beta0.0$ and DGp$5$-$\beta0.0$ schemes, i.e., DG schemes with central fluxes, at different time points. Note that "lines-with-symbols" indicate (physical-mode behavior), where "plain-lines" indicate (true behavior).}
    \label{fig:p2p5_beta0_sdisc_S_err}
\end{figure}

\hfigref{fig:p2p5_beta0_sdisc_S_err} shows the phase error of DGp$2$-$\beta0.0$ and DGp$5$-$\beta0.0$ schemes, i.e., DG schemes with the central flux. From this figure it is evident that the two curves (true, physical) agree well with each other, before and after the spike at $K\approx \pi/4$, keeping in mind that the physical-mode used in these plots follows mode (1) of option (2) in~\hfigref{fig:p2_beta0_wp_compare_sdisc2}. This shows that there is no discontinuity in the true dispersion behavior of DG schemes with a central flux and that the physical-mode may follow two different dispersion curves before and after spike positions. In the fully-discrete analysis, this behavior is more apparent where discontinuities can be seen for dissipation curves as well, and this is discussed later in this paper.  %
In summary, we emphasize that all eigenmodes are significant and they affect the true behavior of the numerical solution. As a result, none should be identified as spurious nor parasite. Nevertheless, at least in the low wavenumber range, the mode that approximates the exact dispersion relation most accurately (the physical-mode)  is dominant especially as the wave moves forward in time. This was illustrated by Asthana et al.~\cite{AsthanaHighOrderFluxReconstruction2015} based on the relative energy content of each mode. The combined-mode analysis revealed that, even in the semi-discrete case, DG schemes always have a slower decaying rate in the high frequency region than what is expected by the physical-mode. This could lead to an accumulation of energy with high and nearly-constant dispersive errors for some time during the simulation even if exact-integration is performed, i.e., no aliasing is permitted. The results in this section were verified for different DG polynomial orders.%
%
\subsection{Semi-discrete analysis of FD schemes}\label{subsec:FDsd_3.2}%
%
Though FD schemes can be analyzed in a similar fashion as DG schemes, another approach is adopted here for simplicity. This approach is the "modified wave number analysis"~\cite{LeleCompactfinitedifference1992,MoinFundamentalsEngineeringNumerical2010}. Let the solution for the linear-advection~\heqref{eqn:lin_advec} be of the form $u_{j} = e^{i k x_{j}} \hat{u}_{j}(t)$ and hence the exact spatial derivative for this solution %
\begin{equation}
\frac{\partial u_{j}}{\partial x} = (ik) u_{j} ,
\end{equation}%
and by substituting into the linear-advection~\heqref{eqn:lin_advec}, we get the following exact relation%
\begin{equation}
\frac{\partial u_{j}}{\partial t} = i(-a k) u_{j} .
\label{eqn:Kex_modifiedanalysis}
\end{equation}%
After that, we substitute the same solution into one of the FD formulas~\heqsref{eqn:FD_cent_4th}{eqn:FD_cent_6th} to get a similar relation%
\begin{equation}
\frac{\partial u_{j}}{\partial t} = i (-a k_{m} ) u_{j} , 
\label{eqn:FD_sdisc_gen1}
\end{equation}%
where $k_{m}$ is the modified wavenumber. For instance, the FD$2$ scheme has a modified wavenumber of the form%
\begin{equation}
K_{m}(k) = k_{m} h =  \sin(K), \quad K=kh,
\label{eqn:FD2_Km_modified_rel}
\end{equation}%
where $K, K_{m}$ are non-dimensional wavenumbers based on the smallest length-scale $h$ that can be captured by a FD scheme. For central schemes, the non-dimensional modified wavenumber $K_{m}$ is a pure real number, whereas for upwind schemes it is in general complex, see~\hyperref[appx:A]{Appendix~A} for more modified wavenumber formulas of FD schemes. \heqref{eqn:FD2_Km_modified_rel} is often called the numerical dispersion relation which when compared with the exact dispersion relation reveals the dispersion/dissipation characteristics of a given scheme. In this case the semi-discrete operator becomes  $\mathcal{A}= -i (a k_{m})=-i(\frac{a}{h} K_{m})$. Consequently, the modified wavenumber $K_{m}$ is compared with $K=kh$, the exact wavenumber, as in the case of DG schemes,~\heqsref{eqn:DG_num_sdisc_disper_rel}{eqn:DG_num_sdisc_dissip_rel}, for assessing dispersion/dissipation behavior of FD schemes. 

\begin{figure}[H]
\begin{subfigure}[h]{0.475\textwidth}
    \centering
    \includegraphics[width=0.97\textwidth]{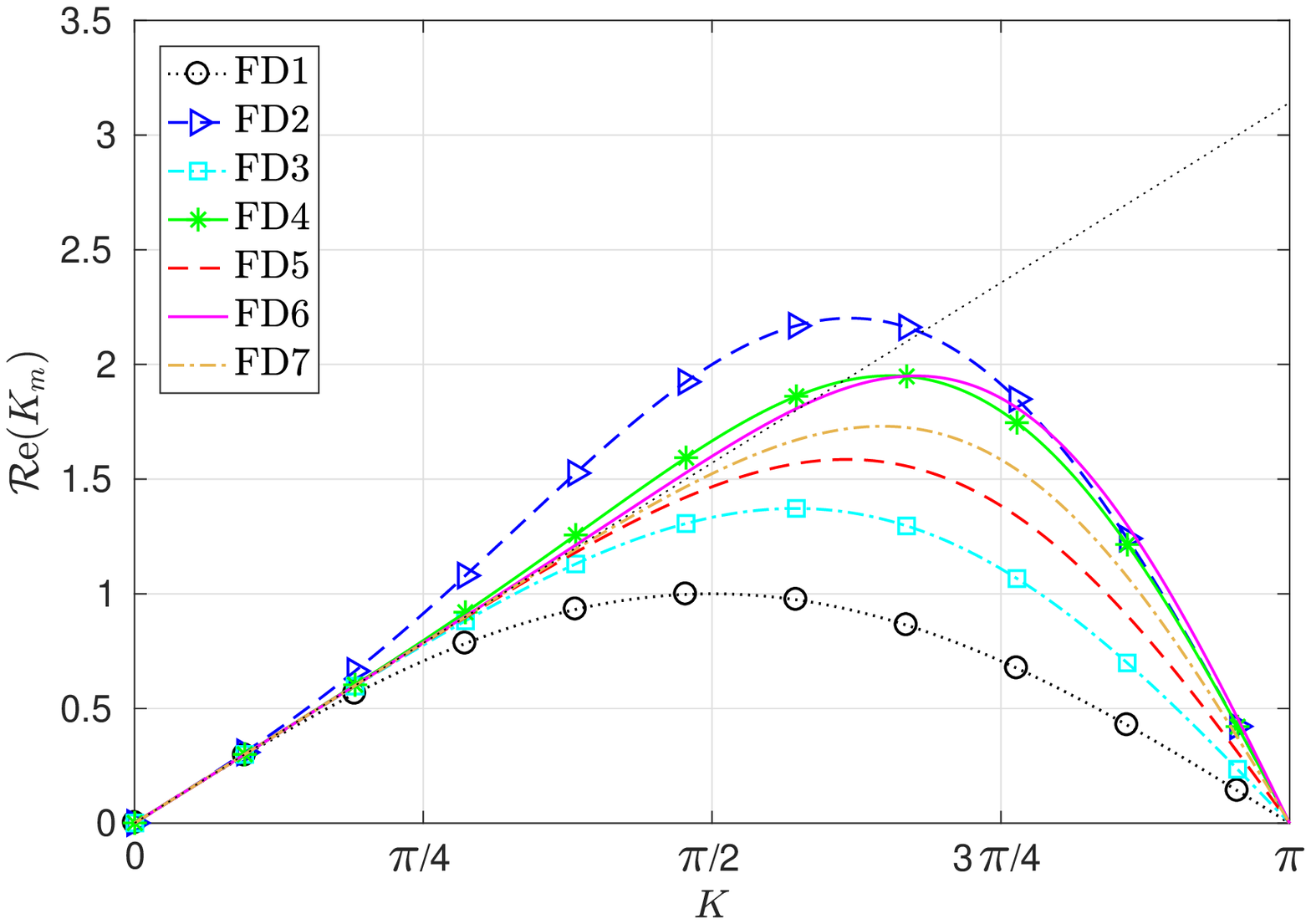} 
    \caption{Dispersion}
    \label{fig:FD_sdisc_compare_bias_zoom}
    \end{subfigure}
    \hspace{0.0125\textwidth}
     \begin{subfigure}[h]{0.475\textwidth}
    \centering
    \includegraphics[width=0.97\textwidth]{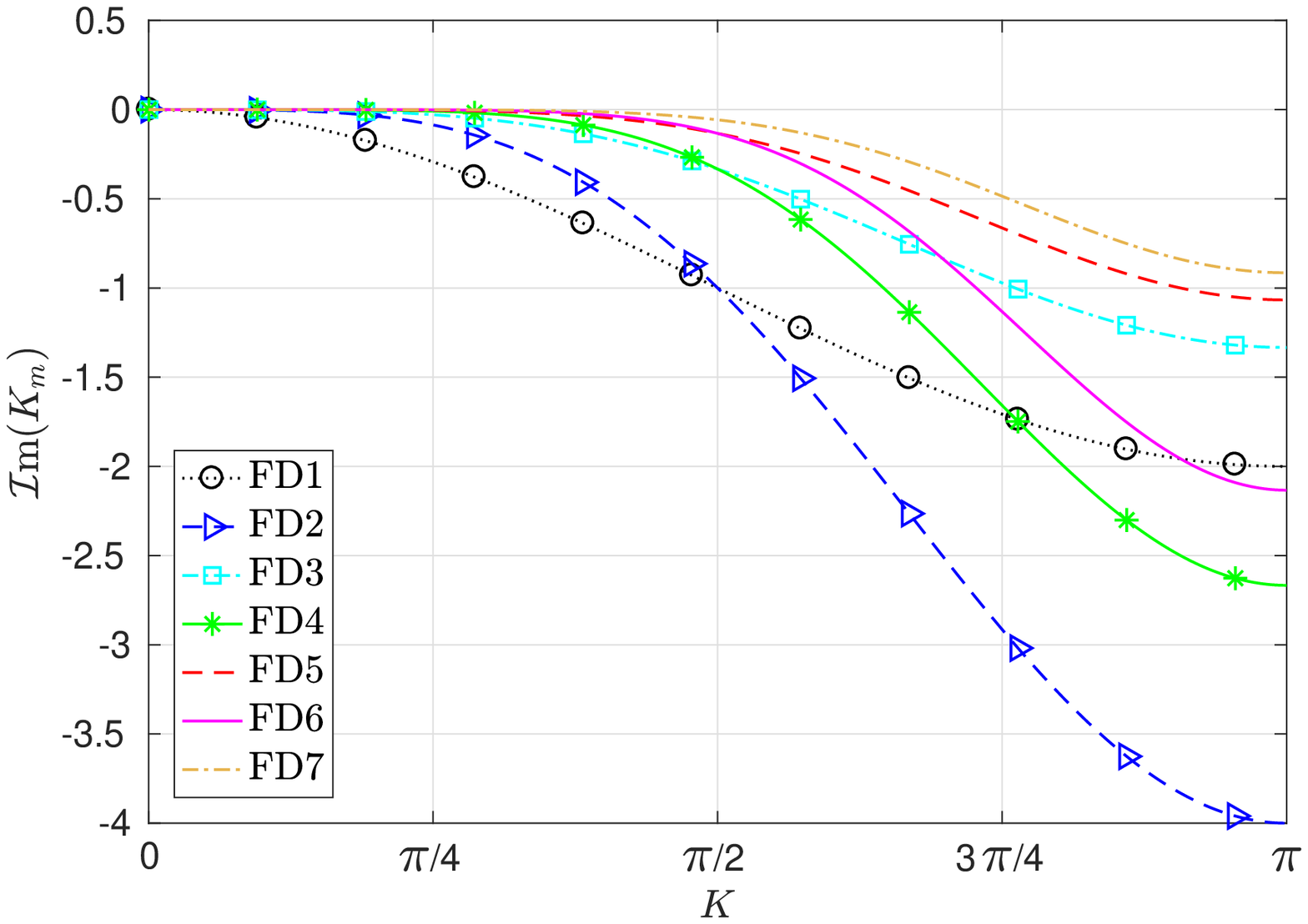}  
    \caption{Dissipation}
    \label{fig:FD_sdisc_compare_bias_a}
    \end{subfigure} 
	\caption{Comparison of the semi-discrete dispersion/dissipation behavior for FD schemes. Note that all the schemes in this figure are upwind-biased. Central schemes has zero dissipation while their dispersion is equivalent to the dispersion of FD schemes with one order lower. }
    \label{fig:FD_sdisc_compare_bias}
\end{figure}

FD schemes have been analyzed extensively, and some major results were proven analytically by Iserles~\cite{IserlesOrderStarsSaturation1982}. Here as a means of verifying our analysis, we review some known results. For example, there is no fully upwind scheme that is stable beyond the $2^{nd}$ order upwind scheme. Instead upwind biasing is needed to achieve stability. \hfigref{fig:FD_sdisc_compare_bias} illustrates the dispersion and dissipation characteristics of various stable upwind and upwind-biased FD schemes. All odd-order schemes have a one-point bias (one more point on the upwind side), while all even-order schemes have a two-point bias. Another interesting property is that all even-order central schemes have the same dispersive behavior as the corresponding upwind-biased scheme of one order lower. This can be seen clearly from the modified wavenumber relation for each scheme included in~\hyperref[appx:A]{Appendix~A}.%
%
\subsection{Semi-discrete analysis of CD schemes}\label{subsec:CDsd_3.3}%
%
In order to perform a Fourier analysis for CD schemes we utilize again the "modified wave number analysis". Let the solution for the linear-advection~\heqref{eqn:lin_advec} be of the form $u_{j} = e^{i k x_{j}} \hat{u}_{j}(t)$. Then we substitute the same assumed solution into~\heqref{eqn:CD_formula} for the compact scheme and perform differentiation to arrive at the following relation for the non-dimensional modified wavenumber%
\begin{equation}
K_{m}(K) = k_{m}h =  \frac{1}{2} \left(\frac{c \: \sin(2 K) + 2 d \: \sin(K)}{1 + 2 \alpha \cos(K)} \right) , \quad K=kh.
\label{eqn:CD_disper_relation}
\end{equation}%
This is what is called the dispersion-relation for compact schemes~\cite{LeleCompactfinitedifference1992} and the semi-discrete operator is given by $\mathcal{A}=-i(\frac{a}{h} K_{m})$. Note that non-dimensionalization is performed here with respect to the smallest length-scale $h$ that can be captured by a CD scheme. Based on the fact that $K_{m}$ is a real number for CD schemes, one expects that $K_{m} \approx K$ for consistent dispersion behavior, while there is no dissipation to be introduced by the spatial scheme. This is due to the nature of difference schemes that are based on central stencils. 

\begin{figure}[H]
	\begin{subfigure}[h]{0.475\textwidth}
    \centering
    \includegraphics[width=0.97\textwidth]{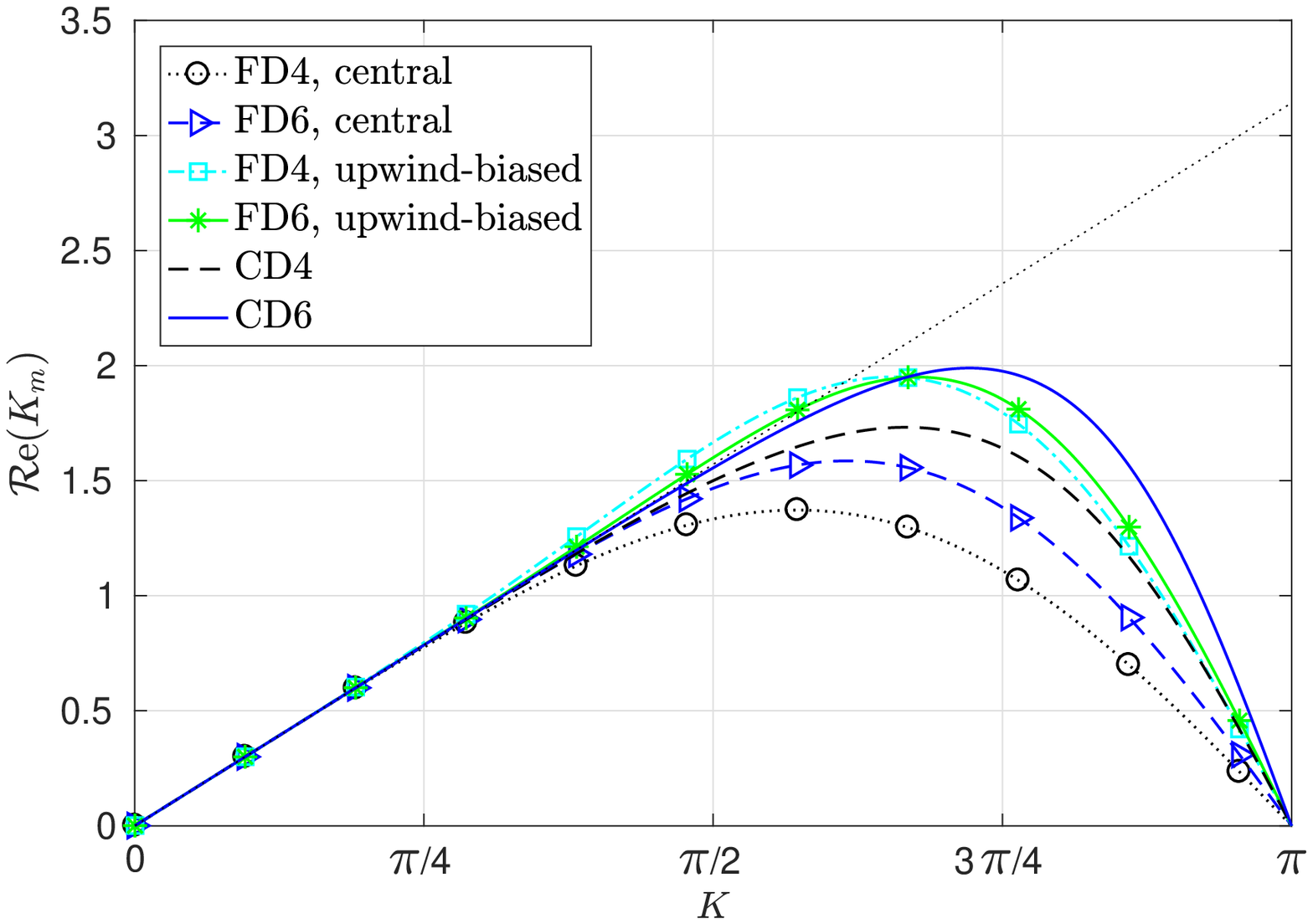} 
    \end{subfigure}
    \hspace{0.0125\textwidth}
    \begin{subfigure}[h]{0.475\textwidth}
    \centering
    \includegraphics[width=0.97\textwidth]{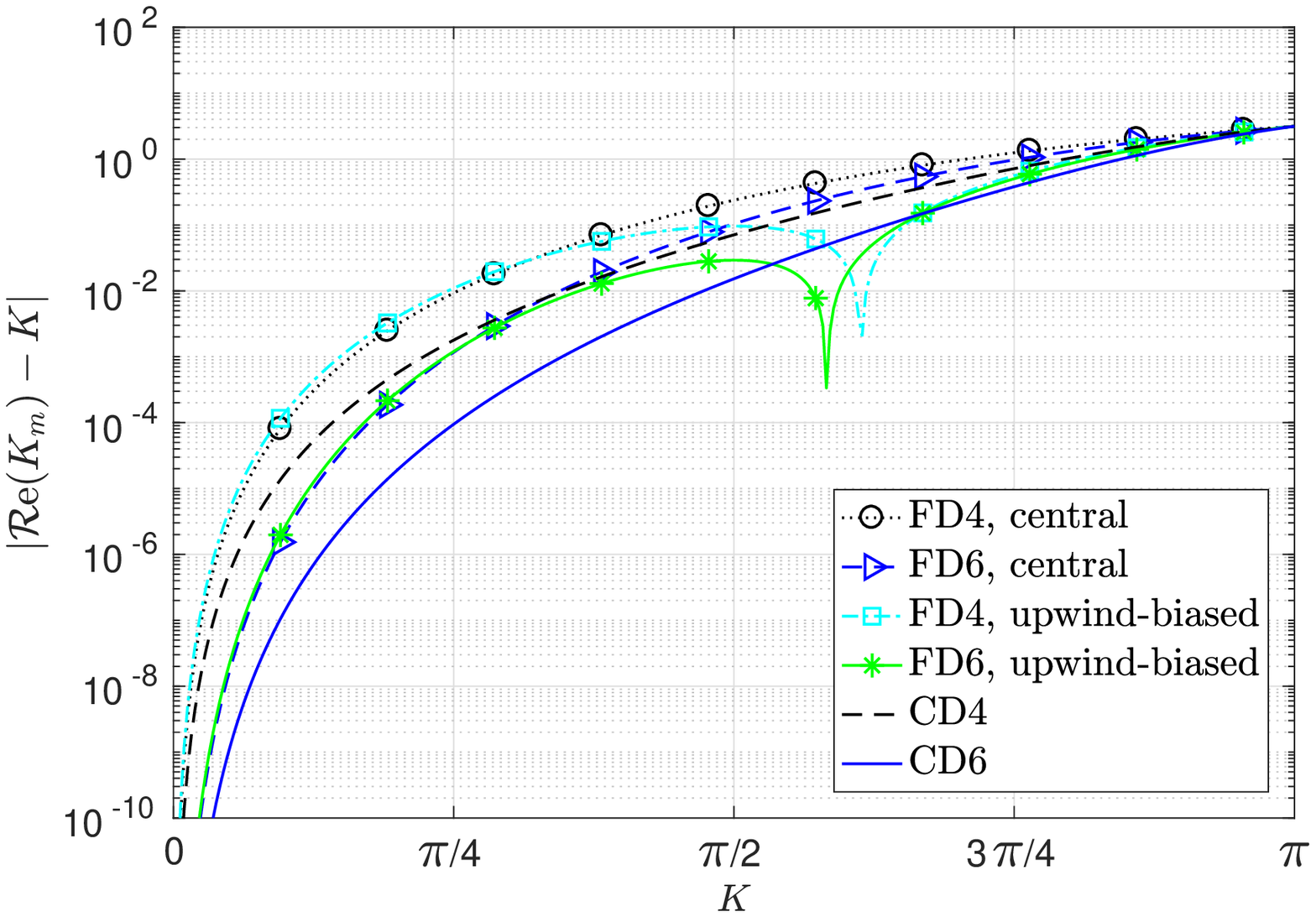} 
    \end{subfigure}
    \caption{Comparison of the semi-discrete dispersion behavior for central finite difference and compact difference schemes.}
    \label{fig:FD_CD_sdisc_wp}
\end{figure}

A comparison of the dispersive behavior of CD schemes and central and upwind-biased FD schemes of the same orders is presented in~\hfigref{fig:FD_CD_sdisc_wp}, where the advantage of the spectral-like resolution of CD schemes can be noticed. It is also inferred that for low wavenumbers the dispersion error of both central and upwind-biased FD schemes of the same order of accuracy is very close, whereas upwind-biased is more accurate for the high wavenumber range. %
%
\subsection{Fully-discrete analysis of DG schemes} \label{subsec:DGfd_3.4}%
%
In order to analyze the dispersion and dissipation properties of a fully-discrete (space-time) scheme, both time and space discretizations are applied to the linear-advection~\heqref{eqn:lin_advec}. Applying RK time discretization to one of the semi-discrete equations in~\secref{sec:num_methods_2} results in an update formula for the solution at $t^{n+1}= (n+1) \Delta t$ of the form%
\begin{equation}
\vec{u}^{n+1} = \mathcal{G} \vec{u}^{n} = \mathcal{G}^{n+1} \vec{u}^{0} , \quad \vec{u}^{0} = \vec{u}(x,0), \quad \mathcal{G} = \mathcal{G}(\mathcal{A}) ,
\label{eqn:fdisc_temp_form} 
\end{equation}%
where $\mathcal{G}$ is the full-discretization operator that is a function of the semi-discrete operator $\mathcal{A}$, and $\vec{u}^{n+1}$ is the vector of unknown DOFs. 

The method used in this section for high-order DG-type methods is similar to the one used by Yang et al.~\cite{YangDispersionDissipationErrors2013}, Vermeire et al.~\cite{Vermeirebehaviourfullydiscreteflux2017}, and Vanharen et al.~\cite{VanharenRevisitingspectralanalysis2017}. We proceed by seeking a wave solution in element $\Omega_{e}$ of the form %
\begin{equation}
\vec{U}^{e,n+1} = \vec{\mu} \: e^{i\left( k x_{e} - \tilde{\omega} t^{n+1} \right)} = e^{- i \tilde{\omega} \Delta t} \vec{U}^{e,n} ,
\label{eqn:blochwave_sol_fulldisc}
\end{equation}%
and substituting in~\heqref{eqn:fdisc_temp_form} yields the following fully-discrete relation for element $\Omega_{e} \in \mathcal{D}$%
\begin{equation}
e^{-i \tilde{\omega} \Delta t} \vec{\mu} = \mathcal{G} \vec{\mu} ,
\label{eqn:DG_full_disc}
\end{equation}%
where $\tilde{\omega}$ is the numerical frequency which is in general complex, and this relation constitutes an eigenvalue problem similar to the case of semi-discrete analysis but with different structure. The eigenvalue problem results in a $p+1$ $(\lambda_{j}, \mu_{j}), \; j=0,...,p$, eigenpairs. The general representation of the vector of spatial solution coefficients can be expressed as a linear combination of the eigenmodes as follows%
\begin{equation}
\vec{U}^{e,n} = \sum_{j=0}^{p} \vartheta_{j} \lambda_{j}^{n} \vec{\mu}_{j} \: e^{i k x_{e} } ,
\label{eqn:lin_combination_eigvec_fulldisc_Un}
\end{equation}%
where the expansion coefficients $\vartheta_{j}$ are again given by~\heqref{eqn:eta_relation1}, and the general solution follows the same~\heqsref{eqn:DG_sdisc_numsol_compact_form}{eqn:DG_sdisc_exactsol_compact_form} with $\Theta_{j} =  \vartheta_{j} \lambda_{j}^{n} = \vartheta_{j} e^{-i \tilde{\omega}_{j} (n \Delta t) } $. To this end, it is clear that the eigenvalues, $\lambda_{j}$, can generally be complex and they are related to the numerical frequency $\tilde{\omega}$ through the following relation%
\begin{equation}
\lambda_{j} = e^{-i \tilde{\omega}_{j} \Delta t} = e^{-i \frac{a \Delta t}{h} \frac{\tilde{\omega}_{j} }{a} \frac{h}{p+1} (p+1)} = e^{-i (p+1) \sigma  K_{m,j}  } , \quad j=0,...,p ,
\label{eqn:full_disc_lambda_K_rel1}
\end{equation}%
where $\sigma$ is the CFL number, and the modified wavenumber $K_{m}=k_{m}h/(p+1)$ can be obtained from%
\begin{equation}
K_{m} = \frac{i \: ln(\lambda)}{(p+1) \sigma} = \operatorname{\mathcal{R}e}(K_{m}) + i \operatorname{\mathcal{I}m} (K_{m}) .
\label{eqn:full_disc_lambda_K_rel2}
\end{equation}%
Therefore, a numerical dispersion relation can be written as %
\begin{equation}
\operatorname{\mathcal{R}e}(K_{m}(K))  \approx K ,
\label{eqn:DG_num_full-disc_disper_relation}
\end{equation}%
where $K = kh/(p+1)$. For stability, the numerical dissipation behavior is required to satisfy%
\begin{equation}
\operatorname{\mathcal{I}m} (K_{m}(K) ) \leq 0.
\label{eqn:DG_num_full-disc_dissipation_relation}
\end{equation}%
Numerical time integration modifies the eigenvalues of the semi-discrete scheme using the same polynomial, see~\heqref{eqn:RK_update_form_DG}, that defines the amplification factor $\mathcal{G}$ as a function of the semi-discrete operator $\mathcal{A}$,%
\begin{equation}
\mathcal{G} = \mathcal{P}(\Delta t \mathcal{A}), \quad \lambda_{\mathcal{G}} = \mathcal{P}(\Delta t \lambda_{\mathcal{A}}) .
\label{eqn:G_Lambda_Poly}
\end{equation}%
Thus the behavior of the fully-discrete scheme depends on both the time-step $\Delta t$ or CFL number besides the form of the polynomial $\mathcal{P}$.

We acknowledge that stability limits for the upwind RKDG schemes with upwind flux ($\beta=1$) were provided in Cockburn et al.~\cite{CockburnRungeKuttaDiscontinuous2001}. Using our Fourier analysis toolbox we obtained the stability limits for a number of RKDG schemes up to order $p=5$ with both upwind ($\beta=1.0$) and central ($\beta=0.0$) fluxes. Stability limits are investigated by assuming a range of $K\in [0,\pi]$ and checking if the $\mathcal{I}\text{m}(K_{m})$ associated with any spatial eigenvalue $\lambda_{j}$ is greater than $0$. This determines the first CFL number that renders the scheme unstable. Approximate stability limits for RKDG schemes for both upwind and central fluxes are provided in~\hyperref[appx:C]{Appendix~C}.%
%
\subsubsection{True behavior of DG schemes through combined-mode fully-discrete analysis }\label{subsubsec:DGfd_true_3.4.2}%
%
Following the same procedure introduced in~\secref{subsubsec:DGsd_true_3.1.2} for investigating the true-behavior of semi-discrete DG schemes, we proceed to apply the same idea to the fully-discrete RKDG schemes. Based on the fact that dispersion and dissipation in the fully-discrete case depends on both the CFL number  and the number of iterations $n$, the amplification factor $G^{phys}(k,t_{n})$ of the physical-mode after $n$ iterations is given by%
\begin{equation}
G^{phys}(k,t_{n}) = e^{\operatorname{\mathcal{I}m} (\tilde{\omega}) n \Delta t } = e^{n \left((p+1) \sigma \operatorname{\mathcal{I}m}(K_{m}) \right)  },
\label{eqn:fdisc_phys_G}
\end{equation}%
and the phase error%
\begin{equation}
\Delta \psi^{phys}(k,t_{n}) = n\sigma | ( \operatorname{\mathcal{R}e}(K_{m})) - K ) | , 
\label{eqn:fdisc_phys_S_err}
\end{equation}%
where $\sigma $ is the CFL number. For the true quantities, the amplification factor $G^{true}(k,t_{n})$ is given by~\heqref{eqn:sdisc_true_G}, the phase shift $\psi^{true}(k,t_{n})$ by~\eqref{eqn:sdisc_true_phase_shift}, and the true fully-discrete phase error can be expressed as%
\begin{equation}
\Delta \psi^{true}(k,t_{n}) = | \psi(k,t_{n}) | /(p+1).
\label{eqn:fdisc_true_S_err}
\end{equation}

Using the above definitions we carry out a combined-mode fully-discrete analysis to verify our observations in the semi-discrete case. The dispersion/dissipation criteria used in this section is similar to what was proposed in~\cite{VanharenRevisitingspectralanalysis2017} to study the fully-discrete behavior of SD schemes coupled with RK time integration schemes for $kh>\pi$.

\begin{figure}[H]
\centering
    \begin{subfigure}[h]{0.475\textwidth}
    \centering
    \includegraphics[width=\textwidth]{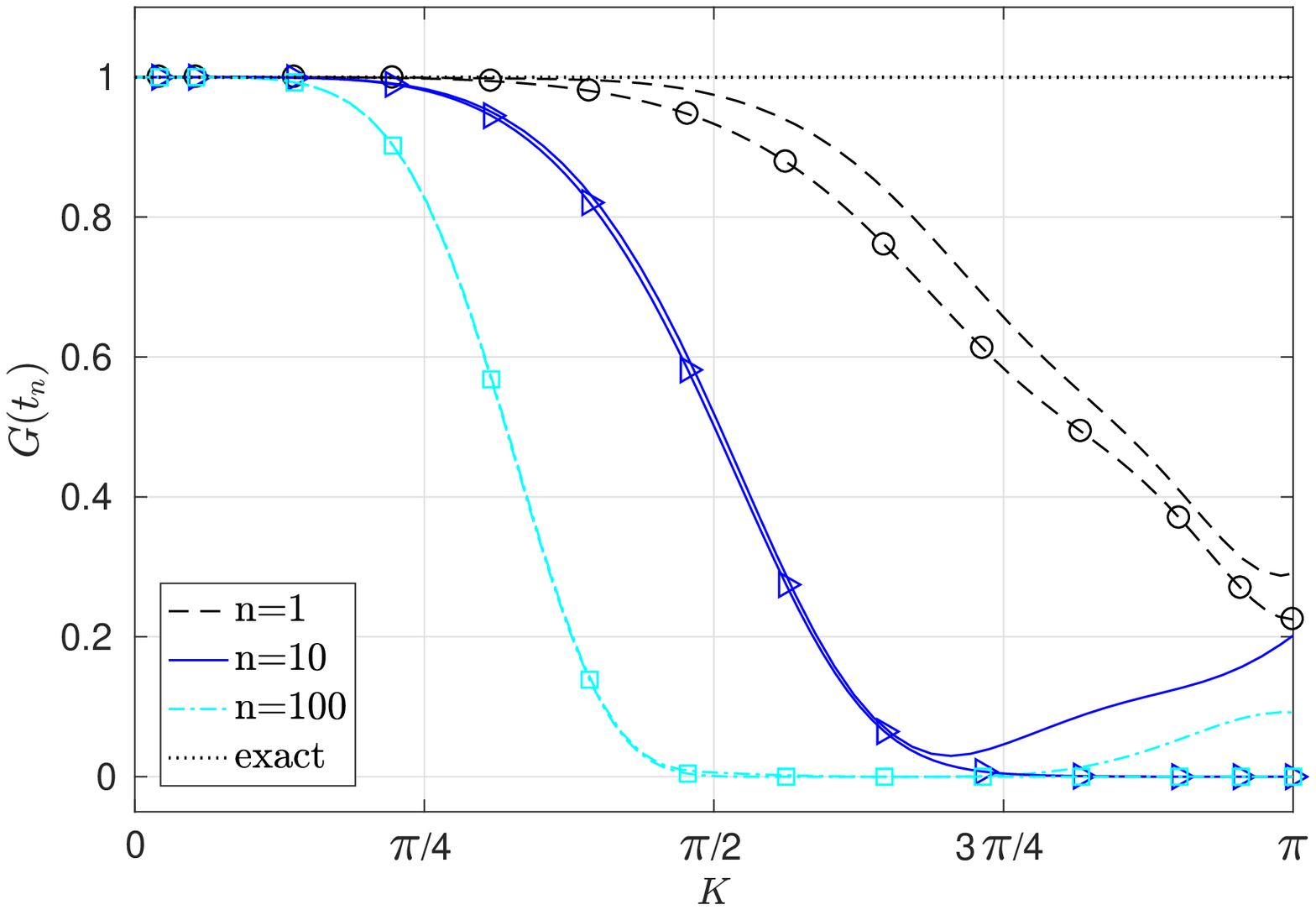}  
     \caption{Amplification factor }
    \label{fig:p2_RK3_fdisc_G_beta1}
    \end{subfigure} 
    \hspace{0.0125\textwidth}
    \begin{subfigure}[h]{0.475\textwidth}
    \centering
   \includegraphics[width=\textwidth]{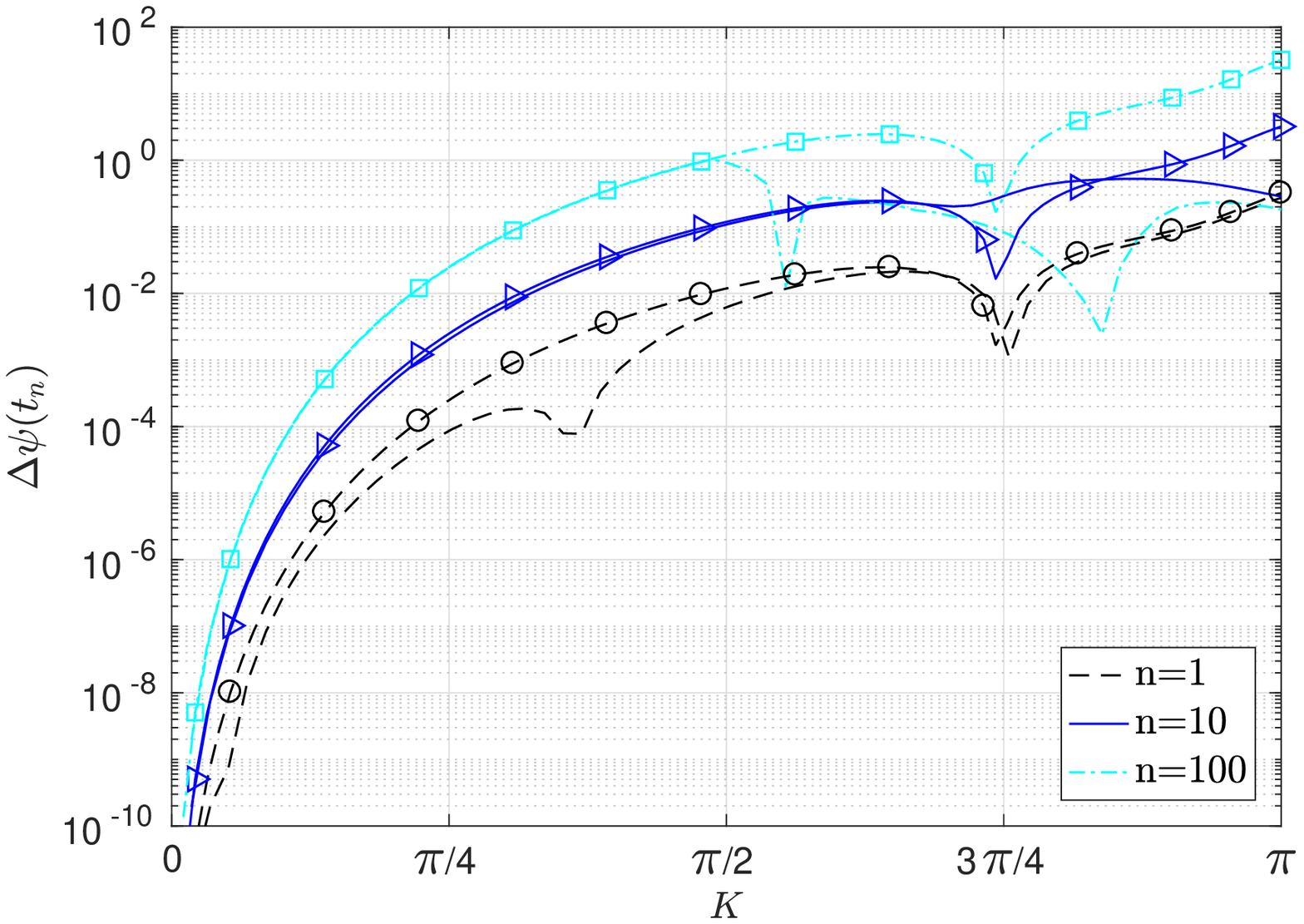}  
   \caption{Phase error}
    \label{fig:p2_RK3_fdisc_S_beta1}
    \end{subfigure}
	\caption{Comparison of the fully-discrete true behavior and physical-mode behavior of DGp$2$-$\beta1.0$ coupled with RK$3$, at CFL$=0.5$CFL$_{max}$. Note that "lines-with-symbols" indicate (physical-mode behavior), while "plain-lines" indicate (true behavior).}
    \label{fig:p2_RK3_fdisc_05cfl_beta1_true}
\end{figure}%
\begin{figure}[H]
\centering
    \begin{subfigure}[h]{0.475\textwidth}
    \centering
    \includegraphics[width=\textwidth]{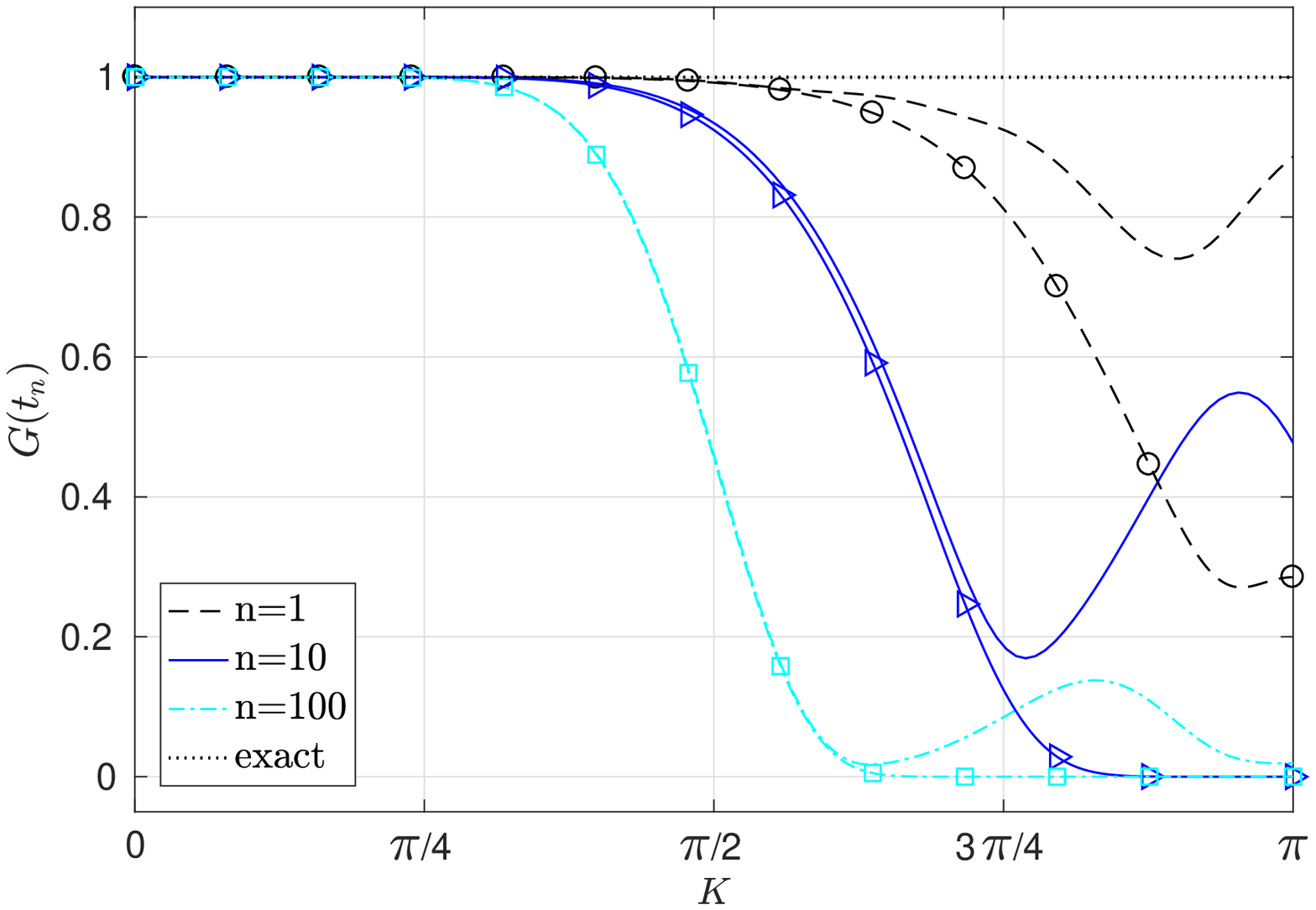}  
    \caption{Amplification factor }
    \label{fig:p5_RK4_fdisc_G_beta1}
    \end{subfigure} 
    \hspace{0.0125\textwidth}
    \begin{subfigure}[h]{0.475\textwidth}
    \centering
   \includegraphics[width=\textwidth]{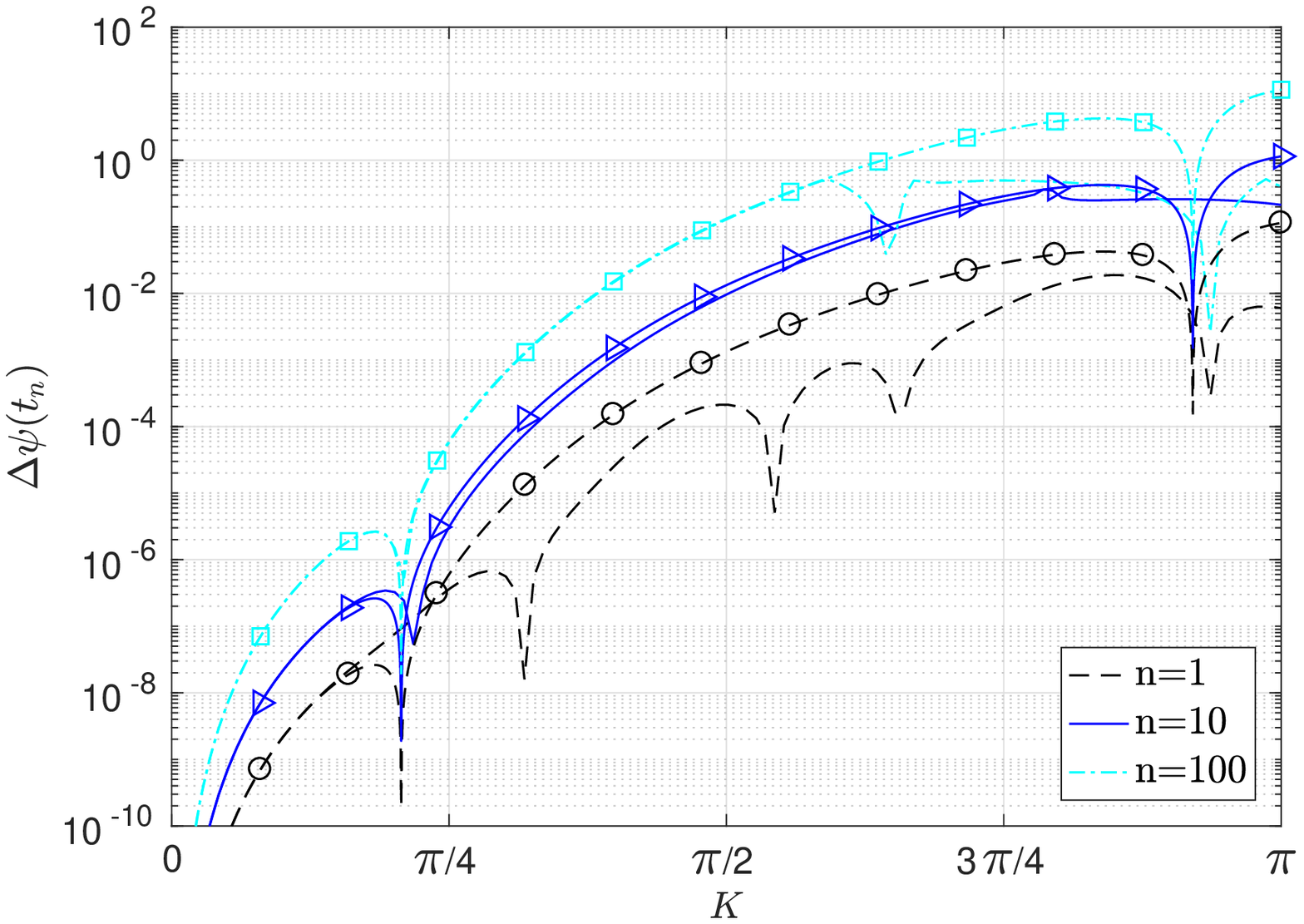}  
    \caption{Phase error}
    \label{fig:p5_RK4_fdisc_S_beta1}
    \end{subfigure}
	\caption{Comparison of the fully-discrete true behavior and physical-mode behavior of DGp$5$-$\beta1.0$ coupled with RK$4$, at CFL$=0.5$CFL$_{max}$. Note that "lines-with-symbols" indicate (physical-mode behavior), while "plain-lines" indicate (true behavior).}
    \label{fig:p5_RK4_fdisc_05cfl_beta1_true}
\end{figure}%
\hfigref{fig:p2_RK3_fdisc_05cfl_beta1_true}, shows the true fully-discrete behavior of DGp$2$-$\beta1.0$ coupled with RK$3$ for time integration. From this figure it is apparent that the true behavior of DG schemes with upwind flux is always less dispersive and dissipative than the asymptotic behavior based on the physical-mode. Thus, secondary modes improve the approximation of RKDG schemes in the low wavenumber range. It is also evident in this figure that the physical-mode approximates the true behavior reasonably well in the low wavenumber range for both dispersion and dissipation. In contrast, in the high wavenumber range, the true behavior is totally different from the physical-mode behavior and a lower decaying rate is observed. This lower decaying rate could lead to an energy accumulation at the high frequency end for some time during the simulation and with high dispersive errors. The results in this section were verified for different CFL numbers,  polynomial orders, and different RK schemes (mainly RK$3$ and RK$4$) and similar observations were found, see for example~\hfigref{fig:p5_RK4_fdisc_05cfl_beta1_true}. 

Note that the true behavior is very similar to the physical-mode behavior in the low-wavenumber regime, while very different in the high-wavenumber regime. For instance,  about $50\%$ of the energy at $K=\pi$ still remains after $10$ iterations according to the true behavior analysis as shown in~\hfigref{fig:p5_RK4_fdisc_05cfl_beta1_true}, whereas no energy is left (for $K > 3\pi/4$) according to the physical-mode analysis. This figure illustrates the usefulness of the combined-mode analysis.%
\begin{figure}[H]
\centering
    \begin{subfigure}[h]{0.475\textwidth}
    \centering
    \includegraphics[width=\textwidth]{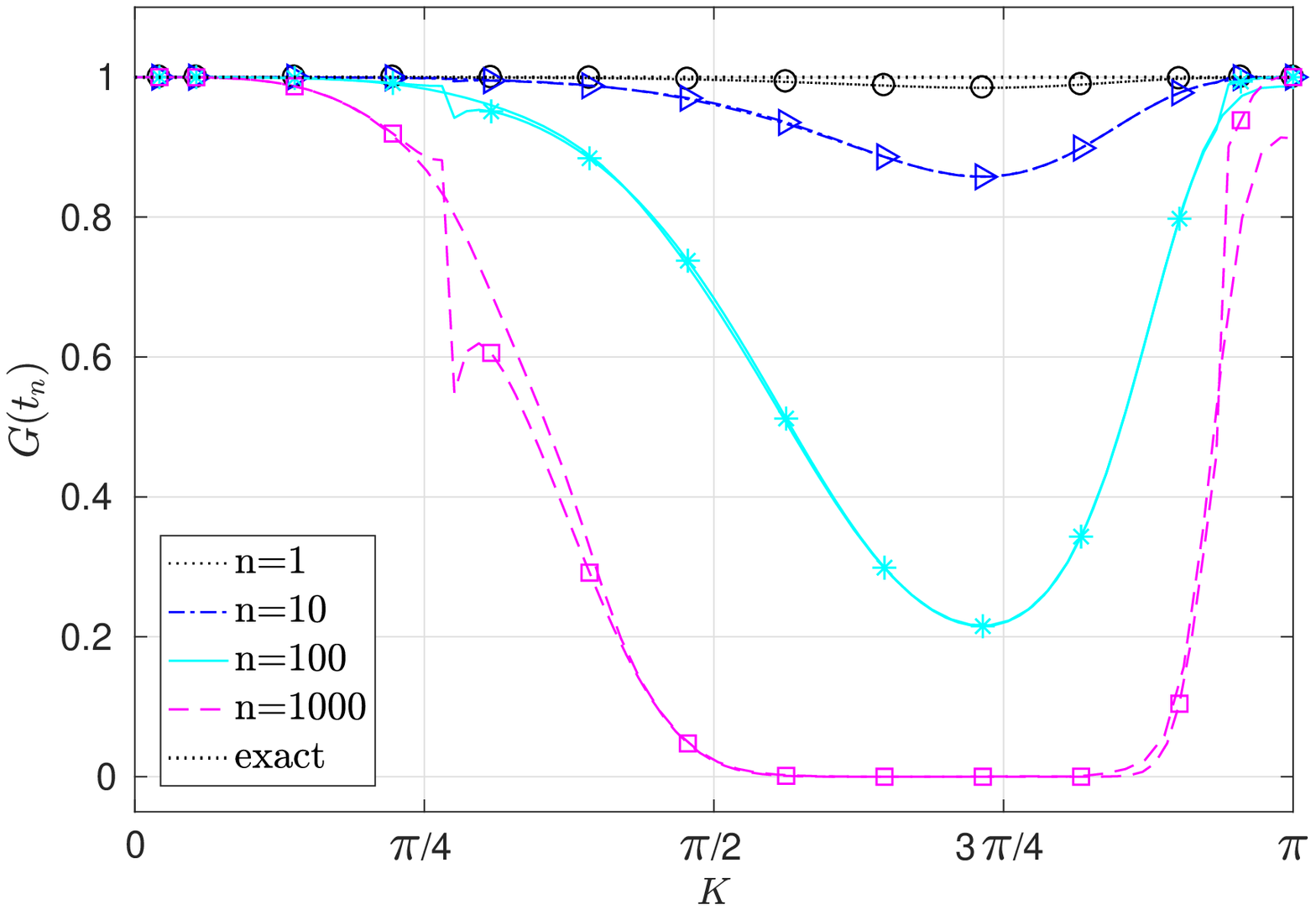}  
    \caption{Amplification factor}
    \end{subfigure} 
    \hspace{0.0125\textwidth}
    \begin{subfigure}[h]{0.475\textwidth}
    \centering
     \includegraphics[width=\textwidth]{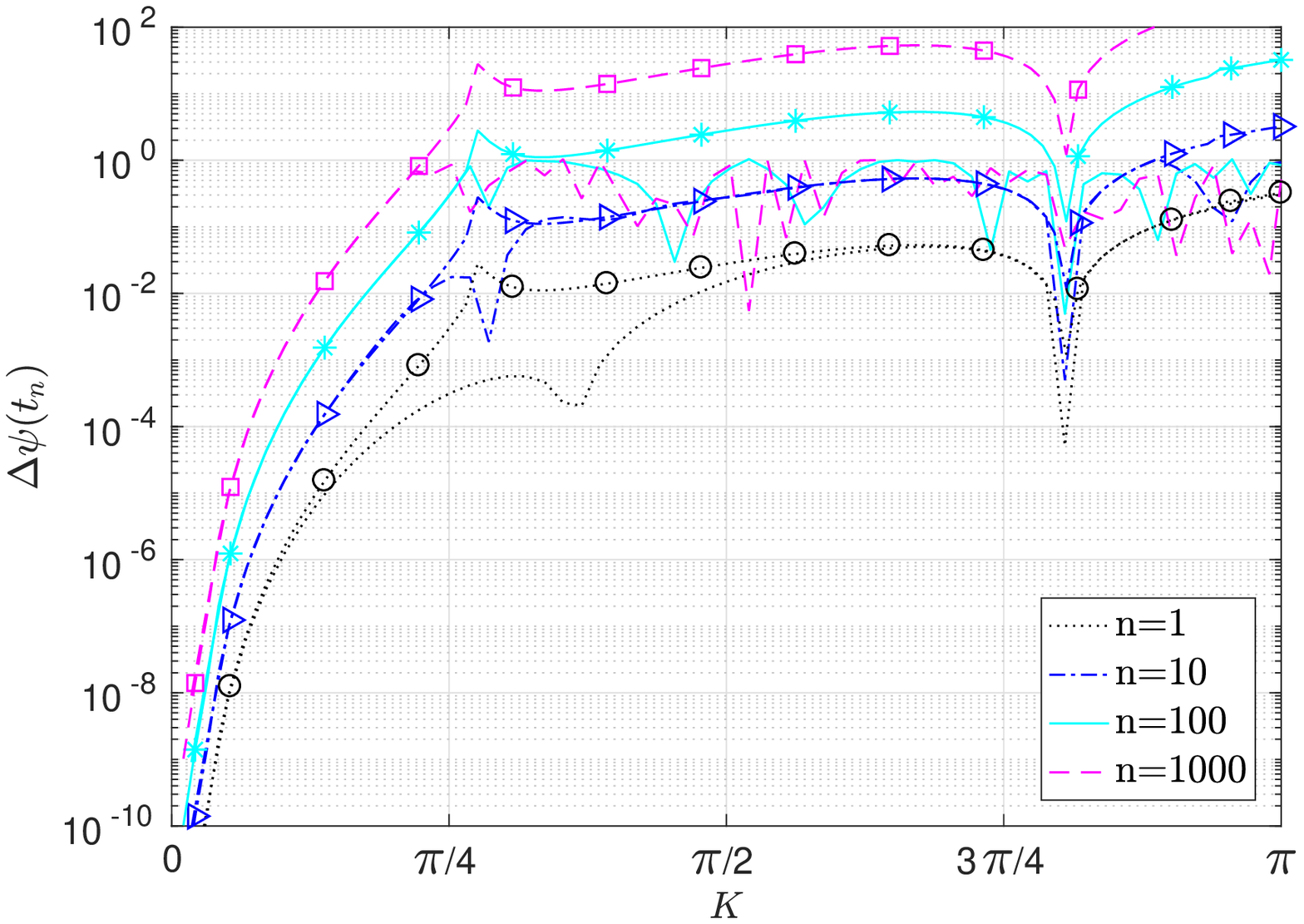}
    \caption{Phase error}
    \end{subfigure}
	\caption{Comparison of the fully-discrete true behavior and physical-mode behavior of DGp$2$-$\beta0.0$ coupled with RK$3$, at CFL$=0.5$CFL$_{max}$. Note that "lines-with-symbols" indicate (physical-mode behavior), while "plain-lines" indicate (true behavior).}
    \label{fig:p2_RK3_fdisc_05cfl_beta0_true}
\end{figure}%
In the case of a central flux, the results are presented in~\hfigref{fig:p2_RK3_fdisc_05cfl_beta0_true}. The transition region around the spike near $K \approx \pi/4$ confirms our observation for this case using the combined-mode semi-discrete analysis. In addition, the physical-mode is not defined by one curve for the entire wavenumber range, but rather it follows different curves before and after the spike positions as in the semi-discrete case. Moreover, at the Nyquist limit $K=\pi$, DG schemes with a central flux have a small amount of dissipation, in contrast to central FD and CD schemes coupled with RK schemes, which have no dissipation at all as is shown next. Unfortunately, this dissipation is too small making DG schemes with a central flux not a reliable choice for ILES.%
%
\subsection{Fully-discrete analysis of FD schemes}\label{subsec:FDfd_3.5}%
%
For FD schemes, we seek a solution at node $j$ and time level of $t^{n+1}= (n+1) \Delta t$ that takes the form%
\begin{equation}
u^{n+1}_{j} = e^{i(k x_{j} - \tilde{\omega} t^{n+1})} = e^{-i \tilde{\omega} \Delta t} u^{n}_{j} .
\label{eqn:fulldisc_FD_Un+1}
\end{equation}%
Substituting the above relation into the fully-discrete~\heqref{eqn:fdisc_temp_form} yields the following relation%
\begin{equation}
e^{-i \tilde{\omega} \Delta t} u^{n}_{j} = \mathcal{G} u^{n}_{j}, \quad \mathcal{G} = \mathcal{P}(\Delta t \mathcal{A}) ,
\label{eqn:FD_fulldisc_dispersion_rel}
\end{equation}%
where $\mathcal{P}(\Delta t \mathcal{A})$ is the time integration polynomial that results from the RK discretization and $\mathcal{A}$ is the semi-discrete operator of the FD scheme. This polynomial can be evaluated for each wavenumber $k$ and hence, $\mathcal{G}$ can be calculated. The value of $\mathcal{G}$ is in general a complex number and thus it can introduce both dispersion and dissipation errors into the solution. For stability the magnitude of $\mathcal{G}$ should satisfy $|\mathcal{G}| \leq 1$, and by defining the non-dimensional modified wavenumber as%
\begin{equation}
K_{m} = \frac{i \: ln(\mathcal{G})}{\sigma} = \operatorname{\mathcal{R}e}(K_{m}) + i \operatorname{\mathcal{I}m} (K_{m}) ,
\label{eqn:FD_fdisc_dissip_rel}
\end{equation}%
we can deduce the dispersion/dissipation properties of FD schemes as in the case of DG schemes. This $K_{m}$ serves as a numerical approximation to $K=kh$, and hence is required to satisfy the same relations as for DG schemes,~\heqsref{eqn:DG_num_full-disc_disper_relation}{eqn:DG_num_full-disc_dissipation_relation}. The stability limits for a number of FD schemes coupled with different RK schemes can be found in~\hyperref[appx:C]{Appendix~C}. %
%
\subsection{Fully-discrete analysis of CD schemes}\label{subsec:CDfd_3.6}%
%
In practice CD schemes are often accompanied by a Pad\'{e}  filter for enhanced stability. In a Fourier analysis, the contribution of the filter is to add only dissipation to the original non-dissipative CD scheme through its transfer function $\mathcal{T}(k)$. This is due to the fact that Pad\'{e} filters are non-dispersive. However, it is worth noting that even if no filter is used, a non-dissipative scheme has some dissipation when coupled with a RK time integration scheme. Similar to FD schemes, the fully-discrete relation for a complete CD scheme with a filter takes the form%
\begin{equation}
e^{-i \tilde{\omega} \Delta t} u^{n}_{j} = \mathcal{G} u^{n}_{j}, \quad \mathcal{G} = \mathcal{T}(K) \: \mathcal{P}(\Delta t  \mathcal{A}) ,
\label{eqn:CD_fulldisc_rel}
\end{equation}%
where $\mathcal{T}(K)$ is the filter transfer function that is obtained by applying the modified wavenumber analysis to the filter~\heqref{eqn:CD_filter_eqn} and $\mathcal{A}$ is the semi-discrete operator of the CD scheme. The transfer function $\mathcal{T}(K)$ for the $8^{th}$ order filter is provided in~\hyperref[appx:B]{Appendix~B}.

The same stability conditions and dispersion relation approximation holds as in the previous section of FD schemes but with the relevant $\mathcal{G}$ definition in~\heqref{eqn:CD_fulldisc_rel}. The stability limits for a number of CD schemes coupled with different RK schemes are provided in~\hyperref[appx:C]{Appendix~C}.

\section{Comparison of the dispersion/dissipation behavior of DG, FD, and CD schemes}\label{sec:DG_FD_CD_comp_4}%
%
In this section we focus on comparing RKDG schemes with other widely used schemes in ILES such as the CD schemes along with the classical central and upwind-biased FD schemes. The CD scheme of interest in this study is the C$6$F$8^{\alpha_{f}}$ scheme, which is of $6^{th}$ order in space coupled with an $8^{th}$ order Pad\'{e} filter with two values for the parameter $\alpha_{f}$. The first value is, $\alpha_{f}=0.49$, chosen to provide the least possible dissipation, and the second value, $\alpha_{f}=0.40$ which was recommended for some ILES simulations by Garmann et al.~\cite{GarmannComparativestudyimplicit2013}. For DG schemes, the DGp$5$-$\beta1.0$ scheme, i.e., DGp$5$ with upwind fluxes, is chosen.  Finally, the $6^{th}$ order central and 2-point upwind-biased FD$6$ schemes are also analyzed, denoted by FD$6$-central and FD$6$-upwind, respectively. In comparing fully-discrete schemes, the time step $\Delta t$ and corresponding CFL number play an important role. Therefore, we have defined two criteria for comparisons. The first one, is by assuming the same CFL ratio $r$ for all schemes, i.e., CFL of a certain scheme is given by%
\begin{equation}
\text{CFL} = \text{r} \times \text{CFL}_{max} ,
\label{eqn:CFL_ratio}
\end{equation}%
where CFL$_{max}$ is the stability limit for each respective scheme. In real world ILES, the time step is usually close to the stability limit to maximize solution efficiency. The second criterion is by requiring the same $\Delta t$ for all schemes so that the error in time integration is comparable. Note that for the latter case, if we fix $\Delta t$ for DGp$5$, C$6$F$8^{\alpha_{f}}$ and FD$6$  have a CFL$=(p+1)$CFL$_{DG}$ for the same nDOFs, and CFL$_{DG}$ is given by~\heqref{eqn:CFL_ratio} with some ratio $r$. All comparisons utilize RK4 for time integration. In order to quantify the dissipation error and resolution of all schemes we restrict our analysis to the low wavenumber range of ($0 \leq K\leq \pi/2$). This range was chosen such that for the DGp$5$-$\beta1.0$ scheme, the dissipation error is bounded by $\approx 10^{-2}$ which is the same criterion used in~\cite{MouraLineardispersiondiffusion2015} for assessing the resolution of DG schemes in ILES/uDNS simulations. In addition, based on our combined-mode analysis for DG schemes, we utilize the physical-mode behavior in this section since it serves as a good approximation for the true behavior of DG schemes in the low wavenumber range. When comparing different schemes, the non-dimensionalization with respect to the smallest length-scale that can be captured by a given scheme is necessary to ensure a fair and consistent comparison of dispersion/dissipation per degree of freedom. For the same nDOFs, $(p+1)$, this length-scale is $h=h_{DG}/(p+1)$, where $h$ is the mesh size of a FD/CD scheme, and $h_{DG}$ is the mesh size of a DG scheme. 

\begin{figure}[H]
\centering
\begin{subfigure}[h]{0.47\textwidth}
    \centering
    \includegraphics[width=\textwidth]{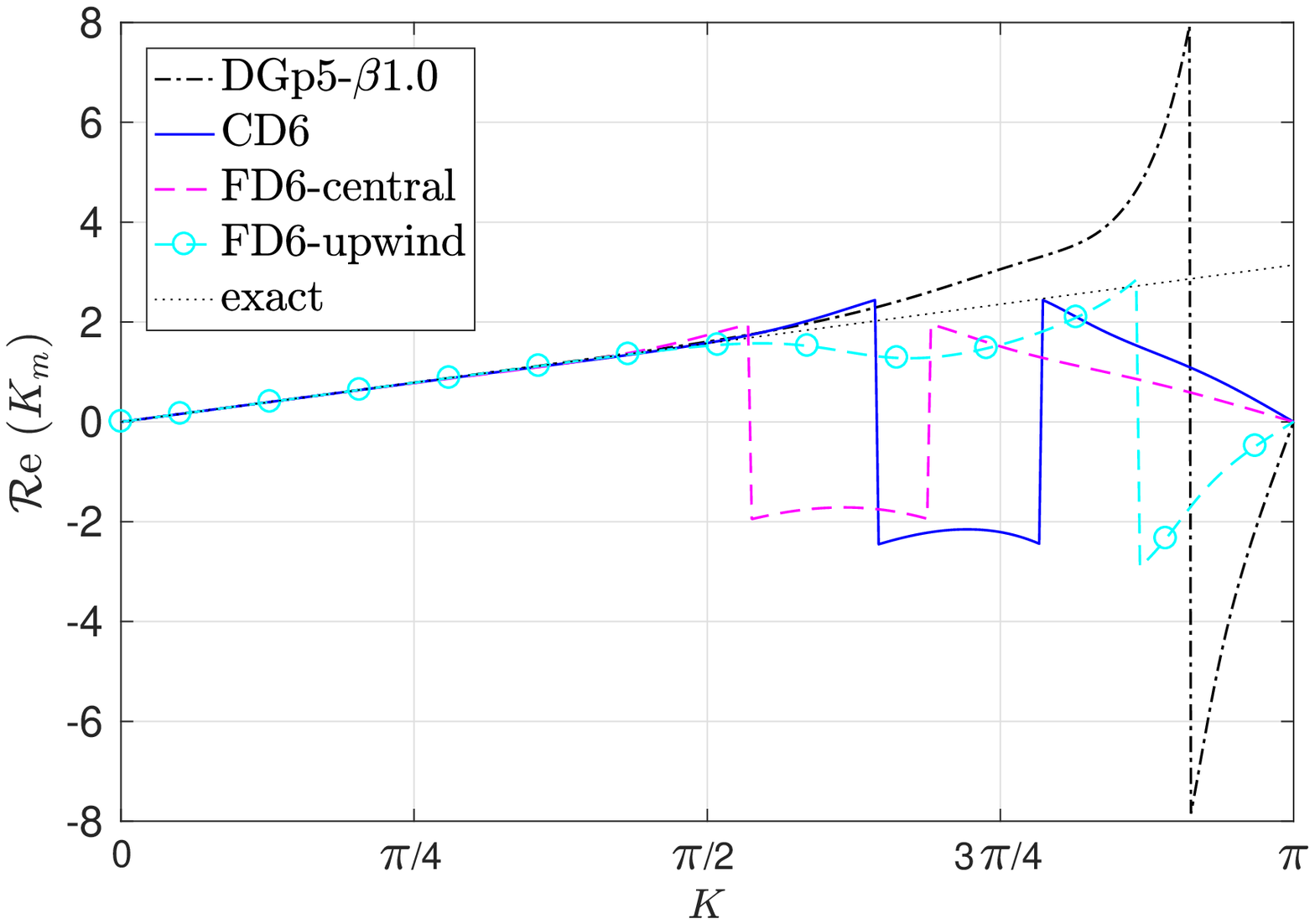} 
    \caption{Dispersion}
    \label{fig:p5cd6_RK4_09CFLmax_wp}
    \end{subfigure}
    \hspace{0.0125\textwidth}
    \begin{subfigure}[h]{0.47\textwidth}
    \centering
    \includegraphics[width=\textwidth]{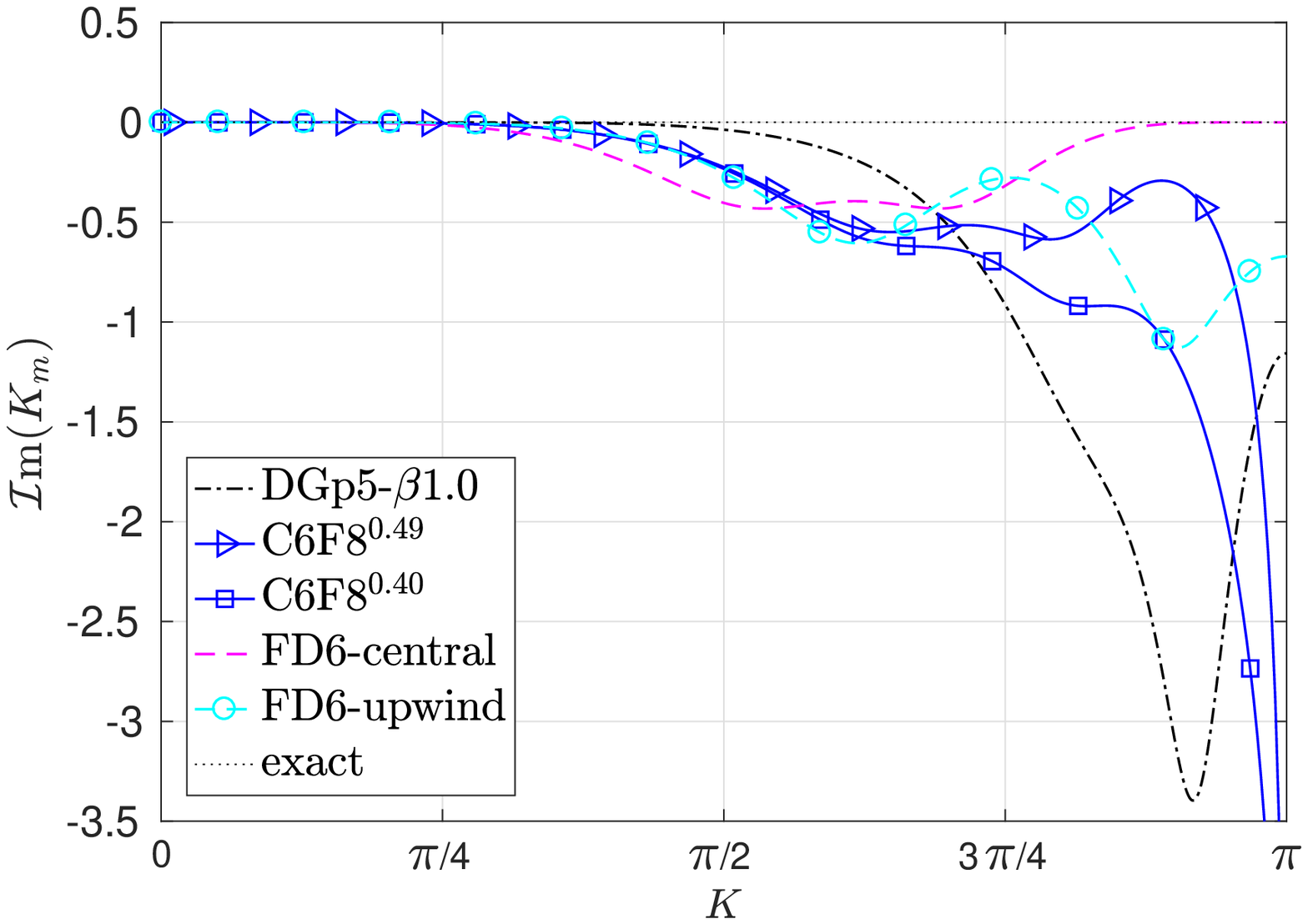}  
    \caption{Dissipation}
    \label{fig:p5cd6_RK4_09CFLmax_wd}
    \end{subfigure} 
	\caption{Comparison of dispersion/dissipation curves for DG, FD, and CD schemes coupled with RK$4$ scheme, and using the same CFL ratio, $r=0.9$. Note that for DG scheme, only the physical-mode is shown.}
    \label{fig:p5cd6_RK4_09CFLmax}
\end{figure}

For a central scheme, the semi-discrete dissipation exponent $\operatorname{\mathcal{I}m}(K_{m})$ is exact and hence when coupled with an explicit time integration scheme such as RK schemes, the dissipation behavior is completely due to the time integration scheme. In~\hfigref{fig:p5cd6_RK4_09CFLmax_wd} we compare the dissipation behavior of all schemes under consideration based on the same CFL ratio $r=0.9$. This value of $r$ is chosen to mimic the practical situation, where one always seeks the maximum possible CFL and hence $\Delta t$. From this figure, it is clear that both FD$6$ and C$6$F$8^{\alpha_{f}}$ schemes possess more dissipation than DGp$5$-$\beta1.0$ in the low to moderate wavenumber range. Only FD$6$-central and CD$6$ (without a filter(not shown)) schemes have $\operatorname{\mathcal{I}m}(K_{m})=0$ at the zero wavenumber and at the Nyquist frequency limit. This behavior of reduced dissipation near and at the highest wavenumber corresponding to $K=\pi$ is undesirable, especially for nonuniform grids and nonlinear problems~\cite{VisbalUseHigherOrderFiniteDifference2002} and often compensated by the use of a filter. 

In~\hfigref{fig:p5cd6_RK4_09CFLmax_wp}, the dispersion curves of the considered schemes are also compared with the exact dispersion relation $\omega=ka$. It is observed that the effect of time integration on the dispersion curves is noticeable and the jumps that were previously noted in~\cite{Vermeirebehaviourfullydiscreteflux2017} for FR-DG schemes are also present for CD and FD schemes, demonstrating that they are mainly due to time integration. However, we emphasize that the true behavior of RKDG schemes in the high wavenumber range is completely different than what is expected by the physical-mode analysis in~\hfigref{fig:p5cd6_RK4_09CFLmax}. %
\vspace{0.13in}%
\vspace{0.12in}
\begin{figure}[H]
\centering
 \begin{subfigure}[h]{0.47\textwidth}
    \centering
    \includegraphics[width=\textwidth]{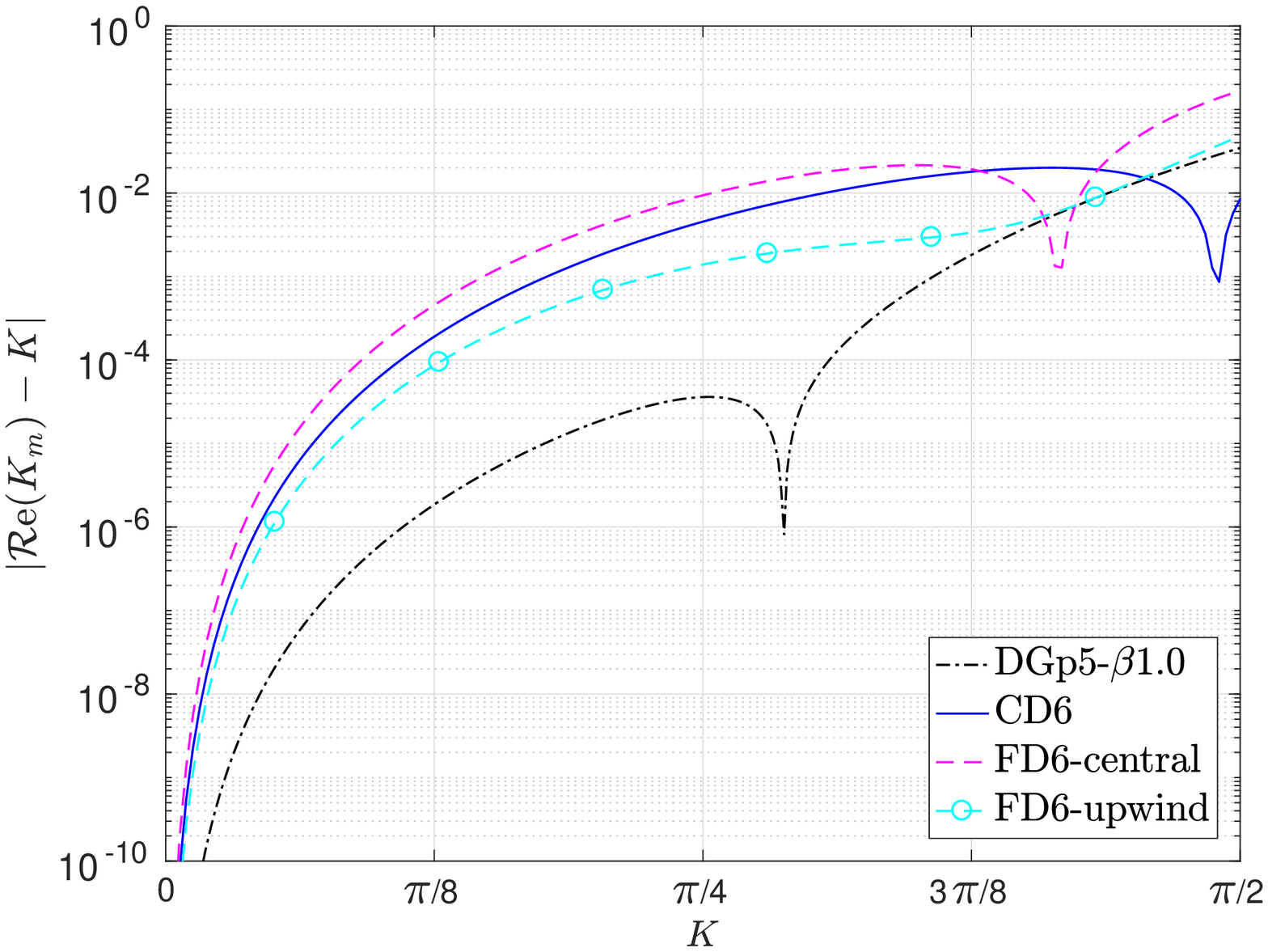} 
    \caption{Dispersion error}
    \label{fig:p5cd6_RK4_09CFLmax_err_wp}
    \end{subfigure}
    \hspace{0.0125\textwidth}
    \begin{subfigure}[h]{0.47\textwidth}
    \centering
    \includegraphics[width=\textwidth]{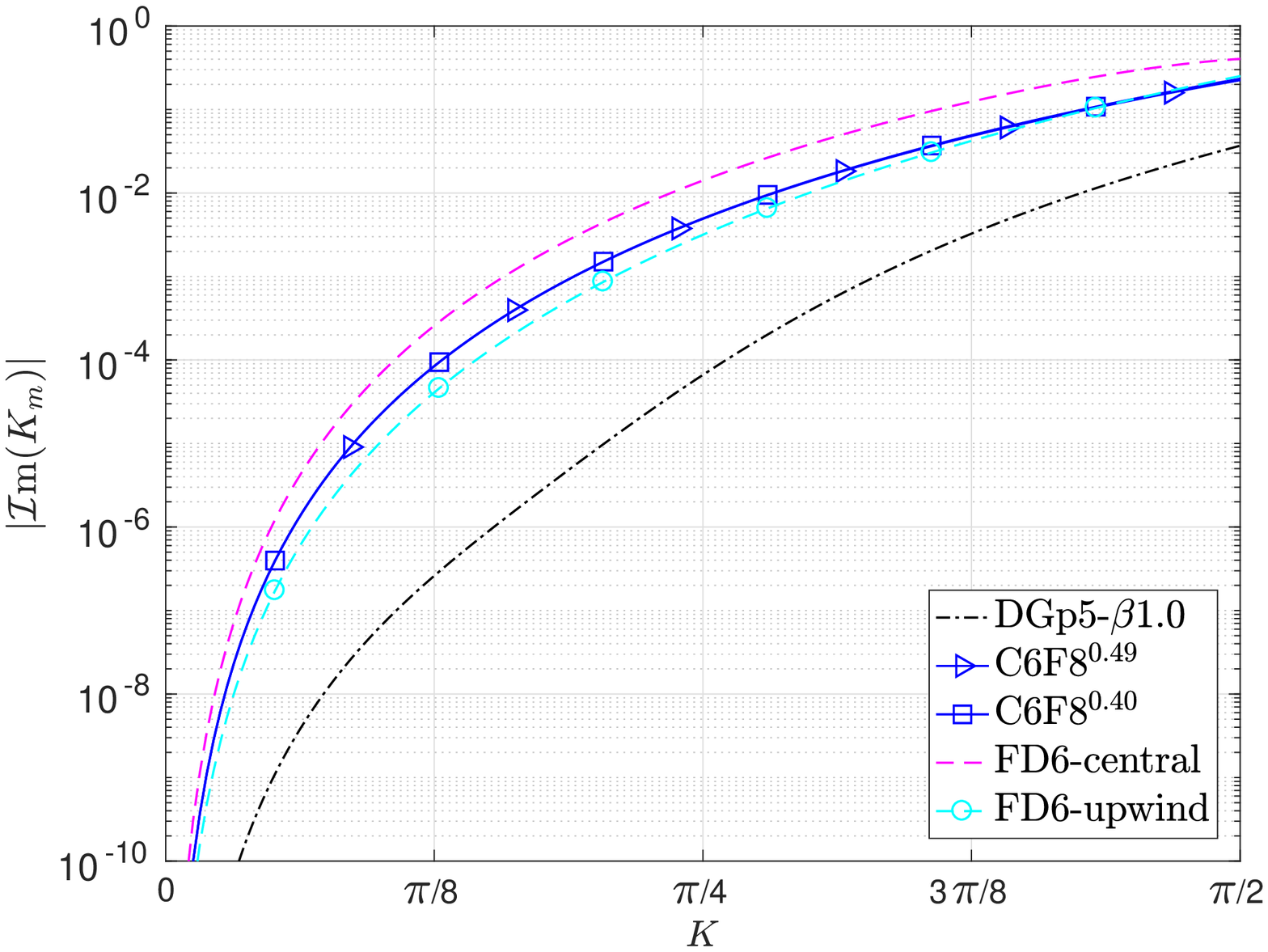}  
    \caption{Dissipation error}
    \label{fig:p5cd6_RK4_09CFLmax_err_wd}
    \end{subfigure} 
	\caption{Comparison of dispersion/dissipation errors for DG, FD, and CD schemes coupled with RK$4$ scheme, and using the same CFL ratio, $r=0.9$.}
    \label{fig:p5cd6_RK4_09CFLmax_err}
\end{figure}%
\vspace{0.1in}%
More detailed information can be obtained by examining the dispersion and dissipation errors in the low wavenumber range. In this study, the dispersion and dissipation errors are defined as%
\begin{align}
\text{dispersion error} &:= \;  |\operatorname{\mathcal{R}e}(K_{m})-K| ,  \\
\text{dissipation error} &:= \; |\operatorname{\mathcal{I}m}(K_{m})| .  
\end{align}%
It is observed in~\hfigref{fig:p5cd6_RK4_09CFLmax_err} that for the same CFL ratio $r=0.9$, the DGp$5$-$\beta 1.0$ scheme has the least dispersion and dissipation errors among all schemes under-consideration. It is also interesting to see that the dispersion and dissipation error of C$6$F$8^{0.49}$ and C$6$F$8^{0.40}$ schemes is much higher than that of DGp$5$-$\beta 1.0$ scheme and very close to the FD$6$-upwind scheme while the FD$6$-central scheme is the most dispersive/dissipative one. This behavior is mainly a result of having a larger $\Delta t$ for CD and FD schemes than the DG scheme since their stability limits are much higher.%
\begin{figure}[H]
\centering
\begin{subfigure}[h]{0.47\textwidth}
    \centering
    \includegraphics[width=\textwidth]{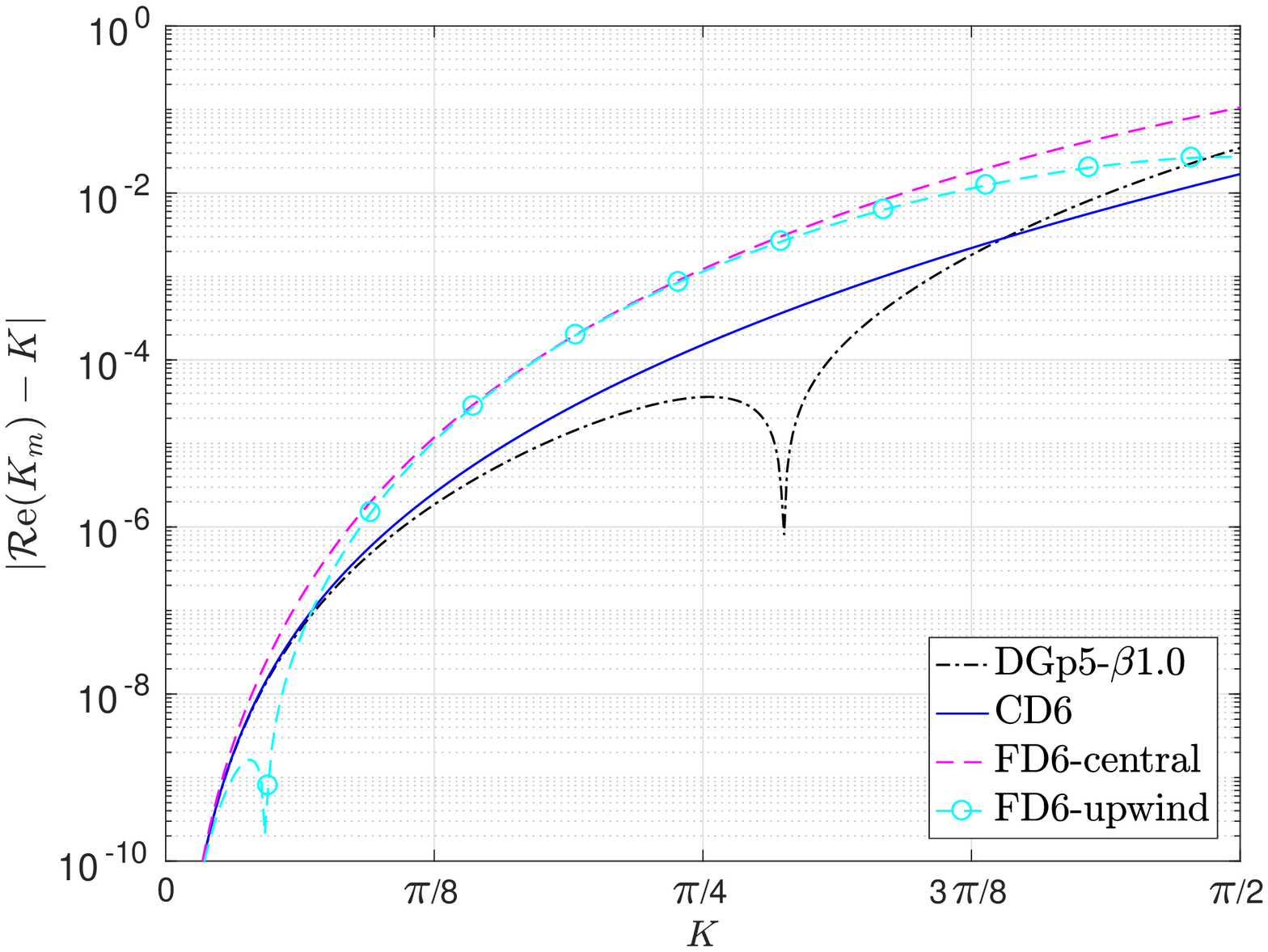} 
    \caption{Dispersion error}
    \label{fig:p5cd6_RK4_09dtmax_err_wp}
    \end{subfigure}
    \hspace{0.0125\textwidth}
    \begin{subfigure}[h]{0.47\textwidth}
    \centering
    \includegraphics[width=\textwidth]{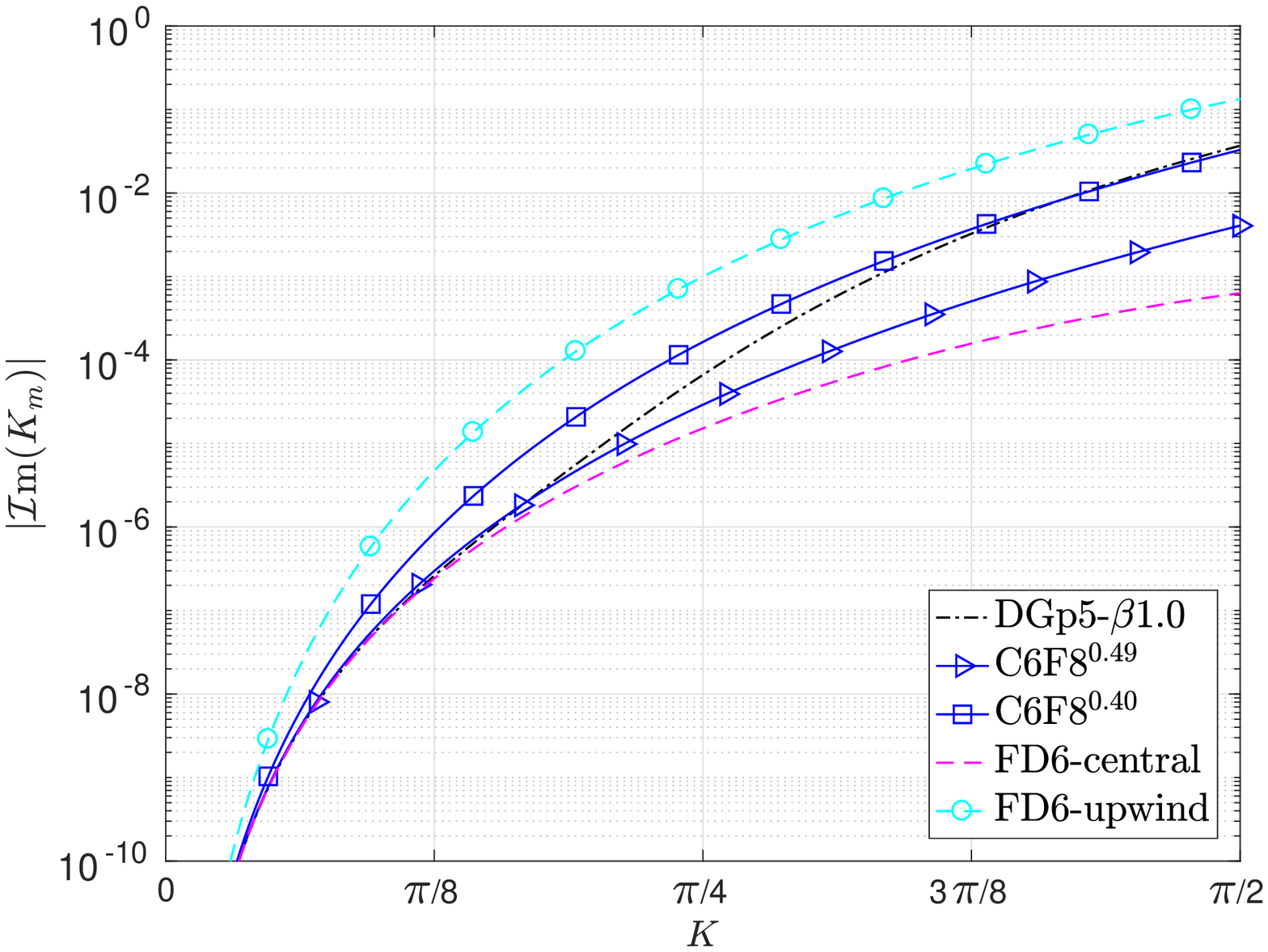}  
    \caption{Dissipation error}
    \label{fig:p5cd6_RK4_09dtmax_err_wd}
    \end{subfigure} 
	\caption{Comparison of dispersion/dissipation errors for DG, FD, and CD schemes coupled with RK$4$ scheme, and using the same $\Delta t$ of DGp$5$-$\beta1.0$ with $r^{DG}=0.9$.}
    \label{fig:p5cd6_RK4_09dtmax_err}
\end{figure}%
\begin{figure}[H]
\centering
 \begin{subfigure}[h]{0.47\textwidth}
    \centering
    \includegraphics[width=\textwidth]{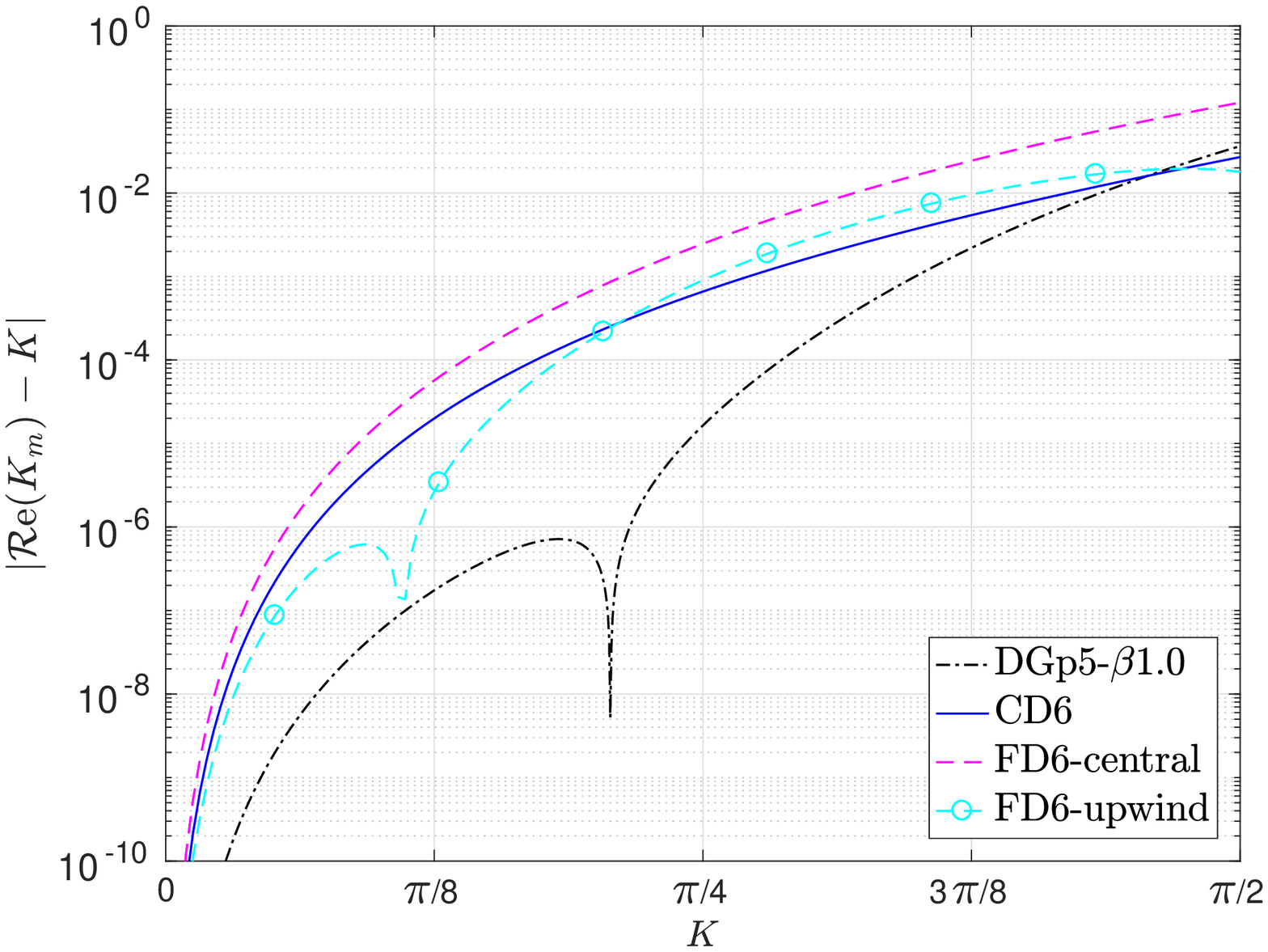} 
    \caption{Dispersion error}
    \label{fig:p5cd6_RK4_05CFLmax_err_wp}
    \end{subfigure}
   \hspace{0.0125\textwidth}
    \begin{subfigure}[h]{0.47\textwidth}
    \centering
    \includegraphics[width=\textwidth]{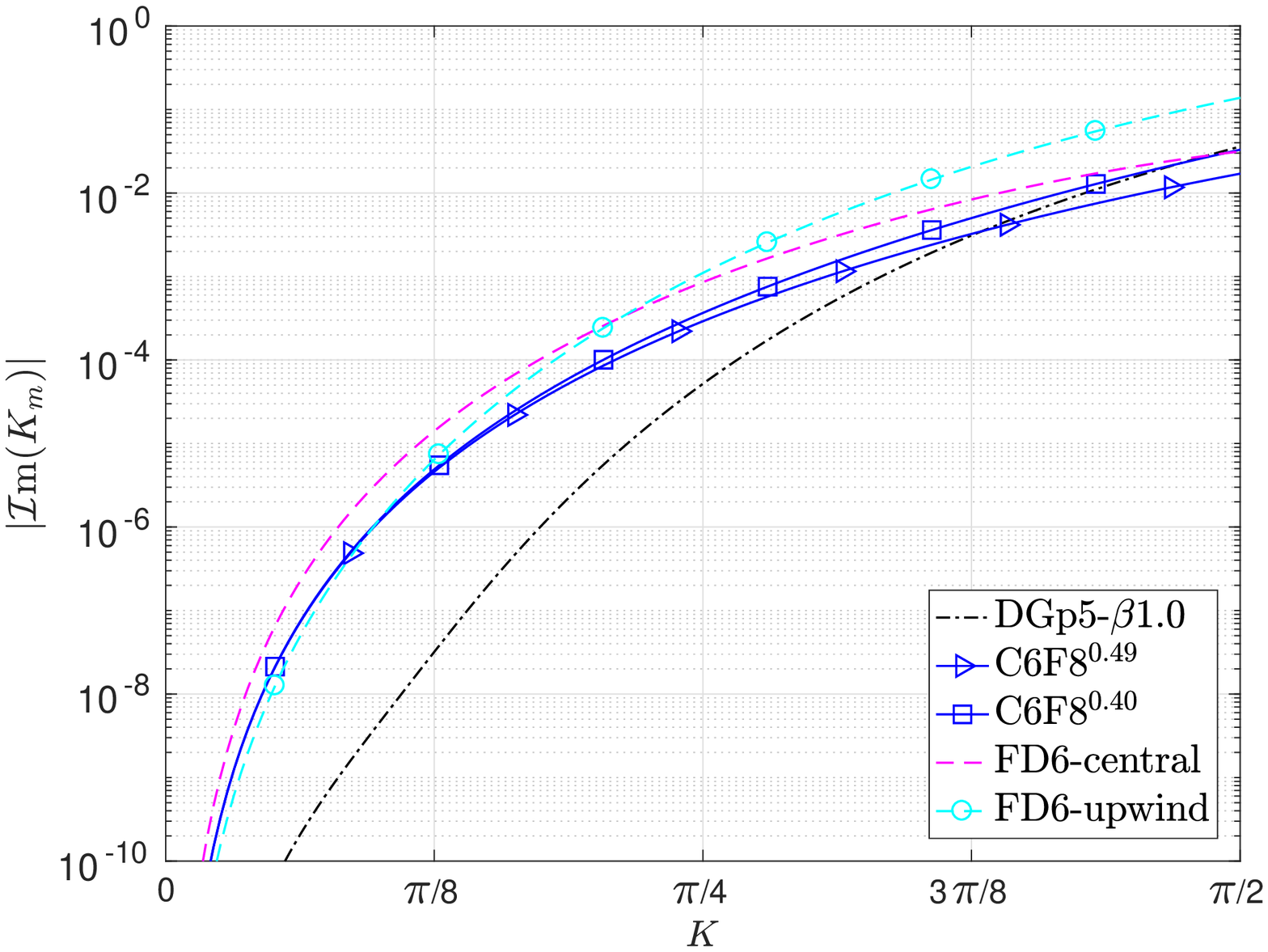}  
    \caption{Dissipation error}
    \label{fig:p5cd6_RK4_05CFLmax_err_wd}
    \end{subfigure} 
	\caption{Comparison of dispersion/dissipation errors for DG, FD, and CD schemes coupled with RK$4$ scheme, and using the same CFL ratio, $r=0.5$.}
    \label{fig:p5cd6_RK4_05CFLmax_err}
\end{figure}%
\begin{figure}[H]
\centering
\begin{subfigure}[h]{0.47\textwidth}
    \centering
    \includegraphics[width=\textwidth]{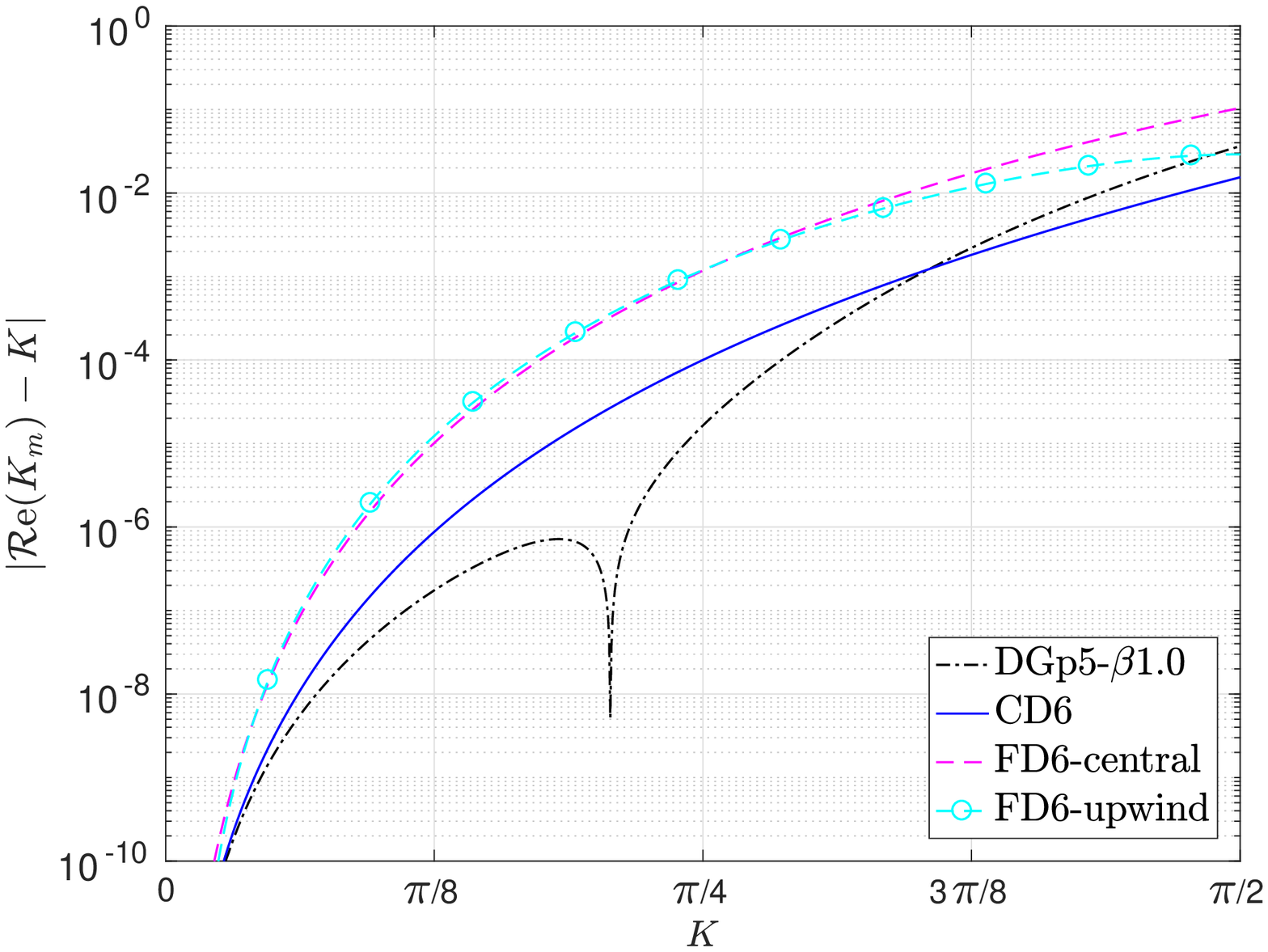} 
    \caption{Dispersion error}
    \label{fig:p5cd6_RK4_05dtmax_err_wp}
    \end{subfigure}
    \hspace{0.0125\textwidth}
    \begin{subfigure}[h]{0.47\textwidth}
    \centering
    \includegraphics[width=\textwidth]{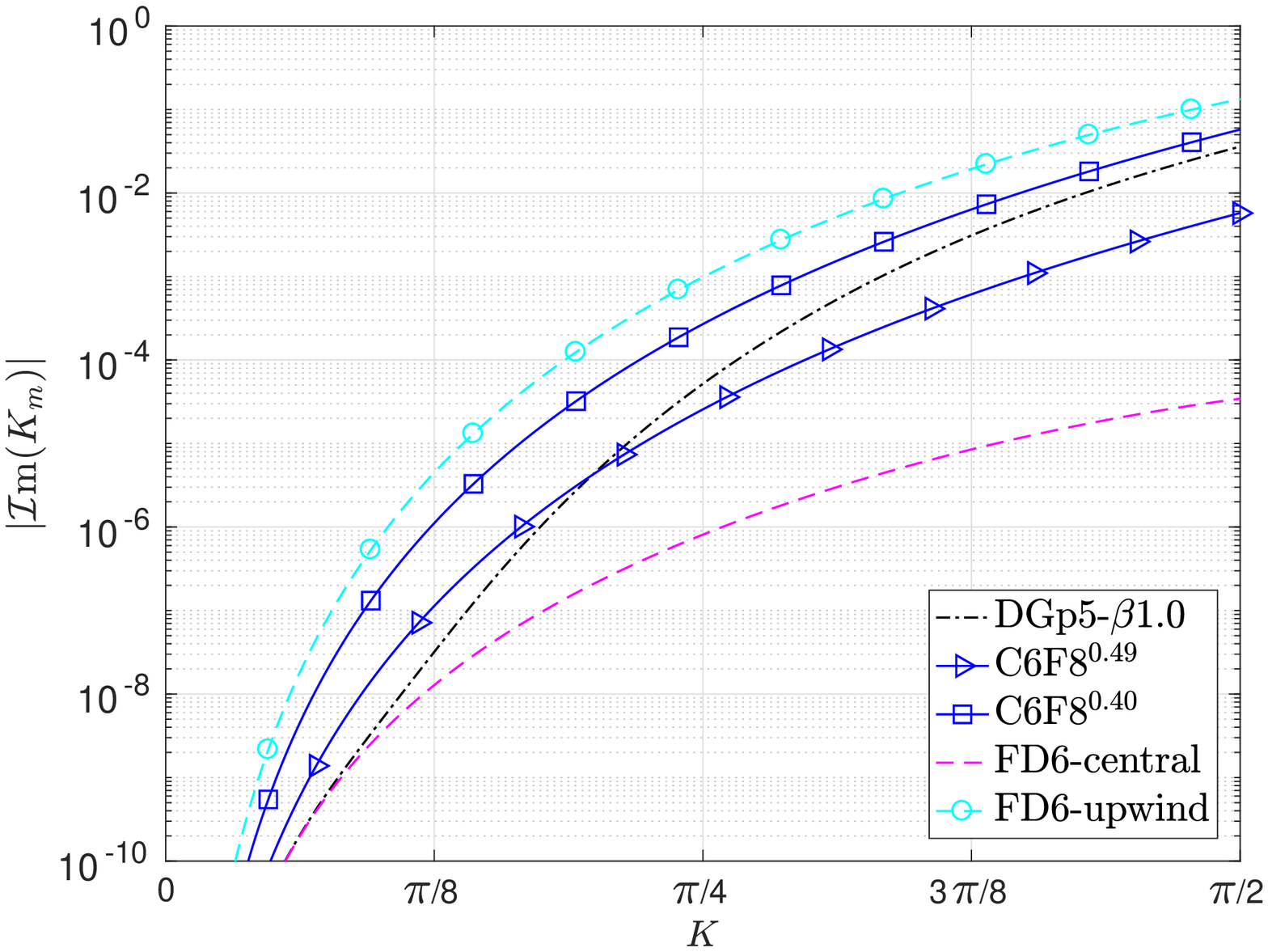}  
    \caption{Dissipation error}
    \label{fig:p5cd6_RK4_05dtmax_err_wd}
    \end{subfigure} 
    \caption{Comparison of dispersion/dissipation errors for DG, FD, and CD schemes coupled with RK$4$ scheme, using the same $\Delta t$ of DGp$5$-$\beta1.0$ with $r^{DG}=0.5$.}
    \label{fig:p5cd6_RK4_05dtmax_err}
\end{figure}%
On the other hand, if FD and CD schemes use the same $\Delta t$ corresponding to $r=0.9$ of  DGp$5$-$\beta 1.0$, the situation changes. Schemes DGp$5$-$\beta 1.0$, C$6$F$8^{0.49}$, and FD$6$-central have a comparable dissipation error until $3\pi/8$, whereas C$6$F$8^{0.40}$ and FD$6$-upwind have higher errors, as shown in~\hfigref{fig:p5cd6_RK4_09dtmax_err_wd}. In contrast, for $K\in \left[ \pi/4,\pi/2 \right]$, DGp$5$-$\beta 1.0$, C$6$F$8^{0.40}$ have a comparable error that is higher than that of C$6$F$8^{0.49}$, while FD$6$-central has the least error among all schemes under-consideration. The most dissipative scheme is always the FD$6$-upwind scheme. For dispersion errors~\hfigref{fig:p5cd6_RK4_09dtmax_err_wp}, the DG scheme is the best, followed by the two CD schemes, and finally by the two FD schemes.

Nearly similar results can be observed for the case of fixed CFL ratio $r=0.5$,~\hfigref{fig:p5cd6_RK4_05CFLmax_err} as in the case of fixed $r=0.9$. If the $\Delta t$ of DGp$5$-$\beta1.0$ scheme with $r=0.5$ is utilized for all schemes~\hfigref{fig:p5cd6_RK4_05dtmax_err}, it can be seen that DGp$5$-$\beta1.0$ scheme is less dissipative than C$6$F$8^{0.40}$ scheme for this entire wavenumber range while it is less dissipative than C$6$F$8^{0.49}$ scheme up to $K \approx \pi/4$. However, the least dissipative scheme in this case is the FD$6$-central scheme while the most dissipative one is the FD$6$-upwind scheme.

In summary, the DGp$5$-$\beta1.0$ has less dispersion/dissipation errors for the entire low wavenumber range than the C$6$F$8^{0.40}$ scheme in all the studied cases. In addition, in some cases the DGp$5$-$\beta1.0$ can indeed have less dissipation than central FD schemes and C$6$F$8^{0.49}$, that is when the same CFL ratio $r$ is utilized. It is shown that DGp$5$-$\beta1.0$ always has the least dispersive error in the low wavenumber range. Finally, we indicate that although the present results in this section are performed with RK$4$, similar trends were observed for RK$3$.
%
%
\section{Numerical Results}\label{sec:num_results_5}%
%
\subsection{Sine wave}\label{subsec:sinewave_5.1}%
%
In this test we choose a smooth sine wave to verify the dissipation of DG, FD, and CD schemes for a single mode. Consider the following IC for the linear-advection~\heqref{eqn:lin_advec} %
\begin{equation}
u(x,0) = \sin \left( \frac{ kx}{L} \right), \quad x \in [0,L] , 
\label{eqn:sine_wave_IC}
\end{equation}%
where $k$ denotes the wavenumber, $L$ is the length of the domain, and $\eta=L$ is the wavelength. In all the test cases in this study we let $L=1, \: a=1$, and hence the period of the wave is $T=1$. For DG schemes, this initial solution is projected using the $L_{2}$ projection onto the space of degree $p$ polynomials on each cell $\Omega_{e}$, while for FD and CD schemes nodal values of this solution are specified at the grid points. 

In order to verify the fully-discrete Fourier analysis, a smooth sine wave of the form in~\heqref{eqn:sine_wave_IC} is simulated using the DGp$5$-$\beta1.0$, FD$6$ (central and upwind), and C$6$F$8^{0.40}$/C$6$F$8^{0.49}$  schemes, all coupled with RK$4$ for time integration. For DGp$5$-$\beta 1.0$, the number of elements $N^{DG}_{e}=4$, while for the same nDOFs, the number of points for the FD$6$ and C$6$F$8^{\alpha_{f}}$ schemes is $(p+1) \times N^{DG}_{e} = 24$. The non-dimensional wavenumber is selected to be $K=k h/(p+1) = \pi/4$ and the wavenumber of the sine wave is $k =((p+1)\pi/4) \times N^{DG}_{e}/L = 6\pi$. Additionally, two time step settings were used, i.e., fixing the CFL ratio to $r=0.9$ for all the schemes, and fixing the time step to the one used by DG.

The dissipation error after $n$ iterations can be defined as%
\begin{equation}
\zeta_{f,n} = |1-G^{n}|  , 
\label{eqn:sinewave_num_dissip_err}
\end{equation}%
where $\zeta_{f}$ denotes the error predicted by the Fourier analysis, and $\zeta_{n}$ denotes what is predicted by the numerical simulation. In order to quantify the dissipation error through numerical simulations, we identify the maximum amplitude of the wave after it travels a certain distance $D_{\eta}$.  Since the amplitude of the initial sine wave is $1$, the predicted amplitude is exactly $G^{n}$ of the final solution. Because each scheme has its own dispersion error, it convects the wave with different numerical speeds $\tilde{a}$ that is generally not equal to the exact one, $a=1$. For a prescribed wavenumber $K$, the numerical wave speed can be determined by %
\begin{equation}
\tilde{a} = \mathcal{R}e(K_{m}) / K , 
\label{eqn:sinewave_num_wavespeed}
\end{equation}%
and consequently for a given distance $D_{\eta}$, the time required for the wave to travel this distance is%
\begin{equation}
t_{\eta} = D_{\eta} / \tilde{a}, \quad D_{\eta}=N_{\eta} \times \eta , 
\label{eqn:sinewave_time_for_wave}
\end{equation}%
where $N_{\eta}$ is the number of wavelengths the wave has traveled. In this study the wavelength $\eta=1$, and we compute the dissipation error for $N_{\eta} = 1, \: 10 $, i.e., after the wave travels either $1$ wavelength or $10$ wavelengths. In this manner, we were able to exclude the dispersion error effects as much as possible. \htablsref{table:sinewave_res_case1_2}{table:sinewave_res_case3_4} show the comparisons of the dissipation error between numerical and Fourier analysis for the all considered cases. From these tables, it can be seen that the numerical and Fourier analysis results agree very well. Note that the accurate point in time may not be reached exactly by the numerical simulation since it has a numerical $\Delta t$ value that does not usually divide the time interval exactly and this introduces small errors in the distance traveled by the wave. %
\begin{table}[H]%
\caption{ Results for the simulation of a sine wave with $k=6\pi$ using DGp$5$-$\beta 1.0$, C$6$F$8^{\alpha_{f}}$, and FD$6$, all combined with RK$4$, and using the same CFL ratio, $r=0.9$.} %
\centering 
\small
\begin{tabular}{|c|c|c|c|c|}
\hline
\multirow{2}{*}{Scheme} & \multicolumn{2}{|c|}{Case$(1)$, $N_{\eta}=1$} & \multicolumn{2}{|c|}{Case$(2)$, $N_{\eta}=10$} \\ 
\cline{2-5}
& $\zeta_{f}$ & $\zeta_{n}$ & $\zeta_{f}$ & $\zeta_{n}$   \\
\hline
\hline
DGp$5$-$\beta1.0$ & $1.55e-03$ & $2.08e-03$  & $1.54e-02$ & $1.60e-02$  \\
 \hline
FD$6$-upwind  & $7.24e-02$ & $8.07e-02$  & $5.35e-01$ & $5.51e-01$ \\
 \hline
FD$6$-central  & $2.98e-01$ & $3.02e-01$  & $9.68e-01$ & $9.68e-01$ \\
 \hline
C$6$F$8^{0.40}$ & $1.13e-01$ & $1.20e-01$  & $7.00e-01$ & $7.15e-01$  \\
 \hline
C$6$F$8^{0.49}$ & $1.12e-01$ & $1.19e-01$  & $6.97e-01$ & $7.12e-01$  \\
 \hline
\end{tabular}
\label{table:sinewave_res_case1_2}
\end{table}%
\begin{table}[H]%
\caption{ Results for the simulation of a sine wave with $k=6\pi$ using DGp$5$-$\beta 1.0$, C$6$F$8^{\alpha_{f}}$, and FD$6$, all combined with RK$4$, and using the same $\Delta t$ of DGp$5$-$\beta 1.0$ with $r^{DG}=0.9$.} %
\centering 
\small
\begin{tabular}{|c|c|c|c|c|}
\hline
\multirow{2}{*}{Scheme} & \multicolumn{2}{|c|}{Case$(3)$, $N_{\eta}=1$} & \multicolumn{2}{|c|}{Case$(4)$, $N_{\eta}=10$} \\ 
\cline{2-5}
& $\zeta_{f}$ & $\zeta_{n}$ & $\zeta_{f}$ & $\zeta_{n}$  \\
\hline
\hline
DGp$5$-$\beta1.0$ & $1.55e-03$ & $2.08e-03$  & $1.54e-02$ & $1.60e-02$   \\
 \hline
FD$6$-upwind  & $2.36e-02$ & $2.37e-02$ & $2.12e-01$ &  $2.17e-01$ \\
 \hline
FD$6$-central  & $3.64e-04$ & $3.89e-04$ & $3.64e-03$ &  $5.96e-03$ \\
 \hline
C$6$F$8^{0.40}$ & $3.93e-03$ & $3.90e-03$ & $3.85e-02$ &  $3.80e-02$   \\
 \hline
C$6$F$8^{0.49}$ & $6.97e-04$ & $7.35e-04$ & $6.94e-03$ &  $7.06e-03$    \\
 \hline
\end{tabular}
\label{table:sinewave_res_case3_4}
\end{table}%
\begin{figure}[H]
  \centering
    \includegraphics[width=0.83\textwidth]{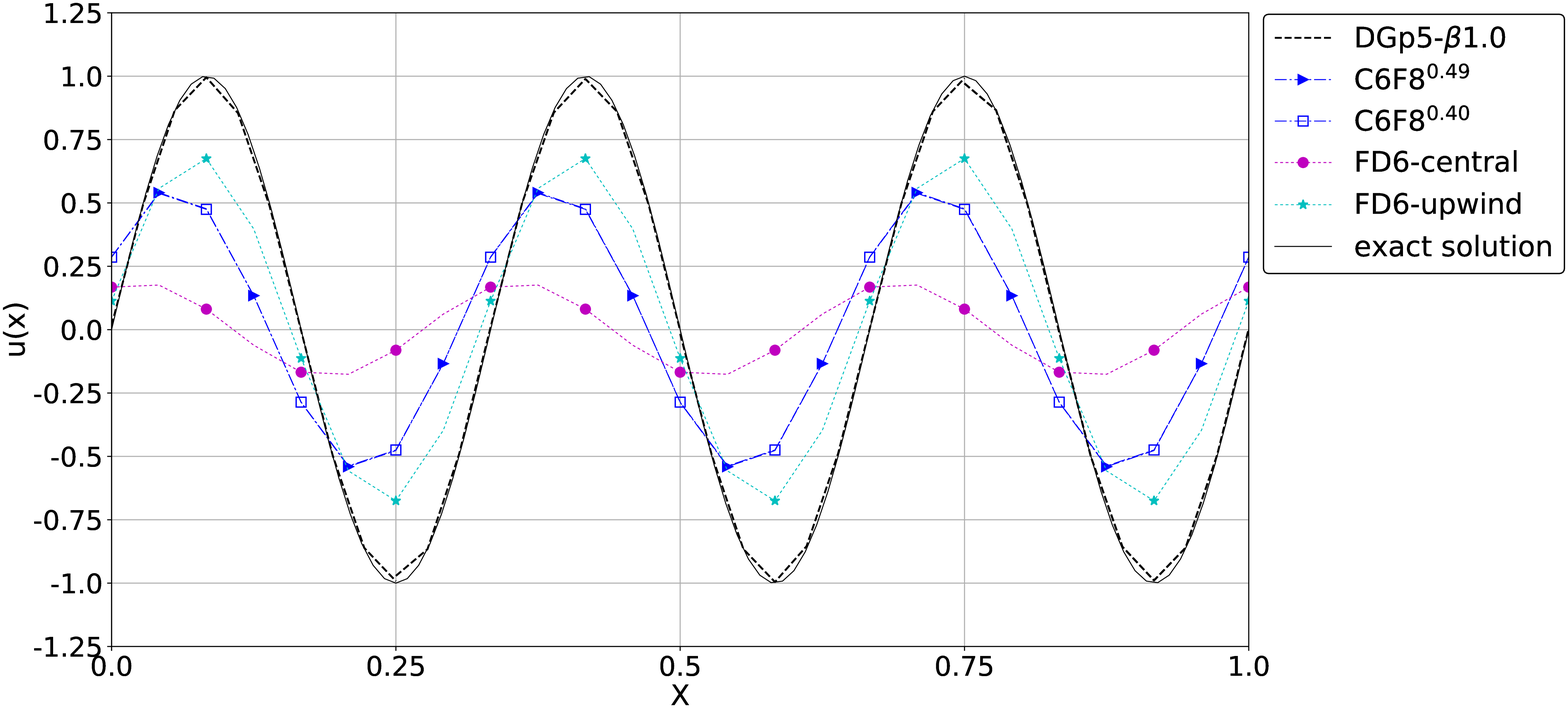} 
	\caption{Comparison of the sine wave solution for DG, FD, and CD schemes coupled with RK$4$ scheme for time integration using the same CFL ratio, $r=0.9$, at $t=5T$. The number of elements for each scheme is; $N^{DG}_{e}=4$, $N_{e}^{CD/FD} = 24$. }
    \label{fig:p5cd6fd6_sinewave_09CFLmax}
\end{figure}%
\begin{figure}[H]
  \centering
    \includegraphics[width=0.83\textwidth]{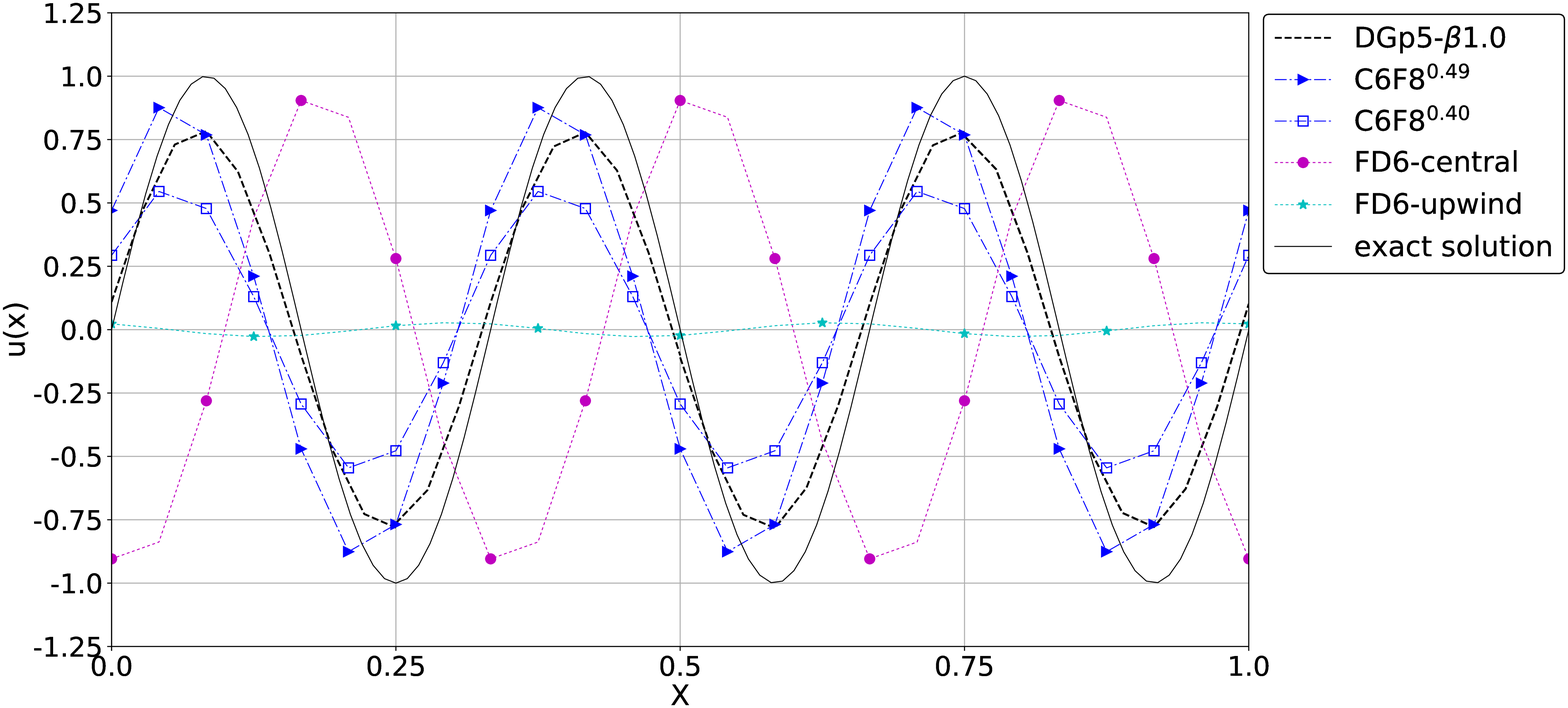} 
	\caption{Comparison of the sine wave solution for DG, FD, and CD schemes coupled with RK$4$ scheme for time integration using the same $\Delta t$ of DGp$5$-$\beta1.0$ with $r^{DG}=0.9$, at $t=150T$. The number of elements for each scheme is; $N^{DG}_{e}=4$, $N_{e}^{CD/FD} = 24$. }
    \label{fig:p5cd6fd6_sinewave_09dtmax}
\end{figure}%
%
Another way to assess the dispersion/dissipation characteristics of all the schemes through numerical simulations is to compare their wave solutions after reaching the same point in time. Utilizing the same CFL ratio $r=0.9$ for all schemes, it is shown in~\hfigref{fig:p5cd6fd6_sinewave_09CFLmax} that indeed the DGp$5$-$\beta1.0$ scheme is the least dispersive/dissipative one amongst all the considered schemes.~\hfigref{fig:p5cd6fd6_sinewave_09dtmax} shows that, for the case of fixed $\Delta t$,  the numerical results again agree very well with the Fourier analysis,  in which the central FD$6$ scheme has the least dissipation, while its dispersion is much larger than the DGp$5$-$\beta1.0$ scheme. In all these cases RK$4$ is employed for time integration.%
%
\subsection{Gaussian wave}\label{subsec:gaussian_5.2}%
%
The Gaussian wave contains a broadband of wave numbers and hence can be used to assess the performance of numerical schemes in a more practical setting. Therefore, in this section we compare the performance of DG, FD, and CD schemes coupled with RK$4$. Consider an initial solution for the linear-advection~\heqref{eqn:lin_advec} of the form%
\begin{equation}
u(x,0) = e^{-38.6 \: x^{2}} \quad, \: x \: \epsilon \: [-L,L] ,
\label{eqn:Gaussian_function_IC}
\end{equation}%
where $2L$ is the length of the domain. This particular case is chosen to be a sufficiently smooth case for our simulations with $L=1,\;a=1$, and hence the period $T=2$. We again test two scenarios. First a fixed CFL ratio is employed for all schemes, i.e., $r=0.9$. Next a fixed $\Delta t$ is used, which is equal to $\Delta t^{DG}$ with $r^{DG}=0.9$. The DG solutions are identical in the two scenarios. %
\begin{figure}[H]
    \centering
    \includegraphics[width=0.63\textwidth]{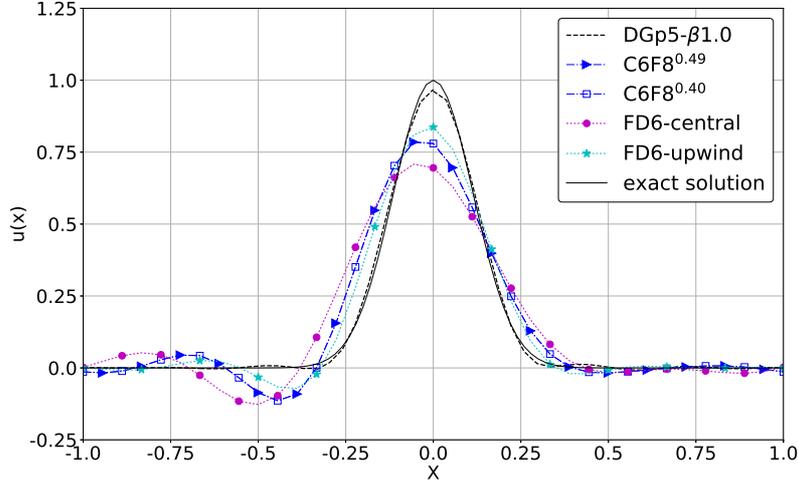}  
	\caption{Comparison of the Gaussian wave solution for DG, FD, and CD schemes coupled with RK$4$ scheme for time integration using the same CFL ratio $r=0.9$, at $t=10T$. The number of elements for each scheme is; $N^{DG}_{e}=6$, $N_{e}^{CD/FD} = 36$. }
    \label{fig:p5cd6fd6_gaussian_cflmax}
\end{figure}%
\begin{figure}[H]
    \centering
    \includegraphics[width=0.63\textwidth]{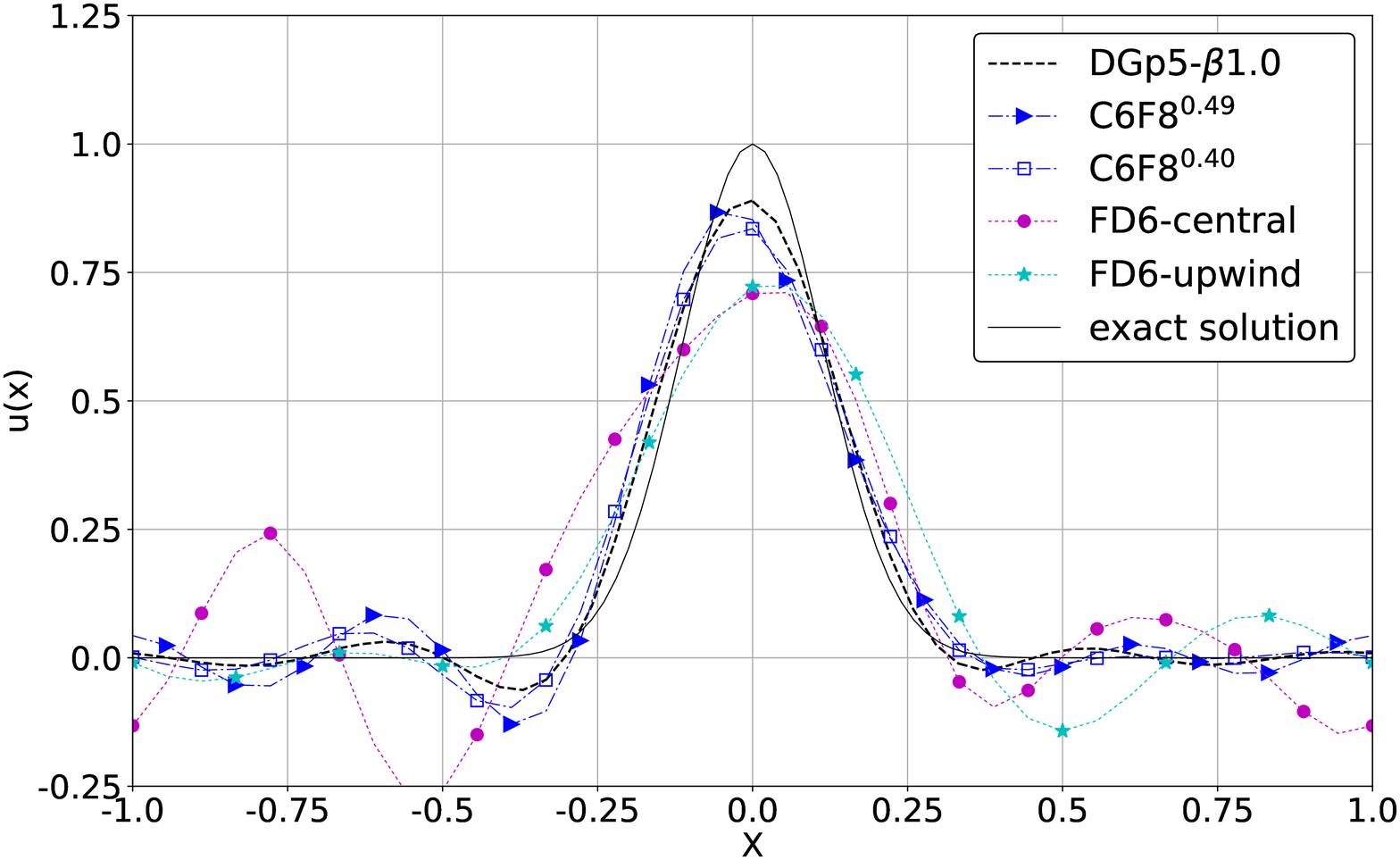} 
	\caption{Comparison of the Gaussian wave solution for DG, FD, and CD schemes coupled with RK$4$ scheme for time integration using the same $\Delta t$ of DGp$5$-$\beta1.0$ with $r^{DG}=0.9$, at $t=200T$. The number of elements for each scheme is; $N^{DG}_{e}=6$, $N_{e}^{CD/FD} = 36$. }
    \label{fig:p5cd6fd6_gaussian_dtmax}
\end{figure}%
\hfigref{fig:p5cd6fd6_gaussian_cflmax} displays the final solutions for fixed CFL ratio at $t=10T$. It can be seen from this figure that the DGp$5$-$\beta1.0$ has the least error in dispersion and dissipation followed by the FD$6$-upwind scheme, the two compact schemes and the central FD scheme, as shown previously. The final solutions with a fixed time step at  $t=100T$ are displayed in \hfigref{fig:p5cd6fd6_gaussian_dtmax}. We can see that DGp$5$-$\beta1.0$ still produced the best overall approximation among all schemes under consideration. While the solution of C$6$F$8^{0.49}$ is close to that of the DGp$5$-$\beta1.0$, it is obviously more dispersive. In addition, C$6$F$8^{0.40}$ is seen to be more dissipative than C$6$F$8^{0.49}$, but less dispersive. The FD$6$-central scheme has the highest dispersion error, whereas FD$6$-upwind is less dispersive than its central counterpart.  We were very surprised that the FD$6$-central scheme smeared the peak so much. To verify the result,  a fast Fourier-transform was performed for all solutions, and we were able to confirm that the FD$6$-central solution indeed has the highest energy content. The under-prediction was indeed due to dispersion errors.%
%
\subsection{Resolution for the Burgers turbulence}\label{subsec:Burgers_5.3}%
%
The Burgers turbulence case was utilized by many researchers as a nonlinear test case to assess the behavior of numerical schemes for ILES~\cite{Liprioriposteriorievaluations2016,SanAnalysislowpassfilters2016,Vermeirebehaviourfullydiscreteflux2017} or uDNS~\cite{MouraLineardispersiondiffusion2015} simulations. The viscous Burgers equation can be written as %
\begin{equation}
\frac{\partial u}{\partial t} + \frac{\partial f(u)}{\partial x} = \gamma \frac{\partial^{2} u}{\partial x^{2}}, \quad f = u^{2}/2 ,
\label{eqn:burgers_eqn}
\end{equation}%
where $\gamma$ is the diffusivity coefficient that is taken to be $\gamma= 2 \times 10^{-4}$. The discretization of the convective flux term in~\heqref{eqn:burgers_eqn} follows our standard DG discretization described earlier in~\secref{sec:num_methods_2}. The viscous term is discretized using the BR$2$ \cite{BassiHigherorderaccuratediscontinuous1997} scheme for DG methods and an exact integration is performed to mitigate aliasing errors. For the CD method, an implicit discretization formula is used for the second derivative\cite{LeleCompactfinitedifference1992} similar to the one introduced for the first derivative in~\secref{sec:num_methods_2}. Formulas for the discretization of the second derivative using the FD method can be found in~\hyperref[appx:A]{Appendix~A} and using the CD method in~\hyperref[appx:B]{Appendix~B}.   

In our study we conduct the simulation for an initial solution of a decaying Burgers turbulence case similar to the case introduced in~\cite{SanAnalysislowpassfilters2016,DeStefanoSharpcutoffsmooth2002} in a periodic domain, with $x \in [ 0,2 \pi ]$. The initial energy spectrum is defined according to the following equation%
\begin{equation}
E(k,0) = E_{0}(k) = A k^{4} \rho^{5} e^{-k^{2} \rho^{2} } ,
\label{eqn:energ_spect_omersan}
\end{equation}%
where $k$ is the prescribed wavenumber,  $\rho$ is a constant of 10 to control the position of the maximum energy, and $A$ is a constant given by%
\begin{equation}
A = \frac{2}{3\sqrt{ \pi}} ,
\end{equation}%
which yields a spectrum that reaches its maximum at $k=13$. The initial velocity field can be derived from the initial energy spectrum $E_{0}(k)$, assuming that it has a Gaussian distribution with random phases as follows%
\begin{equation}
\hat{v}(k) = \sqrt{2 E(k)} e^{i 2 \pi \Phi(k)} ,
\label{eqn:burger_velocity_fourierspace}
\end{equation}%
where $\Phi(k)$ is a random phase angle that is uniformly distributed in $ \left[ 0,1 \right] $ for each wavenumber $k$. In order to compute a real velocity field in the physical space an inverse Fourier transform is employed analytically with the random phase angle satisfying $\Phi(k) = - \Phi(-k)$. Consequently, the velocity field in the physical space can be written as%
\begin{equation}
v(x) = \sum_{j=0}^{n_{k}} \sqrt{2 E(k_{j})} \cos \left( k_{j} x + 2 \pi \Phi(k_{j}) \right) + v_{m} ,
\label{eqn:burger_velocity_physicalspace}
\end{equation}%
where $v_{m}$ is the mean velocity to be specified, $n_{k}$ is the number of prescribed wavenumbers that is defined by the maximum wavenumber of $k_{max}=2048$ so that $n_{k}=2048$, and $k_{j}$ is an integer wavenumber. In the present work, $v_{m}=75$ is selected to have a turbulence intensity of $\approx 0.67\%$ ~\cite{Liprioriposteriorievaluations2016} in assessing the behavior of high-order methods for the Burgers case. For DG methods this initial condition is projected into the solution polynomial space of degree $p$, whereas for FD and CD methods a simple nodal value is used. In addition, the Burgers equation is solved for a 64 randomly generated samples of the initial velocity field and then the energy spectrum is computed using an ensemble averaged fast Fourier transform (FFT) of these samples. The same randomly generated samples is utilized by all schemes under consideration in order to have a fair and consistent comparison. %
\begin{figure}[H]
    \begin{subfigure}[h]{\textwidth}
    \centering
    \includegraphics[width=0.556\textwidth]{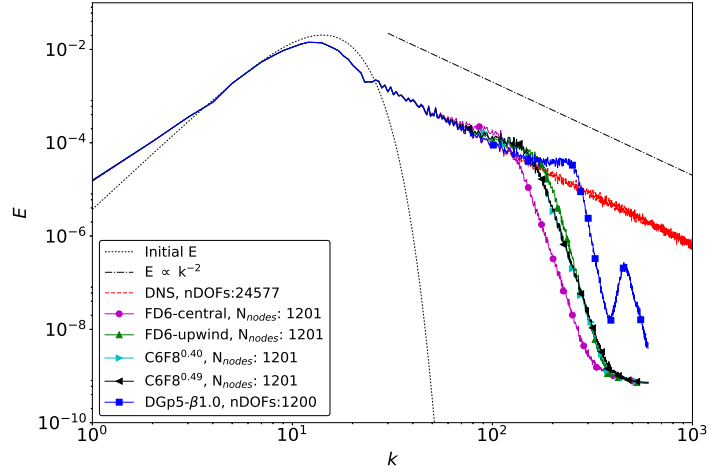} 
    \caption{Fixed CFL ratio $r=0.9$.\vspace{0.05in}}
    \label{fig:p5cd6RK4_um75_09CFLmax_01t_KE}
    \end{subfigure}
    \\ 
    \begin{subfigure}[h]{\textwidth}
    \centering
    \includegraphics[width=0.556\textwidth]{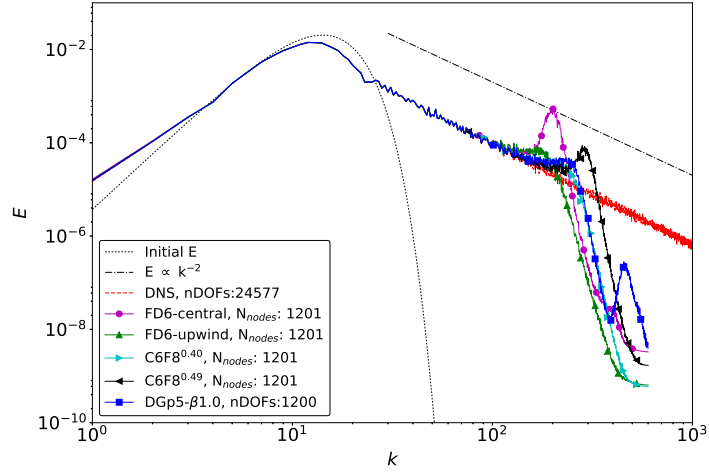}  
    \caption{Fixed $\Delta t=2e-5$.\vspace{0.05in}}
    \label{fig:p5cd6RK4_um75_dt2e-05_01t_KE}
    \end{subfigure} 
        \\ 
    \begin{subfigure}[h]{\textwidth}
    \centering
    \includegraphics[width=0.556\textwidth]{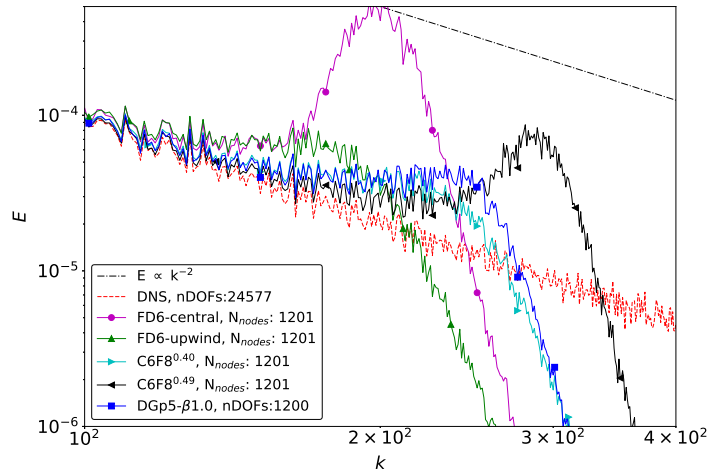}  
    \caption{Zoom on fixed $\Delta t=2e-5$ case.}
    \label{fig:p5cd6RK4_um75_dt2e-05_01t_KE_zoom}
    \end{subfigure} 
	\caption{Comparison of the energy spectrum at t$=0.1$ for DGp$5$-$\beta1.00$, FD$6$, C$6$F$8^{0.40}$, and C$6$F$8^{0.49}$ schemes. For time integration RK$4$ is utilized.}
    \label{fig:p5cd6RK4_KE}
\end{figure}%
We compare DGp$5$-$\beta1.0$, C$6$F$8^{0.40}$, C$6$F$8^{0.49}$, and FD$6$ central and upwind-biased schemes coupled with RK$4$  for time integration. The number of nodes of the FD and CD schemes are set to be $N_{n}=1201$, and for the same number of degrees of freedoms, the DG scheme has  $N_{e}=200$ elements. Because this is a nonlinear case, we adopted a slightly modified settings for this problem. We define the CFL number based on  the advective wave only, i.e., CFL$=|u_{max}| \frac{\Delta t}{\Delta x}$, where $|u_{max}|$ is the maximum eigenvalue of the initial solution. In this way, we use a nearly constant CFL number for all the points/elements, which is close to a linear case. Afterwards, the energy spectrum of all schemes is compared at $t=0.1$, long enough so that the solution reaches a statistically steady state and agrees very well with the theoretical expected energy slope of $E \propto k^{-2}$.

\hfigref{fig:p5cd6RK4_KE} presents the energy spectrum for two simulation settings for all schemes along with the expected energy spectrum of $E \propto k^{-2}$ for the decaying Burgers turbulence~\cite{BecBurgersturbulence2007}. This figure also includes a reference DNS solution where a DGp$5$-$\beta1.0$ simulation was conducted with $N_{e}=4096$. From this figure, it is noticed that for the case of CFL ratio $r=0.9$, the DGp$5$-$\beta1.0$ scheme is able to capture a wider spectrum than all other schemes, followed by FD$6$-upwind scheme which is very close to the compact schemes, C$6$F$8^{0.40}$/C$6$F$8^{0.49}$, whereas FD$6$-central captures the narrowest spectrum among all schemes under-consideration. Note that all schemes display mild energy pile-up at the highest wave number. On the other hand, if a fixed $\Delta t=2e-5$ is adopted as in part(b) of~\hfigref{fig:p5cd6RK4_KE}, then DGp$5$-$\beta1.0$ scheme, C$6$F$8^{0.40}$ and C$6$F$8^{0.49}$, all capture almost the same width of the spectrum, and reasonably accurate. However, there is a significant energy pileup in the result of  C$6$F$8^{0.49}$  indicating a lack of dissipation at the highest wavenumber. This figure justifies the use of C$6$F$8^{0.40}$ for ILES. The FD$6$-upwind scheme is indeed the most dissipative whereas the FD$6$-central scheme has the most severe energy pileup at the highest wave number. 

These observed severe energy pileups indicate a lack of dissipation at large wavenumbers, and can cause a non-linear simulation to blow up. The present analysis explains why central FD schemes need sub-grid scale models to provide further dissipation to stabilize LES.  

Finally, we indicate that although the present results in this section utilize only RK$4$ for time integration, similar trends were observed for  RK$3$  with $4^{th}$ order accurate spatial schemes.%
%
\section{Conclusions}\label{sec:conclusions}
%
In this paper, the dispersion/dissipation behavior of DG, FD and CD methods was studied through both semi-discrete and fully-discrete Fourier analysis. For DG schemes, it was verified that the physical-mode can serve as a good approximation for the true behavior in the low wavenumber regime using a combined-mode Fourier analysis. However, in the high wavenumber regime, no single mode can characterize the behavior of DG schemes solely and DG schemes always have slower decaying rate in this range than what is expected by the physical-mode. Secondary modes appear to always improve the accuracy of the scheme in the low wavenumber regime. 

In comparing different methods, it was found that time integration schemes (e.g., Range-Kutta) have a significant effect on the overall numerical dispersion and dissipation. If we fix the ratio of CFL to the maximum, the CD and FD schemes have larger dispersion and dissipation errors than DG schemes of the same order. On the other hand, if the same $\Delta t$ is utilized for all schemes, central FD schemes have the minimum dissipation while the upwind-biased schemes have the most dissipation, and the DG scheme is between the CD schemes with different filtering parameters. In addition, it was found that DGp$5$-$\beta1.0$  has a lower dispersion error in the low wavenumber range than FD$6$ and C$6$F$8^{0.40}$/C$6$F$8^{0.49}$. For the Gaussian wave simulation, it was demonstrated that the numerical dispersion of the FD$6$-central scheme causes it to severely under-predict the peak value. This case highlights the impact of dispersion errors. The Burgers turbulence case revealed that the two best schemes for ILES are DGp$5$-$\beta1.0$ and C$6$F$8^{0.40}$ because of their overall resolution and high wavenumber damping. FD$6$-central and C$6$F$8^{0.49}$ appear to have insufficient high wavenumber damping, while FD$6$-upwind has too much dissipation.  %
%
\section*{Acknowledgements}%
%
The research outlined in the present paper has been supported by AFOSR under grant FA9550-16-1-0128, and US Army Research Office under grant W911NF-15-1-0505. %
\addcontentsline{toc}{section}{Acknowledgements}%
%
\section*{Appendix A. FD schemes and modified wavenumber formulas}\label{appx:A} %
\addcontentsline{toc}{section}{Appendix A. FD schemes and modified wavenumber formulas}%
%
\noindent FD formulas for the first derivative,%
%
\begin{alignat}{2}
\text{FD}5\text{-\textit{1point-biased,}} & \quad u^{\prime} = \frac{ -1.5 u_{j+2} + 15 u_{j+1} + 10 u_{j} -30 u_{j-1} + 7.5 u_{j-2} -  u_{j-3} }{ 30 h} + O(h)^{5}.
\label{eqn:FD5_1p-upwbiased} \\
\text{FD}6\text{-\textit{2point-biased,}} & \quad u^{\prime} = \frac{ -2 u_{j+2} + 24 u_{j+1} + 35 u_{j} -80 u_{j-1} + 30 u_{j-2} - 8 u_{j-3} + u_{j-4} }{ 60 h} + O(h)^{6}.
\label{eqn:FD6_2p-upwbiased}
\end{alignat}%
\noindent FD modified wavenumbers for the linear-advection equation,%
%
\begingroup
\allowdisplaybreaks
\begin{alignat}{2}
\text{FD}1\text{-\textit{fully-upwind,}}   \hspace{0.05in} & \quad K_{m}(K) =  \sin (K) - i \left( 1 -\cos (K) \right). 
\label{eqn:FD1_Km} \\
\text{FD}2\text{-\textit{central,}}   \hspace{0.375in} & \quad K_{m}(K) =  \sin (K). 
\label{eqn:FD2_Km1} \\
\text{FD}3\text{-\textit{1point-biased,}} & \quad K_{m}(K) =  \left( 8 \sin (K) - \sin (2K)  \right) / 6 + i \left( 4 \cos (K) - \cos(2K)  -3 \right)  / 6.
\label{eqn:FD3_Km} \\
\text{FD}4\text{-\textit{central,}}  \hspace{0.375in} & \quad K_{m}(K) = \left(  8 \sin (K) - \sin (2K) \right) / 6.
\label{eqn:FD4_Km} \\
\text{FD}5\text{-\textit{1point-biased,}} & \quad K_{m}(K) =  \left( 45 \sin (K) - 9 \sin (2K) + \sin (3K) \right) / 30 \nonumber \\ & \qquad \qquad \quad+ i \left( -6\cos(2K) + 15\cos(K) +\cos(3K) - 10  \right )  / 30 . 
\label{eqn:FD5_1p-biased_Km} \\
\text{FD}6\text{-\textit{central,}} \hspace{0.375in} & \quad K_{m}(K) = \left( 45 \sin (K) - 9 \sin (2K) + \sin (3K) \right) / 30. 
\label{eqn:FD6_Km}\\
\text{FD}6\text{-\textit{2point-biased,}} & \quad K_{m}(K) =  \left( - 32 \sin(2K) + 104 \sin(K) + 8 \sin(3K) -
\sin(4K) \right) /60 \nonumber \\ & \qquad \qquad \quad+ i \left( -28\cos(2K) +56 \cos(K) + 8\cos(3K) - \cos(4K) -35  \right )  / 60 . 
\label{eqn:FD6_2p-biased_Km} 
\end{alignat}
\endgroup%
%
\noindent FD formulas for the second derivative,%
%
%
\begin{alignat}{2}
\text{FD}4\text{-\textit{central,}} & \quad u^{\prime \prime}  = \frac{- u_{j+2} + 16 u_{j+1} -30 u_{j} + 16 u_{j-1} - u_{j-2}}{ 12 h} + O(h)^{4}. 
\label{eqn:df2_FD_cent_4th} \\
\text{FD}6\text{-\textit{central,}} & \quad u^{\prime \prime}  = \frac{u_{j+3} - 13.5 u_{j+2} + 135 u_{j+1} -245 u_{j} + 135 u_{j-1} - 13.5 u_{j-2} + u_{j+3}}{ 90 h} + O(h)^{6}.
\label{eqn:df2_FD_cent_6th}
\end{alignat}%
%
%
\section*{Appendix B. CD schemes and Pad\'{e} filter}\label{appx:B}%
\addcontentsline{toc}{section}{Appendix B. CD schemes and Pad\'{e} filter}
%
%
\noindent CD implicit equation for the second derivative,%
%
\begin{equation}
\alpha u^{\prime \prime}_{m-1} + u_{m}^{\prime \prime} + \alpha u_{m+1}^{\prime \prime} = c \frac{u_{m+2}-2u_{m}+u_{m-2}}{4 h^{2}} + d \frac{u_{m+1}-2u_{m}+u_{m-1}}{h^{2}} , 
\end{equation}%
where $\alpha=2/11$, $c=3/11$, and $d=12/11$ for the CD$6$ scheme.  

\noindent The spatial coefficients for the $8^{th}$ order Pad\'{e} filter are given by%
\begin{equation}
\begin{array}{c c c}
d_{0} = \frac{93 + 70 \alpha_{f}}{128},  & \; \;  d_{1} = \frac{7 + 18 \alpha_{f}}{16}, & \; \;  d_{2} = \frac{-7 + 14 \alpha_{f}}{32},  \nonumber \\
d_{3} = \frac{1 -2 \alpha_{f}}{16},  &  \; \; d_{4} = \frac{-1 + 2 \alpha_{f}}{128} , & \; \; 
\end{array}
\end{equation}%
and the transfer function%
\begin{equation}
\mathcal{T}(K) = \frac{ d_{0} + d_{1} \cos(K) + d_{2} \cos(2K) +d_{3} \cos(3K) + d_{4} \cos(4K) }{1+ 2 \alpha_{f} \cos (K)} .
\label{eqn:Pade_F8_T}
\end{equation}%
%
\section*{Appendix C. Stability limits for DG, FD, and CD schemes}\label{appx:C}%
\addcontentsline{toc}{section}{Appendix C. Stability limits for DG, FD, and CD schemes}%
%
%
\begin{table}[H]%
\caption{Stability limits for DG schemes with different numerical fluxes and Runge-Kutta time integration schemes for the linear advection equation. Unstable schemes are indicated by $*$.} %
\centering 
\small
\begin{tabular}{|c|c|c|c|}
\hline
DG scheme, $p$ & RK scheme, $s$ & $\beta$ & CFL \\ 
\hline
\hline
 \multirow{6}{*}{$1$} & \multirow{2}{*}{$2$} & $0.00$ & $*$ \\ 
 \cline{3-4}
 & & $1.00$ & $0. \overline{333}$ \\ 
 \cline{2-4}
& \multirow{2}{*}{$3$} & $0.00$ & $0.433$ \\ 
 \cline{3-4}
 &  & $1.00$ & $0.409$\\ 
 \cline{2-4}
& \multirow{2}{*}{$4$} & $0.00$ & $0.707$ \\ 
\cline{3-4}
  & & $1.00$ & $0.464$\\ 
\hline
 \multirow{4}{*}{$2$} & \multirow{2}{*}{$3$} & $0.00$ & $0.210$ \\ 
 \cline{3-4}
 & & $1.00$ & $0.209$\\ 
 \cline{2-4}
& \multirow{2}{*}{$4$} & $0.00$ & $0.349$ \\ 
\cline{3-4}
 & & $1.00$ & $0.235$ \\ 
 \hline
 \multirow{4}{*}{$3$} & \multirow{2}{*}{$3$} & $0.00$ & $0.130$ \\ 
 \cline{3-4}
  & & $1.00$ & $0.130$\\ 
 \cline{2-4}
& \multirow{2}{*}{$4$} & $0.00$ & $0.210$ \\ 
\cline{3-4}
  & & $1.00$ & $0.145$ \\ 
 \hline
 \multirow{4}{*}{$4$} & \multirow{2}{*}{$3$} & $0.00$ & $0.088$ \\ 
 \cline{3-4}
  & & $1.00$ & $0.089$\\ 
 \cline{2-4}
& \multirow{2}{*}{$4$} & $0.00$ & $0.100$ \\ 
\cline{3-4}
  & & $1.00$ & $0.100$ \\ 
 \hline
 \multirow{4}{*}{$5$} & \multirow{2}{*}{$3$} & $0.00$ & $0.063$ \\ 
 \cline{3-4}
  & & $1.00$ & $0.066$\\ 
 \cline{2-4}
& \multirow{2}{*}{$4$} & $0.00$ & $0.103$ \\ 
\cline{3-4}
  & & $1.00$ & $0.073$ \\ 
 \hline
\end{tabular}
\label{table:stability_limits_DG_advection}
\end{table}%
%
%
\begin{table}[H]%
\caption{Stability limits for FD and CD schemes with Runge-Kutta time integration schemes for the linear advection equation.} %
\centering 
\small
\begin{tabular}{|c|c|c|}
\hline
Spatial scheme & RK scheme, $s$ &  CFL \\ 
\hline
\hline
\multirow{2}{*}{FD$2$-\textit{central}} & $2$ & $*$ \\
\cline{2-3}
& $3$ & $1.732$ \\
\cline{2-3} 
&  $4$ &  $2.828$ \\
\hline
\multirow{2}{*}{FD$3$-\textit{1point-biased}} & $3$ & $1.625$ \\
\cline{2-3}
&  $4$ & $1.745$ \\
\hline
\multirow{2}{*}{FD$4$-\textit{central}} & $3$ & $1.262$ \\
\cline{2-3}
& $4$ & $2.062$ \\
\hline
\multirow{2}{*}{FD$6$-\textit{central}}  & $3$ & $1.092$ \\
\cline{2-3}
& $4$ & $1.783$ \\
\hline
\multirow{2}{*}{FD$6$-\textit{2point-biased}}  & $3$ & $1.069$ \\
\cline{2-3}
& $4$ & $1.199$ \\
\hline
\multirow{2}{*}{CD$4$} & $3$ & $1.000$  \\
\cline{2-3}
&  $4$ &  $1.632$ \\
\hline
\multirow{2}{*}{CD$6$} & $3$ & $0.870$ \\
\cline{2-3}
& $4$ & $1.421$ \\
\hline
\end{tabular}
\label{table:stability_limits_FD_advection}
\end{table}%
%
%
\section*{References}
%
\bibliography{references}%
\addcontentsline{toc}{section}{References}%
\end{document}